\pdfoutput=1
\documentclass[pdflatex,sn-mathphys-ay]{sn-jnl}

\usepackage{graphicx}%
\usepackage{multirow}%
\usepackage{amsmath,amssymb,amsfonts}%
\numberwithin{equation}{section}
\usepackage{amsthm}%
\usepackage{mathrsfs}%
\usepackage[title]{appendix}%
\usepackage{xcolor}%
\usepackage{textcomp}%
\usepackage{manyfoot}%
\usepackage{booktabs}%
\usepackage{algorithm}%
\usepackage{algorithmicx}%
\usepackage{algpseudocode}%
\usepackage{listings}%
\usepackage{bm}
\usepackage{newtxtext}
\usepackage{natbib}
\usepackage{setspace}
\usepackage{ulem} 
\usepackage{accents} 
\usepackage{subfigure}
\theoremstyle{thmstyleone}%
\theoremstyle{thmstyletwo}%

\theoremstyle{thmstylethree}%

\begin{document}

\title[Article Title]{Quadruple decomposition of the boundary vorticity flux}


\author*[1]{\fnm{Tao} \sur{Chen}}\email{chentao2023@njust.edu.cn}

\affil*[1]{\orgdiv{School of Physics}, \orgname{Nanjing University of Science and Technology}, \orgaddress{ \city{Nanjing}, \postcode{210094}, \country{China}}}

\author[2]{\fnm{Tianshu} \sur{Liu}}

\affil[2]{\orgdiv{Department of Mechanical and Aerospace Engineering}, \orgname{Western Michigan University}, \orgaddress{\city{Kalamazoo}, \postcode{49008}, \state{Michigan}, \country{USA}}}


\abstract{First introduced by Lighthill in 1963 for two-dimensional flows and later generalized by Jie-Zhi Wu to three-dimensional scenarios since 1986, 
	the boundary vorticity flux (BVF) is the cornerstone of boundary vorticity dynamics, which quantifies the vorticity source strength on a solid boundary. Recent advances in vorticity and vortex dynamics have revealed both the rigid-rotation and spin modes of vorticity from multiple perspectives. In the present study, beyond the conventional single-vorticity research framework, we propose a novel quadruple decomposition of the BVF on a stationary solid wall, which essentially uncovers the boundary creation rates of the elementary vorticity modes for both the tangential and wall-normal BVF components, respectively. The theory is further extended to the decomposition of the boundary enstrophy flux (BEF), which reveals the boundary sources of various enstrophy constituents. The proposed framework is illustrated through skin-friction and surface-pressure measurements for flow over a hill model in a low-speed wind tunnel, revealing a set of intriguing patterns exhibited by the BVF and BEF constituents. These theoretical results are expected to be valuable for global surface flow diagnostics when combined with experiments, as well as for understanding the formation mechanisms of near-wall coherent structures and flow-induced noise.}


\keywords{Boundary vorticity dynamics; Boundary vorticity flux; Boundary enstrophy flux; Quadruple decomposition; Global surface flow diagnostics}



\maketitle
\tableofcontents
\section{Introduction}\label{Introduction}
Vorticity, which is defined as the curl of the velocity field $(\bm{\omega}\equiv\bm{\nabla}\times\bm{u})$, equals twice the angular velocity of a fluid volume element~\citep{Helmholtz1858,Truesdell1954}. 
It has been widely used to investigate coherent structures that are regarded as the sinews and muscles of complex fluid motions~\citep{Kuchemann1965report,Saffman1992}. The vorticity-based description has demonstrated its capacity to yield deeper insight and a more intuitive physical picture for the generation, mutual interaction, and spatiotemporal evolution of vortical flow structures~\citep{Batchelor1967}. Boundary vorticity dynamics is a vital branch of fluid mechanics that investigates the physical mechanisms of vorticity production on solid boundaries or fluid-fluid interfaces, as well as the interactions between vorticity and boundaries~\citep{WuWu1996,Wu2006vorticity}.

It has long been recognized that the most important source of vorticity is a solid boundary~\citep{Lamb1932hydrodynamics}. 
However, the concept of vorticity creation from a boundary did not become a formal theory until~\citet{Lighthill1963}, who laid the first cornerstone of boundary vorticity dynamics. For a two--dimensional (2D) incompressible viscous flow interacting with a stationary flat wall,~\citet{Lighthill1963} formally introduced the boundary vorticity flux (BVF) to measure the vorticity source strength (i.e., the rate of vorticity generation per unit area), defined as $\bm{\sigma}\equiv\nu\left[\partial_{n}\bm{\omega}\right]_{w}=\nu\bm{n}_{w}\bm{\cdot}\left[\bm{\nabla\omega}\right]_{w}$, where $\nu$ is the kinematic viscosity of fluid, $\bm{n}_{w}$ is the unit wall-normal vector directed into the fluid, and the subscript ``$w$'' indicates the restriction to the wall. By applying the Navier--Stokes (NS) equations at the wall and using the no-slip boundary condition, he derived an exact on-wall relation revealing vorticity creation by the surface pressure gradient, with its production rate characterized by the BVF. About two decades later,~\citet{Panton1984} extended Lighthill's original definition of the BVF to a flat wall in three--dimensional (3D) space. The BVF expression on a curved wall was first established by~\citet{Wu1986} for 3D incompressible viscous flow.
The full expression of the BVF was eventually derived by~\citet{Wu1993} for 3D compressible viscous flow interacting with an arbitrarily moving and deformable wall, and later generalized to a material interface separating two immiscible fluids~\citep{Wu1995}. A comprehensive review was provided in~\citet{WuWu1996}. It should be noted that \citet{Hornung1989} independently developed a generalized BVF theory valid for any continuous medium by utilizing the Stokes~--~Helmholtz decomposition of the divergence of the total stress tensor. Rather than employing $\bm{\sigma}$ to quantify the boundary vorticity creation rate,~\citet{Lyman1990vorticity} proposed an alternative BVF, $\bm{\sigma}_{L}\equiv-\nu\left[\bm{n}\times\left(\bm{\nabla}\times\bm{\omega}\right)\right]_{w}$ (referred to as the Lyman flux), which directly incorporates the purely 3D viscous contribution into the definition of $\bm{\sigma}$. Although $\bm{\sigma}$ and $\bm{\sigma}_{L}$ differ in describing the local rate of vorticity creation on a boundary, their integrals over a closed surface are identical. It is worth noting that these two BVFs arise from distinct definitions of the viscous vorticity current tensor in the vorticity transport equation. Consequently, they offer two different yet equally valid dynamical interpretations of vorticity and enstrophy dynamics: the Lighthill--Panton--Wu and Lyman--Huggins interpretations~\citep{Terrington2023LH,Chen2021features}.
The kinematic expressions of the BVF have been derived in an orthogonal coordinate system aligned with the surface principal directions~\citep{Xie2020theory}. Recent BVF-related studies have focused on vorticity dynamics in the presence of a deformable boundary~\citep{WuWangWu2007,ChenXie2016}, the boundary enstrophy flux (BEF) and surface footprints~\citep{LiuTS2016,Chen2019Relations,Chen2021features,LiuTS2019PAS,Miozzi2019}, the Lamb vector divergence and its boundary flux~\citep{Chen2023PhysicaD}, as well as the Lie derivatives of fundamental surface quantities with respect to the celerity of near-wall coherent structures~\citep{Chen2023Lie}. Further studies include interfacial vorticity dynamics in the Navier-Stokes-Korteweg system with a diffuse interface possessing finite thickness~\citep{Chen2024IJMF}, the generalized interface circulation dynamics~\citep{Terrington2022JFM}, and the connection with Josephson-Anderson relation~\citep{Eyink2021,DuZaki2025}.

Recent studies have revealed two distinct modes embedded within vorticity, i.e., the rigid-rotation mode and the spin/shear mode, either
through characteristic algebraic approaches~\citep{Kolar2007Vortex,Li2014APS,Liu2018Rortex,Chen2026Kinematic,Chen2026Lagrange} or by employing material and field descriptions~\citep{Chen2025AMM,Chen2025a,Chen2026General}. These studies have been proven useful in characterizing the fundamental forms of vortex structures, their interactions, and mutual conversion during spatiotemporal evolution processes. In the present work, by adopting a differential--geometric approach (\S\S~\ref{Rational} and~\ref{Differental}), after a brief overview of the BVF in \S~\ref{Expression}, we propose a quadruple decomposition of the BVF in~\S\S~\ref{Decomposition of the tangential BVF} and~\ref{Decomposition of the wall-normal BVF} to quantify the boundary sources of each vorticity mode. This naturally leads to the decomposition of the total force exerted on a closed solid boundary in \S~\ref{force_decomposition}. The theory is applied to global surface flow diagnostics of flow over a hill model in~\S\S~\ref{Global surface flow diagnostics in a wind tunnel} and~\ref{Results and analysis}. In \S\ref{BEF_decomp}, the decomposition is extended to the BEF. Conclusions are made in \S\ref{Conclusions and discussions}. Appendix~\ref{geodesic} gives a novel representation on the geodesic curvature of a skin friction line. Appendix~\ref{Near-wall Taylor-series expansion and the decomposition of the wall-normal BVF} discusses the relationship between the near-wall Taylor-series expansion and the wall-normal BVF decomposition. Appendix~\ref{geometry} provides a detailed geometry analysis of the hill model, showing its spatial configuration and geometric characteristics. Appendix~\ref{Transformation of vorticity components} derives the transformation of vorticity components across different bases. Appendix~\ref{literature} clarifies the relationship between the present work and the recent study by~\citet{Chen2025AMM}.

\section{Surface geometry and related differential operators}
\subsection{Rational description of the boundary and its neighborhood}\label{Rational}
Consider a stationary solid wall with the surface $\bm{\Sigma}_{w}$ locally parameterized by a curvilinear coordinate system, $\bm{x}=(x^{1},x^{2})\in\mathscr{D}_{\bm{x}}\subset\mathbb{R}^{2}\mapsto\bm{\Sigma}_{w}(\bm{x})\in\mathbb{R}^{3}$~\citep{Aris1962vectors,carmo2016differential,Chen2017DG}, where the domain of definition $\mathscr{D}_{\bm{x}}$ is a 2D open set, and the subscript ``$w$'' denotes the wall. At any point $P(\bm{x})\in\bm{\Sigma}_{w}$, the local covariant basis vectors are given by $\widehat{\bm{g}}_{\alpha}(\bm{x})\equiv{\partial_{\alpha}\bm{\Sigma}_{w}}(\bm{x})~(\alpha=1,2)$, which represent the tangent vectors of the coordinate curves (i.e., $x^\alpha$--curves), with $\partial_{\alpha}\equiv\partial/\partial x^\alpha$ denoting the partial derivative. The tangent plane at $P(\bm{x})$ is $T_{P}\bm{\Sigma}_{w}\equiv{\rm Span}\left\{\widehat{\bm{g}}_{1},\widehat{\bm{g}}_{2}\right\}$. The surface metric tensor (i.e., the first fundamental form of $\bm{\Sigma}_{w}$) is given by
${\bm{G}_w}\equiv\widehat{g}_{\alpha\beta}\widehat{\bm{g}}^{\alpha}\widehat{\bm{g}}^{\beta}$, where the inner products $\widehat{g}_{\alpha\beta}\equiv\widehat{\bm{g}}_{\alpha}\bm{\cdot}\widehat{\bm{g}}_{\beta}~(\alpha,\beta=1,2)$ define its covariant components. The contravariant basis vectors are defined by $\widehat{\bm{g}}^{\alpha}\equiv\widehat{g}^{\alpha\beta}\widehat{\bm{g}}_{\beta}$, where the contravariant components are uniquely determined via compatibility relation $\widehat{g}^{\alpha\beta}\widehat{g}_{\beta\gamma}=\delta_{\gamma}^{\alpha}$. Here, $\delta_{\gamma}^{\alpha}$ is the Kronecker symbol. The unit normal vector of $\bm{\Sigma}_{w}$ is uniquely determined through $\bm{n}_{w}(\bm{x})\equiv\widehat{\bm{g}}_{1}\times\widehat{\bm{g}}_{2}/\lVert\widehat{\bm{g}}_{1}\times\widehat{\bm{g}}_{2}\rVert$.
If the surface coordinate system is orthogonal, a surface-attached orthonormal triad $\left\{\widehat{\bm{e}}_{1},\widehat{\bm{e}}_{2},\widehat{\bm{e}}_{3}\right\}$ can be introduced on $\bm{\Sigma}_{w}$, with
$\widehat{\bm{e}}_{\alpha}\equiv{\widehat{\bm{g}}_{\alpha}}/{\lVert\widehat{\bm{g}}_{\alpha}\rVert}~(\alpha=1,2)$ and $\widehat{\bm{e}}_{3}\equiv\bm{n}_{w}$.
The surface curvature tensor (i.e., the
second fundamental form of $\bm{\Sigma}_{w}$) is given by ${\bm{K}_w}\equiv \widehat{b}_{\alpha\beta}\widehat{\bm{g}}^{\alpha}\widehat{\bm{g}}^{\beta}$, where $\widehat{b}_{\alpha\beta}=\partial_{\beta}\widehat{\bm{g}}_{\alpha}\bm{\cdot}\bm{n}_{w}$, and $\widehat{b}_{\beta}^{\alpha}=\widehat{\bm{g}}_{\beta}\bm{\cdot}{\bm{K}_w}\bm{\cdot}\widehat{\bm{g}}^{\alpha}=\widehat{g}^{\alpha\gamma}\widehat{b}_{\beta\gamma}$. The scalar ${K}_{w}\equiv{\rm tr}({\bm{K}_w})=\widehat{b}^{\alpha}_{\alpha}$ is twice the mean curvature.
\begin{figure}[htbp]
	\centering
	\includegraphics[width=0.9\columnwidth,trim={0cm 4cm 0cm 3.5cm},clip]{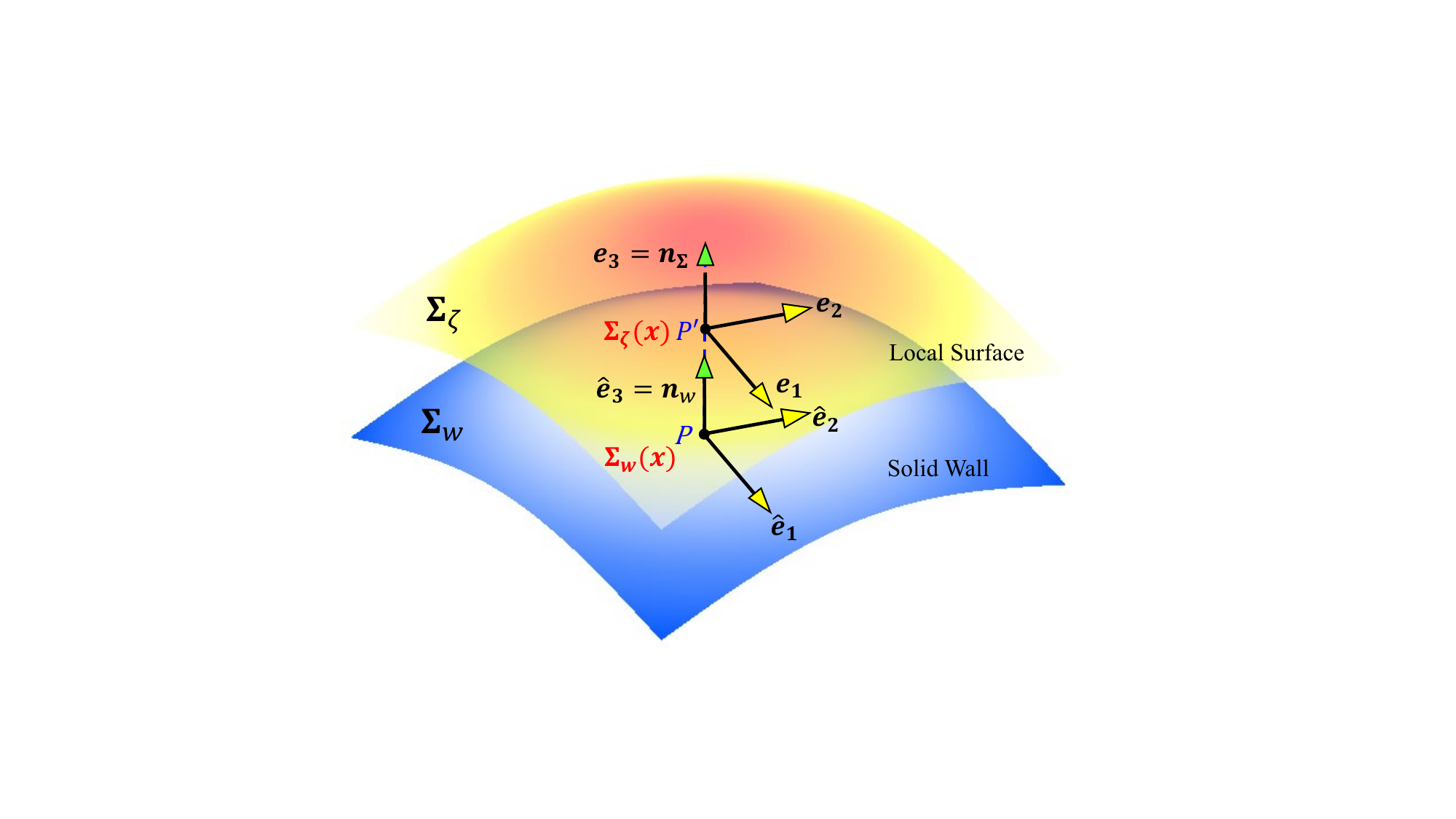}
	\caption{Schematic of the local surface, $\bm{\Sigma}_{\zeta}(\bm{x})=\bm{\Sigma}_{w}(\bm{x})+\zeta\bm{n}_{w}(\bm{x})$, in a small neighborhood of the solid wall $\bm{\Sigma}_{w}$ (base surface), where $P\in\bm{\Sigma}_{w}$ and $P^{\prime}\in\bm{\Sigma}_{\zeta}$. The unit surface-normal vectors satisfy $\bm{n}_{\Sigma}=\bm{n}_{w}$.} 
	\label{boundary_surface}
\end{figure}

By performing the normal extension into a small neighborhood of $\bm{\Sigma}_{w}$ (figure~\ref{boundary_surface}), a 3D orthogonal coordinate system $(\bm{x},x^3)=(x^1,x^2,\zeta)\in\mathscr{D}_{\bm{x}}\times(-\delta,\delta)$ is constructed such that $\bm{\Sigma}(\bm{x},\zeta)=\bm{\Sigma}_{w}(\bm{x})+\zeta\bm{n}_{w}(\bm{x})$, where $\delta$ is a sufficiently small positive number~\citep{huang2003tensoranalysis,Xie2020theory}. The local covariant basis vectors of $(\bm{x},\zeta)$ are evaluated as ${\bm{g}}_{\alpha}(\bm{x},\zeta)\equiv\partial_{\alpha}\bm{\Sigma}(\bm{x},\zeta)=(\delta_{\alpha}^{\beta}-\zeta\widehat{b}_{\alpha}^{\beta})\widehat{\bm{g}}_{\beta}(\bm{x})$, and ${\bm{g}}_{3}(\bm{x},\zeta)\equiv{\partial_{\zeta}\bm{\Sigma}}(\bm{x},\zeta)=\bm{n}_{w}(\bm{x})=\bm{n}_{\Sigma}(\bm{x})$ is the unit normal of the local surface $\bm{\Sigma}_{\zeta}(\bm{x})\equiv\bm{\Sigma}(\bm{x},\zeta)$ at a fixed $\zeta$. The corresponding local contravariant basis is $\left\{\bm{g}^{i}\vert i=1,2,3\right\}=\left\{\bm{g}^{\alpha}\vert \alpha=1,2\right\}\cup\bm{n}_{\Sigma}$. Obviously, $\bm{\Sigma}_{\zeta}$ approaches $\bm{\Sigma}_{w}$ as $\zeta\rightarrow0^+$. 
In general, the orthogonality of $\widehat{\bm{g}}_{1}$ and $\widehat{\bm{g}}_{2}$ does not imply that of ${\bm{g}}_{1}$ and ${\bm{g}}_{2}$, unless $\widehat{\bm{g}}_{1}$ and $\widehat{\bm{g}}_{2}$ align with the principal directions of $\bm{\Sigma}_{w}$, in which case they also align with those of $\bm{\Sigma}_{\zeta}$. If so, one can introduce on $\bm{\Sigma}_{\zeta}$ that ${\bm{e}}_{\alpha}\equiv{{\bm{g}}_{\alpha}}/{\lVert{\bm{g}}_{\alpha}\rVert}=\widehat{\bm{e}}_{\alpha}~(\alpha=1,2)$, and $\bm{e}_{3}\equiv\bm{n}_{\Sigma}=\bm{n}_{w}=\widehat{\bm{e}}_{3}$.
The metric and curvature tensors of $\bm{\Sigma}_{\zeta}$ are denoted by $\bm{G}=g_{\alpha\beta}\bm{g}^{\alpha}\bm{g}^{\beta}$ and $\bm{K}=b_{\alpha\beta}\bm{g}^{\alpha}\bm{g}^{\beta}$, respectively.

\subsection{Differential operators on the local surface}\label{Differental}
\subsubsection{Surface gradient operator}
On the local surface $\bm{\Sigma}_{\zeta}$, the surface gradient operator is introduced as
\begin{eqnarray}\label{mf1}
	\bm{\nabla}_{\pi}\equiv\bm{g}^{\alpha}\partial_{\alpha}.
\end{eqnarray}
From~\eqref{mf1}, it follows that the surface curvature tensor $\bm{K}$ and the scalar curvature $K$ can be expressed intrinsically as
\begin{equation}\label{KK}
	\bm{K}=-\bm{\nabla}_{\pi}\bm{n}_{\Sigma},~K\equiv{\rm tr}(\bm{K})=-\bm{\nabla}_{\pi}\bm{\cdot}\bm{n}_{\Sigma}.
\end{equation}
Furthermore, the full spatial gradient operator admits the orthogonal decomposition
\begin{equation}
\bm{\nabla}=\bm{\nabla}_{\pi}+\bm{n}_{\Sigma}\partial_{n},
\end{equation}
where $\partial_{n}=\partial_{\zeta}=\bm{n}_{\Sigma}\bm{\cdot}\bm{\nabla}=\partial/\partial{n}$ denotes the surface-normal derivative operator. 
\subsubsection{Levi-Civita gradient operator}\label{Sec_LCC}
Since $(\bm{\Sigma}_{\zeta},\bm{G})$ can be viewed as a 2D Riemannian submanifold of $\mathbb{R}^{3}$, the Levi-Civita connection (also termed the induced connection) ${\nabla}_{\partial_{\alpha}}$ can be introduced on $\bm{\Sigma}_{\zeta}$, being effective only to the indices corresponding to the tangent bundle $T\bm{\Sigma}_{\zeta}$~\citep{Dubrovin1992modern}. For the tangent and cotangent vector fields $\{\partial_{\alpha},dx^{\beta}\}$, two elegant properties then follow: $	{\nabla}_{\partial_{\alpha}}\partial_{\beta}=\Gamma_{\alpha\beta}^{\gamma}\partial_{\gamma}$ and ${\nabla}_{\partial_{\alpha}}d{x}^{\beta}=-\Gamma_{\alpha\gamma}^{\beta}d{x}^{\gamma},$
where $\Gamma_{\alpha\beta,\gamma}$ and
$\Gamma_{\alpha\beta}^{\gamma}\equiv{g}^{\gamma\mu}\Gamma_{\alpha\beta,\mu}$ denote the Christoffel connection coefficients of the first and second kinds, respectively. It is noted that $\Gamma_{\alpha\beta,\gamma}$ are determined by the covariant components $g_{\alpha\beta}$ of the metric tensor~\citep{Chen2017DG}:
\begin{equation}
\Gamma_{\alpha\beta,\gamma}
=\frac{1}{2}\left(
\partial_{\alpha}g_{\beta\gamma}
+\partial_{\beta}g_{\alpha\gamma}
-\partial_{\gamma}g_{\alpha\beta}
\right).
\end{equation}
By virtue of the Weingarten mapping for the unit normal $\bm{n}_{\Sigma}$, treating $\{\partial_{\alpha},dx^{\beta}\}$ as equivalent to $\{\bm{g}_{\alpha},\bm{g}^{\beta}\}$ results in the frame-movement equations, known in differential geometry as the Gauss-Weingarten-Codazzi (GWC) formulas~\citep{Dubrovin1992modern,Chen2017DG,Xie2020theory}. Then, the Levi-Civita gradient operator on $\bm{\Sigma}_{\zeta}$ is defined as~\citep{XIE2013688}
\begin{eqnarray}\label{LC}
	\bm{\nabla}_{C}\equiv\bm{g}^{\alpha}{\nabla}_{\partial_\alpha}.
\end{eqnarray}
On $\bm{\Sigma}_{w}$, the two differential operators are denoted by $\widehat{\bm{\nabla}}_{\pi}$ and $\widehat{\bm{\nabla}}_{C}$, respectively. Equation~\eqref{KK} becomes
\begin{equation}\label{KK1}
\bm{K}_{w}=-\widehat{\bm{\nabla}}_{\pi}\bm{n}_{w},~K_{w}\equiv{\rm tr}(\bm{K}_{w})=-\widehat{\bm{\nabla}}_{\pi}\bm{\cdot}\bm{n}_{w}.
\end{equation}

\section{Quadruple decomposition of the boundary vorticity flux}
\subsection{Boundary vorticity flux on a stationary no-slip wall}\label{Expression}
On a solid wall $\bm{\Sigma}_{w}$, by exploiting the solenoidal property of vorticity, the BVF can generally be split into two parts~\citep{Chen2021features}:
\begin{eqnarray}\label{a88}
	\bm{\sigma}\equiv\nu\left[\partial_{n}\bm{\omega}\right]_{w}=\nu\left[(\bm{n}\times\bm{\nabla})\times\bm{\omega}\right]_{w}-\nu\left[\bm{n}\times\left(\bm{\nabla}\times\bm{\omega}\right)\right]_{w}.
\end{eqnarray}
Restricting the NS equations to the wall and using the velocity adherence condition $(\bm{u}_{w}=\bm{0})$, the evaluation of the two terms on the right-hand side of~\eqref{a88} yields the expression of the BVF~\citep{Wu1993,Wu2006vorticity}:
\begin{subequations}\label{q15abc}
	\begin{gather}
		\bm{\sigma}=\bm{\sigma}_{L}+\bm{\sigma}_{vis}, \label{q15a} \\
		\bm{\sigma}_{L}\equiv-\nu\left[\bm{n}\times\left(\bm{\nabla}\times\bm{\omega}\right)\right]_{w}
		=\bm{n}_{w}\times\widehat{\bm{\nabla}}_{\pi}\widetilde{P}_{w},~\widetilde{P}\equiv (p-\mu_{\vartheta}\vartheta)/\rho, \label{q15b} \\
		\bm{\sigma}_{vis}\equiv\nu[(\bm{n}\times\bm{\nabla})\times\bm{\omega}]_{w}
		=\nu\bm{K}_{w}\bm{\cdot}\bm{\omega}_{w}
		-\nu(\widehat{\bm{\nabla}}_{\pi}\bm{\cdot}\bm{\omega}_{w})\bm{n}_{w}, \label{q15c}
	\end{gather}
\end{subequations}
where $p$ is the pressure, $\rho$ is the fluid density, and $\vartheta\equiv\bm{\nabla}\bm{\cdot}\bm{u}$ is the dilatation; $\mu_{\vartheta}\equiv\mu_{b}+(4/3)\mu$, with $\mu_{b}$ the bulk viscosity and $\mu$ the dynamic viscosity. In~\eqref{q15b}, the Lyman flux
$\bm{\sigma}_{L}$~\citep{Lyman1990vorticity} unravels a local dynamic causal mechanism of boundary vorticity creation due to the surface pressure gradient and the surface acceleration. The latter vanishes due to $\bm{u}_{w}=\bm{0}$.
In~\eqref{q15c}, $\bm{\sigma}_{vis}$ describes the 3D viscous contributions to the BVF, arising from the boundary geometry--vorticity coupling, and the surface vorticity divergence (or the skin-friction curl).  
From~\eqref{q15abc},  we obtain an orthogonal decomposition of $\bm{\sigma}$ as (where $\bm{\sigma}_{n}=\sigma_{n}\bm{n}_{w}$)
\begin{subequations}\label{BVF_pi_n}
	\begin{gather}
		\bm{\sigma}=\bm{\sigma}_{\pi}+\bm{\sigma}_{n}, \label{BVF0} \\
		\bm{\sigma}_{\pi}\equiv\nu\left[\partial_{n}\bm{\omega}_{\pi}\right]_{w}
		=\bm{n}_{w}\times\widehat{\bm{\nabla}}_{\pi}\widetilde{P}_{w}
		+\nu\bm{K}_{w}\bm{\cdot}\bm{\omega}_{w}, \label{BVF1} \\
		\bm{\sigma}_{n}\equiv\nu\left[\partial_{n}\bm{\omega}_{n}\right]_{w}
		=-\nu(\widehat{\bm{\nabla}}_{\pi}\bm{\cdot}\bm{\omega}_{w})\bm{n}_{w}=\nu(\bm{n}_{w}\bm{\cdot}\widehat{\bm{\nabla}}_{\pi}\times\bm{\xi})\bm{n}_{w}. \label{BVF2}
	\end{gather}
\end{subequations}
The tangential BVF $\bm{\sigma}_{\pi}$ arises dynamically from the surface pressure gradient and the geometry--vorticity coupling, a process known as the ``ascending mechanism'': owing to diffusion, vorticity lines
infinitesimally close to the surface spread out from, and remain parallel to, the body surface~\citep{Lighthill1963,Wu1993,WuWu1996}. The wall-normal BVF $\bm{\sigma}_{n}$ originates from the solenoidal property of the vorticity field and represents
the ``upturning mechanism'', which is characterized by the surface vorticity divergence: vorticity lines that lie close to, and are originally parallel to, the solid surface tend to turn toward the normal direction~\citep{Wu1993}. This places $\bm{\sigma}_{n}$ at the intersection of the kinematics and dynamics of vorticity on boundaries. 
Typically, around a focus on $\bm{\Sigma}_{w}$ associated with a tornado-like vortex, $\bm{\sigma}_{n}$ appears together with converging boundary vorticity lines $(\bm{\omega}_{w}\text{-lines})$ and swirling skin-friction lines $(\bm{\tau}\text{-lines})$.

\subsection{Decomposition of the tangential boundary vorticity flux}\label{Decomposition of the tangential BVF}
On the local surface $\bm{\Sigma}_{\zeta}$, the tangential vorticity component $\bm{\omega}_{\pi}$ admits the decomposition~\citep[Eqs. (4.13) and (4.14)]{Chen2026General}:
\begin{subequations}\label{SEB}
	\begin{gather}
		\bm{\omega}_{\pi}=\bm{R}_{\Sigma}(\bm{n}_{\Sigma})+\bm{S}_{\Sigma}(\bm{n}_{\Sigma}), \label{ZZ1} \\
		\bm{R}_{\Sigma}(\bm{n}_{\Sigma})\equiv-2\bm{n}_{\Sigma}\times\left(\bm{A}\bm{\cdot}\bm{n}_{\Sigma}\right),~
		\bm{S}_{\Sigma}(\bm{n}_{\Sigma})\equiv2\bm{n}_{\Sigma}\times\left(\bm{D}\bm{\cdot}\bm{n}_{\Sigma}\right). \label{ss10a}
	\end{gather}
\end{subequations}
Here, $\bm{R}_{\Sigma}(\bm{n}_{\Sigma})$ is referred to as the surface-element-based rigid-rotation mode, and $\bm{S}_{\Sigma}(\bm{n}_{\Sigma})$ the surface-element-based spin/shear mode. It can be shown that $\bm{R}_{\Sigma}(\bm{n}_{\Sigma})=2\bm{W}_{\Sigma}^{\rm eff}(\bm{n}_{\Sigma})$, where $\bm{W}_{\Sigma}^{\rm eff}(\bm{n}_{\Sigma})\equiv-\bm{n}_{\Sigma}\times\left(\bm{A}\bm{\cdot}\bm{n}_{\Sigma}\right)$ is the effective angular velocity of the unit normal $\bm{n}_{\Sigma}$, satisfying $D\bm{n}_{\Sigma}/Dt=\bm{W}_{\Sigma}^{\rm eff}(\bm{n}_{\Sigma})\times\bm{n}_{\Sigma}$.
Using the generalized Caswell formula~\citep{Caswell1967,WuJZ2005JFM,Wu2006vorticity},
\citet[Eq. (6.3)]{Chen2026General} showed that ${\bm{S}}_{\Sigma}(\bm{n}_{\Sigma})=\bm{\omega}_{r}$, where the relative vorticity $\bm{\omega}_{r}$ determines the surface shear stress $\bm{\tau}_{\Sigma}=\mu\bm{\omega}_{r}\times\bm{n}_{\Sigma}$. Note that $\bm{A}\equiv\bm{\nabla}\bm{u}$ is the velocity gradient tensor (VGT), and $\bm{D}\equiv\mathscr{S}[\bm{A}]=(\bm{A}+\bm{A}^{\rm T})/2$ is the strain-rate tensor, where $\mathscr{S}$ denotes the symmetrization operator. On $\bm{\Sigma}_{w}$, $\bm{u}_{w}=\bm{0}$ implies $[\bm{R}_{\Sigma}]_{w}=\bm{0}$, and hence $\bm{\omega}_{w}=[\bm{S}_{\Sigma}]_{w}$.

Using the differential-geometric techniques presented in~\S\S~\ref{Rational} and~\ref{Differental} and enforcing the NS equations on the no-slip wall, the wall-normal derivative of the VGT $\bm{A}$ is derived as
\begin{eqnarray}\label{L16}
	\left[\partial_{n}\bm{A}\right]_{w}&=&\bm{n}_{w}\bm{n}_{w}(\left[\partial_{n}\vartheta\right]_{w}+K_{w}\vartheta_{w}-\widehat{\bm{\nabla}}_{\pi}\bm{\cdot}\bm{\xi})+\bm{n}_{w}(K_{w}\bm{\xi}+\nu^{-1}\widehat{\bm{\nabla}}_{\pi}\widetilde{P}_{w})\nonumber\\
	& &+(\widehat{\bm{\nabla}}_{\pi}\vartheta_{w})\bm{n}_{w}+\bm{n}_{w}(\widehat{\bm{\nabla}}_{\pi}\vartheta_{w})-\vartheta_{w}\bm{K}_{w}+\widehat{\bm{\nabla}}_{\pi}\bm{\xi}.
\end{eqnarray}
Here, $\bm{\xi}\equiv\bm{\omega}_{w}\times\bm{n}_{w}$ denotes the vector orthogonal to $\bm{\omega}_{w}$ on $\bm{\Sigma}_{w}$, and $\bm{\tau}=\mu\bm{\xi}$ is the skin-friction vector. $\vartheta_{w}\equiv\left[\bm{\nabla}\bm{\cdot}\bm{u}\right]_{w}=[\partial_{n}u_{n}]_{w}$ is the dilatation at the wall, and $[\partial_{n}\vartheta]_{w}$ is the boundary dilatation flux. Evaluating the wall-normal derivatives of~\eqref{ZZ1} and~\eqref{ss10a} at the wall and using~\eqref{L16},  
we obtain a double decomposition of the tangential BVF $\bm{\sigma}_{\pi}$ in~\eqref{BVF1}:	
\begin{subequations}\label{L17ab}
	\begin{gather}
	\bm{\sigma}_{\pi}=\bm{\sigma}_{\pi}^{R}+\bm{\sigma}_{\pi}^{S},~\bm{\sigma}_{\pi}^{R}\equiv\nu\left[\partial_{n}\bm{R}_{\Sigma}\right]_{w},~\bm{\sigma}_{\pi}^{S}\equiv\nu\left[\partial_{n}\bm{S}_{\Sigma}\right]_{w}, \label{L17a} \\
		\bm{\sigma}_{\pi}^{R}
		=-2\nu\bm{n}_{w}\times\widehat{\bm{\nabla}}_{\pi}\vartheta_{w}
		-2\nu\breve{\bm{K}}_{w}\bm{\cdot}\bm{\omega}_{w}, \label{L17b} \\
		\bm{\sigma}_{\pi}^{S}
		=2\nu\bm{n}_{w}\times\widehat{\bm{\nabla}}_{\pi}\vartheta_{w}
		+2\nu\breve{\bm{K}}_{w}\bm{\cdot}\bm{\omega}_{w}
		+\nu\bm{K}_{w}\bm{\cdot}\bm{\omega}_{w}
		+\bm{n}_{w}\times\widehat{\bm{\nabla}}_{\pi}\widetilde{P}_{w}. \label{L17c}
	\end{gather}
\end{subequations}
In~\eqref{L17ab}, ${\bm{\sigma}}_{\pi}^{R}$ is referred to as the surface-element-based boundary rigid-rotation flux (or simply the boundary $\bm{R}_{\Sigma}$ flux), and ${\bm{\sigma}}_{\pi}^{S}$ analogously as the surface-element-based boundary spin flux (or the boundary $\bm{S}_{\Sigma}$ flux for short). We notice that the conjugate
curvature tensor $\breve{\bm{K}}_{w}\equiv K_{w}\bm{G}_{w}-\bm{K}_{w}$ can be introduced to compactify the expressions; it plays an essential role in constructing a pair of fundamental differential operators on curved surfaces and in the shape-driving forces on biomembrane surfaces~\citep{Yin2008symmetrical,Yin2013}. Both ${\bm{\sigma}}_{\pi}^{R}$ and ${\bm{\sigma}}_{\pi}^{S}$ can be generated by the surface dilatation gradient $(\pm2\nu\bm{n}_{w}\times\widehat{\bm{\nabla}}_{\pi}\vartheta_{w})$ and the vorticity~--~curvature interaction ($\pm2\nu\breve{\bm{K}}_{w}\bm{\cdot}\bm{\omega}_{w}$ and $\nu{\bm{K}}_{w}\bm{\cdot}\bm{\omega}_{w}$), whereas the surface pressure gradient (i.e., the Lyman flux $\bm{\sigma}_{L}\equiv\bm{n}_{w}\times\widehat{\bm{\nabla}}_{\pi}\widetilde{P}_{w}$) contributes solely to ${\bm{\sigma}}_{\pi}^{S}$. These physical mechanisms can be categorized as ``ascending mechanisms''. It is emphasized that~\citet{Wu1993} identified only one ascending mechanism, $\nu{\bm{K}}_{w}\bm{\cdot}\bm{\omega}_{w}$ (of the ${\bm{K}}_{w}-\bm{\omega}_{w}$ type), induced by the wall curvature for $\bm{\sigma}_{\pi}$. By contrast,~\eqref{L17b} and~\eqref{L17c} further reveal a new ascending mechanism of the $\breve{\bm{K}}_{w}-\bm{\omega}_{w}$ type (i.e., the terms $\pm2\nu\breve{\bm{K}}_{w}\bm{\cdot}\bm{\omega}_{w}$) for both $\bm{\sigma}_{\pi}^{R}$ and $\bm{\sigma}_{\pi}^{S}$; the ${\bm{K}}_{w}-\bm{\omega}_{w}-$type ascending mechanism of~\citet{Wu1993} contributes solely to $\bm{\sigma}_{\pi}^{S}$.  For incompressible flow $(\vartheta=0)$, ${\bm{\sigma}}_{\pi}^{R}$ exists only in the presence of a curved wall, which also holds for steady flows because the continuity equation implies $\vartheta_{w}=-\rho_{w}^{-1}\partial_{t}\rho_{w}=0$. In the following discussion, for convenience, we denote
\begin{equation}\label{sigma_K}
\bm{\sigma}_{K}\equiv2\nu\breve{\bm{K}}_{w}\bm{\cdot}\bm{\omega}_{w}+\nu\bm{K}_{w}\bm{\cdot}\bm{\omega}_{w}=2\nu{K}_{w}\bm{\omega}_{w}-\nu\bm{K}_{w}\bm{\cdot}\bm{\omega}_{w}.
\end{equation}

\subsection{Decomposition of the wall-normal boundary vorticity flux}\label{Decomposition of the wall-normal BVF}
From an intrinsic perspective, as understood by an observer living on $\bm{\Sigma}_{\zeta}$ who cannot perceive the normal dimension, one can operate entirely within the 2D surface geometry and consider only the surface-parallel velocity field $\bm{u}_{\pi}$ on $\bm{\Sigma}_{\zeta}$ when evaluating $\bm{\omega}_{n}=\omega_{n}\bm{n}_{\Sigma}$. At each point $P(\bm{x})\in\bm{\Sigma}_{\zeta}$, we introduce an orthonormal triad $(\bm{t},\bm{t}_{\perp},\bm{n}_{\Sigma})$, where $\bm{t}\equiv\bm{u}_{\pi}/\tilde{q}$ is the unit tangent vector of a streamline $\mathscr{C}\subset\bm{\Sigma}_{\zeta}$, with $\tilde{q}\equiv\lVert\bm{u}_{\pi}\rVert$ denoting the velocity magnitude, and its orthogonal vector is $\bm{t}_{\perp}\equiv\bm{n}_{\Sigma}\times\bm{t}$.
Then, based on the Levi-Civita gradient operator $\bm{\nabla}_{C}$, we decompose $\bm{\omega}_{n}$ as
\begin{subequations}\label{q8abff}
	\begin{gather}
		\bm{\omega}_{n}=\bm{R}_{L}(\bm{t})+\bm{S}_{L}(\bm{t}), \label{q80ff} \\
		\bm{R}_{L}(\bm{t})=2\bm{t}\times(\bm{t}\bm{\cdot}\bm{\nabla}_{C}\bm{u}_{\pi}),~
		\bm{S}_{L}(\bm{t})=-2\bm{t}\times(\bm{t}\bm{\cdot}\mathscr{S}[\bm{\nabla}_{C}\bm{u}_{\pi}]). \label{q8aff}
	\end{gather}
\end{subequations}
Here, $\bm{R}_{L}(\bm{t})$ and $\bm{S}_{L}(\bm{t})$ are referred to as the streamline-based rigid-rotation and spin modes, respectively. The values of $\bm{R}_{L}$ and $\bm{S}_{L}$ on $\bm{\Sigma}_{w}$ are defined as their limiting values for $\zeta\rightarrow 0^+$. Both are generally parallel to $\bm{n}_{\Sigma}$ on $\bm{\Sigma}_{\zeta}$ and vanish on $\bm{\Sigma}_{w}$. The  $\bm{n}_{\Sigma}$-component of $\bm{R}_{L}(\bm{t})$ is $\bm{R}_{L}(\bm{t})\bm{\cdot}\bm{n}_{\Sigma}=2\bm{t}\bm{\cdot}\bm{\nabla}_{C}\bm{u}_{\pi}\bm{\cdot}\bm{t}_{\perp}=2{\kappa}_{g,t}\tilde{q}$, which equals twice the angular velocity of a local circular motion with the speed $\tilde{q}$ and the geodesic curvature $\kappa_{g,t}$ within $T_{P}\bm{\Sigma}_{\zeta}$. Hence, $\bm{R}_{L}(\bm{t})$ may also be referred to as the orbital-rotation mode. For a pair of instantaneous orthogonal infinitesimal material line elements lying on $\bm{\Sigma}_{\zeta}$ with unit vectors $(\bm{t},\bm{t}_{\perp})$, the instantaneous rate of change of their intersection angle in $\mathbb{R}^{3}$ is given by the sum of a term $\bm{S}_{L}(\bm{t})\bm{\cdot}\bm{n}_{\Sigma}=-2\bm{t}\bm{\cdot}\mathscr{S}[\bm{\nabla}_{C}\bm{u}_{\pi}]\bm{\cdot}\bm{t}_{\perp}$ and an additional term 
$2u_{n}\tau_{g}(\bm{e})=2u_{n}(\bm{t}\bm{\cdot}\bm{K}\bm{\cdot}\bm{t}_{\perp})$ arising from the coupling between the normal flow and the geodesic torsion $\tau_{g}(\bm{e})\equiv\bm{t}\bm{\cdot}\bm{K}\bm{\cdot}\bm{t}_{\perp}$ along $\bm{e}$. This additional term vanishes on $\bm{\Sigma}_{w}$ owing to the stationary and impermeable wall boundary condition. Such a contribution in $\mathbb{R}^3$ remains undetectable to a 2D observer confined to the local surface $\bm{\Sigma}_{\zeta}$. In addition, we note that $\bm{\nabla}_{C}\bm{u}_{\pi}$ can be replaced by $\bm{\nabla}_{\pi}\bm{u}_{\pi}$ in~\eqref{q8aff} for the evaluation of $\bm{R}_{L}(\bm{t})\bm{\cdot}\bm{n}_{\Sigma}$ and $\bm{S}_{L}(\bm{t})\bm{\cdot}\bm{n}_{\Sigma}$, owing to the identity $\bm{\nabla}_{\pi}\bm{u}_{\pi}=\bm{\nabla}_{C}\bm{u}_{\pi}+(\bm{K}\bm{\cdot}\bm{u}_{\pi})\bm{n}_{\Sigma}$. Equations~\eqref{q80ff} and~\eqref{q8aff} extend the direction-dependent vorticity decomposition (DVD), originally formulated for general flows in 2D Euclidean space $\mathbb{R}^2$~\citep{Chen2025a}, to a 2D curved surface in $\mathbb{R}^3$ based on the Levi-Civita connection.

On a stationary wall $\bm{\Sigma}_{w}$, skin friction $\bm{\tau}$ and surface vorticity $\bm{\omega}_{w}$ form an orthogonal pair at a regular point satisfying $\bm{\omega}_{w}\neq\bm{0}$ (or equivalently, $\bm{\xi}\neq\bm{0}$ or $\bm{\tau}\neq\bm{0}$), and one can introduce at such a point the right-handed $\bm{\tau}$-frame $\left(\bm{e}_{\tau},\bm{e}_{\omega}\right)\equiv\left(\bm{\tau}/\lVert\bm{\tau}\rVert,\bm{\omega}_{w}/\lVert\bm{\omega}_{w}\rVert\right)$, which satisfies $\bm{n}_{w}=\bm{e}_{\tau}\times\bm{e}_{\omega}$~\citep{Wu2000POF}. As ${\zeta}\rightarrow0^+$, $(\bm{t},\bm{t}_{\perp})$ approaches $(\bm{e}_{\tau},\bm{e}_{\omega})$ asymptotically. Then, using~\eqref{q80ff} and~\eqref{q8aff}, we propose a double decomposition of the wall-normal BVF $\bm{\sigma}_{n}$:
\begin{subequations}\label{L230ab}
	\begin{gather}
		\bm{\sigma}_{n}={\bm{\sigma}}_{n}^{R}+{\bm{\sigma}}_{n}^{S},~{\bm{\sigma}}_{n}^{R}\equiv\nu\left[\partial_{n}\bm{R}_{L}\right]_{w},~{\bm{\sigma}}_{n}^{S}\equiv\nu\left[\partial_{n}\bm{S}_{L}\right]_{w}, \label{L230} \\
		\bm{\sigma}_{n}^{R}=2\nu\bm{e}_{\tau}\times(\bm{e}_{\tau}\bm{\cdot}\widehat{\bm{\nabla}}_{C}\bm{\xi})=2\nu(\bm{e}_{\tau}\bm{\cdot}\widehat{\bm{\nabla}}_{\pi}\bm{\xi}\bm{\cdot}\bm{e}_{\omega})\bm{n}_{w}, \label{L23a} \\
		\bm{\sigma}_{n}^{S}=-2\nu\bm{e}_{\tau}\times(\bm{e}_{\tau}\bm{\cdot}\mathscr{S}[\widehat{\bm{\nabla}}_{C}\bm{\xi}])=-2\nu(\bm{e}_{\tau}\bm{\cdot}\mathscr{S}[\widehat{\bm{\nabla}}_{\pi}\bm{\xi}]\bm{\cdot}\bm{e}_{\omega})\bm{n}_{w}. \label{L23b}
	\end{gather}
\end{subequations}
Here, we denote by $\bm{\sigma}_{n}^{R}$ and $\bm{\sigma}_{n}^{S}$ the streamline-based boundary rigid-rotation flux (i.e., the boundary $\bm{R}_{L}$ flux) and the boundary spin flux (i.e., the boundary $\bm{S}_{L}$ flux), respectively. The integral lines of the $\bm{\tau}$-frame $(\bm{e}_{\tau},\bm{e}_{\omega})$ form a dense network on $\bm{\Sigma}_{w}$, in which the wall-normal components of $(\bm{\sigma}_{n}^{R},\bm{\sigma}_{n}^{S})$ are expressed as
\begin{subequations}\label{jj123}
	\begin{gather}
		\sigma_{n}=-\nu\widehat{\bm{\nabla}}_{\pi}\bm{\cdot}\bm{\omega}_{w}
		=-\nu(\partial_{\omega}\lVert\bm{\omega}_{w}\rVert-\kappa_{g,\xi}\lVert\bm{\omega}_{w}\rVert),
		\label{jj1}\\
		\sigma_{n}^{R}=2\nu\kappa_{g,\xi}\lVert\bm{\omega}_{w}\rVert,
		\label{jj2}\\
		\sigma_{n}^{S}=-\nu\partial_{\omega}\lVert\bm{\omega}_{w}\rVert
		-\nu\kappa_{g,\xi}\lVert\bm{\omega}_{w}\rVert,
		\label{jj3}
	\end{gather}
\end{subequations}
where $\kappa_{g,\tau}(\equiv\kappa_{g,\xi})$ is the geodesic curvature of a $\bm{\tau}$-line, and $\partial_{\omega}\equiv\partial/\partial{s}_{\omega}$, with $s_{\omega}$ the arc-length parameter of an $\bm{\omega}_{w}$-line. These mechanisms create nonzero wall-normal rigid-rotation and spin components in the near-wall region, which can be referred to as ``upturning mechanisms''. To demonstrate the role of the unit vector field $\bm{e}_{\tau}\equiv\bm{\tau}/\lVert\bm{\tau}\rVert$, $\sigma_{n}$ is expanded as
\begin{equation}\label{vv1}
\sigma_{n}=\nu\bm{n}_{w}\bm{\cdot}(\widehat{\bm{\nabla}}_{\pi}\lVert\bm{\omega}_{w}\rVert\times\bm{e}_{\tau})+\nu\lVert\bm{\omega}_{w}\rVert\bm{n}_{w}\bm{\cdot}(\widehat{\bm{\nabla}}_{\pi}\times\bm{e}_{\tau}).
\end{equation}
On the right-hand side of~\eqref{vv1}, the first term is determined by the tangential variation rate of the surface vorticity magnitude along a surface vorticity line, which evaluates to $-\nu\partial_{\omega}\lVert\bm{\omega}_{w}\rVert$. The second term, arising from the curl of $\bm{e}_{\tau}$, simplifies to $\nu\kappa_{g,\xi}\lVert\bm{\omega}_{w}\rVert$ where it is shown in Appendix~\ref{geodesic} that
\begin{equation}
\kappa_{g,\xi}=\bm{n}_{w}\bm{\cdot}(\widehat{\bm{\nabla}}_{\pi}\times\bm{e}_{\tau})=-\widehat{\bm{\nabla}}_{\pi}\bm{\cdot}\bm{e}_{\omega}.
\end{equation}
Therefore, the condition $\kappa_{g,\xi}>\partial_{\omega}\operatorname{ln}\lVert\bm{\omega}_{w}\rVert$ corresponds to the existence of a positive wall-normal vorticity component in a small neighborhood adjacent to the wall. Even in the absence of appreciable $\kappa_{g,\xi}$ (i.e., when the skin-friction line is approximately straight and exhibits no discernible spiraling), $\sigma_{n}^{S}$ may dominate $\sigma_{n}$ owing to a non-zero $\nu\partial_{\omega}\lVert\bm{\omega}_{w}\rVert$.
In practical experimental measurements, the critical points of a skin friction field are generally discretely distributed on a test surface. We may still perform the calculation of $(\bm{\sigma}_{n},\bm{\sigma}_{n}^R,\bm{\sigma}_{n}^{S})$ directly using~\eqref{L230ab}, and a finite number of critical points exert negligible influence on the observed surface patterns of the wall-normal BVF constituents. See Appendix~\ref{Near-wall Taylor-series expansion and the decomposition of the wall-normal BVF} for a general discussion based on the near-wall Taylor-series expansion of the NS equations. 

\subsection{Boundary vorticity flux and the total force on a closed boundary}\label{force_decomposition}
Consider an incompressible viscous flow interacting with a stationary rigid body $\mathscr{B}$ with the closed surface $\partial\mathscr{B}$ in $\mathbb{R}^3$. Since $\vartheta=0$, it holds that $\widetilde{P}=p/\rho$. According to~\citet{Wu1993}, the total force $\bm{F}$ acting on a closed boundary $\partial\mathscr{B}$ can be expressed solely in terms of the BVF:
\begin{equation}\label{total_force}
\bm{F}=\oint_{\partial\mathscr{B}}\rho\bm{X}\times\left(\frac{1}{2}\bm{\sigma}_{p}+\bm{\sigma}_{vis}\right)dS,
\end{equation}
where $\bm{X}$ is the position vector of a point on $\bm{\Sigma}_{w}$. In this case, the BVF component $\bm{\sigma}_{p}\equiv\bm{n}_{w}\times\widehat{\bm{\nabla}}_{\pi}({p}_{w}/\rho)$ equals the Lyman flux $\bm{\sigma}_{L}$ in~\eqref{q15b}, and $\bm{\sigma}_{vis}$ is given in~\eqref{q15c}. It can be shown that $\bm{\sigma}_{vis}=(\bm{n}_{w}\times\widehat{\bm{\nabla}}_{\pi})\bm{\cdot}(\bm{\tau}\bm{n}_{w})$ where $\bm{\tau}=\mu\bm{\xi}$ is the skin friction vector. This result indicates that the tangential variation of the surface pressure and vorticity along a boundary (i.e., the vorticity-creation rate) is directly responsible for the total force exerted on it. The surface distributions of $\bm{\sigma}_{p}$ and $\bm{\sigma}_{vis}$ are usually more localized than those of $p$ and $\bm{\omega}_{w}$~\citep{Wu1993}. Using~\eqref{q15abc},~\eqref{BVF_pi_n},~\eqref{L17ab},~\eqref{sigma_K} and~\eqref{L230ab},~\eqref{total_force} can be decomposed as
\begin{subequations}
\begin{equation}\label{total_force_decomp}
	\bm{F}=\bm{F}[\bm{\sigma}_{\pi}^{S}]+\bm{F}[\bm{\sigma}_{\pi}^{R}]+\bm{F}[\bm{\sigma}_{n}^{S}]+\bm{F}[\bm{\sigma}_{n}^{R}],
\end{equation}
\begin{equation}\label{total_force1}
	\bm{F}[\bm{\sigma}_{\pi}^{S}]=\oint_{\partial\mathscr{B}}\rho\bm{X}\times\left(\frac{1}{2}\bm{\sigma}_{p}+\bm{\sigma}_{K}\right)dS,
\end{equation}
\begin{equation}\label{total_force2}
	\bm{F}[\bm{\sigma}_{\pi}^{R}]=\oint_{\partial\mathscr{B}}\rho\bm{X}\times\bm{\sigma}_{\pi}^{R}dS,
\end{equation}
\begin{equation}\label{total_force3}
	\bm{F}[\bm{\sigma}_{n}^{S}]=\oint_{\partial\mathscr{B}}\rho\bm{X}\times\bm{\sigma}_{n}^{S}dS,
\end{equation}
\begin{equation}\label{total_force4}
	\bm{F}[\bm{\sigma}_{n}^{R}]=\oint_{\partial\mathscr{B}}\rho\bm{X}\times\bm{\sigma}_{n}^{R}dS.
\end{equation}	
\end{subequations}
Here, the four force constituents on the right-hand side of~\eqref{total_force_decomp} arise respectively from the four BVF constituents enclosed in square brackets. These formulas enable a more detailed diagnosis of aerodynamic force, separating contributions from boundary fluxes associated with distinct rigid-body rotation and spin modes. Note that the sum of~\eqref{total_force3} and~\eqref{total_force4} yields
\begin{equation}
	\bm{F}[\bm{\sigma}_{n}]\equiv\oint_{\partial\mathscr{B}}\rho\bm{X}\times\bm{\sigma}_{n}dS=\bm{F}[\bm{\sigma}_{n}^R]+\bm{F}[\bm{\sigma}_{n}^S].
\end{equation}

\section{Global surface flow measurements in a wind tunnel}\label{Global surface flow diagnostics in a wind tunnel}
Global luminescent oil film (GLOF) skin friction measurements were conducted on flow
over a NASA (National Aeronautics and Space Administration) FAITH
(Fundamental Aeronautics Investigates The Hill) hill model~\citep{Bell2012,Husen2018} in a low-speed wind tunnel at Western Michigan University. The model was manufactured using resin-based stereolithography (SLA) 3D printing. 
The height function is given by $z(r)=a\cos(\pi r/3a)+a$, where $r$ is the radial distance measured from the axis of rotational symmetry, and $a=30~{\rm mm}$.
This yields a maximum height $h_{\rm max}=2a$ and a base diameter $D_{\rm base}=6a$ (or a base radius $R_{\rm base}=3a$). Geometric analysis of the hill surface is provided in Appendix~\ref{geometry}. During testing, the model was flush-mounted onto a polycarbonate (PC) plate measuring 400 mm wide, 1000 mm long, and 15 mm thick. To ensure smooth flow over the flat plate, a rounded leading edge and a sharp trailing edge were added. Measurements were performed at a freestream velocity of $15~{\rm m/s}$, corresponding to a Reynolds number of $Re_{h}=6.08\times10^4$ based on the maximum height. The experimental system consisted of a Basler charge-Coupled device (CCD) camera equipped with a long-pass optical filter, ultraviolet (UV) lamps, and a personal computer for image acquisition and processing.
The luminescent oil was prepared by blending a 200-centistoke (cSt) silicone oil with a UV-excitable fluorescent tracer dye (DFSB-K175, Risk Reactor). Before each test, the oil was evenly applied onto the model surface using a foam brush to form a uniform thin film. Under UV illumination, the luminescent oil emits radiation at approximately 550 nm due to the Stokes shift. To suppress background luminescence, the model surface was first coated with a black Mylar underlayer. All experiments were conducted in a darkened environment. We adopt the optical flow method (OFM) to reconstruct a high-resolution skin friction field $(\bm{\tau}\text{-field})$ from a time sequence of GLOF images~\citep{LiuShen2008,LiuTS2019PAS}. The surface pressure field $p_{w}$ can then be extracted from the $\bm{\tau}$-field, either via a variational method~\citep{LiuCai2024} or through pressure-sensitive paint (PSP) measurements~\citep{Liu2021PSP}. The sensitivity of the extracted skin friction and surface pressure fields is evaluated over a range of the Lagrange multiplier $\lambda_{\mathscr{L}}=200-10,000$. The relative variation in skin friction, defined as $\lVert\bm{\tau}-\langle\bm{\tau}\rangle\rVert/{\rm max}(\lVert\bm{\tau}\rVert)$, is approximately $5\%$ for the entire $\bm{\tau}$-field, where $\langle\bm{\tau}\rangle$ denotes the skin friction averaged across the range of $\lambda_{\mathscr{L}}$. Likewise, the relative variation in surface pressure is around $0.5\%$. The transformation of the vorticity components between different coordinate triads is discussed in Appendix~\ref{Transformation of vorticity components}. 
\begin{figure}[t]
	\centering
	\subfigure[$(\bm{\tau},\lVert\bm{\tau}\rVert)$]{
		\begin{minipage}[t]{0.5\linewidth}
			\centering
			\includegraphics[width=1.0\columnwidth,trim={0.7cm 1.5cm 0.7cm 1.4cm},clip]{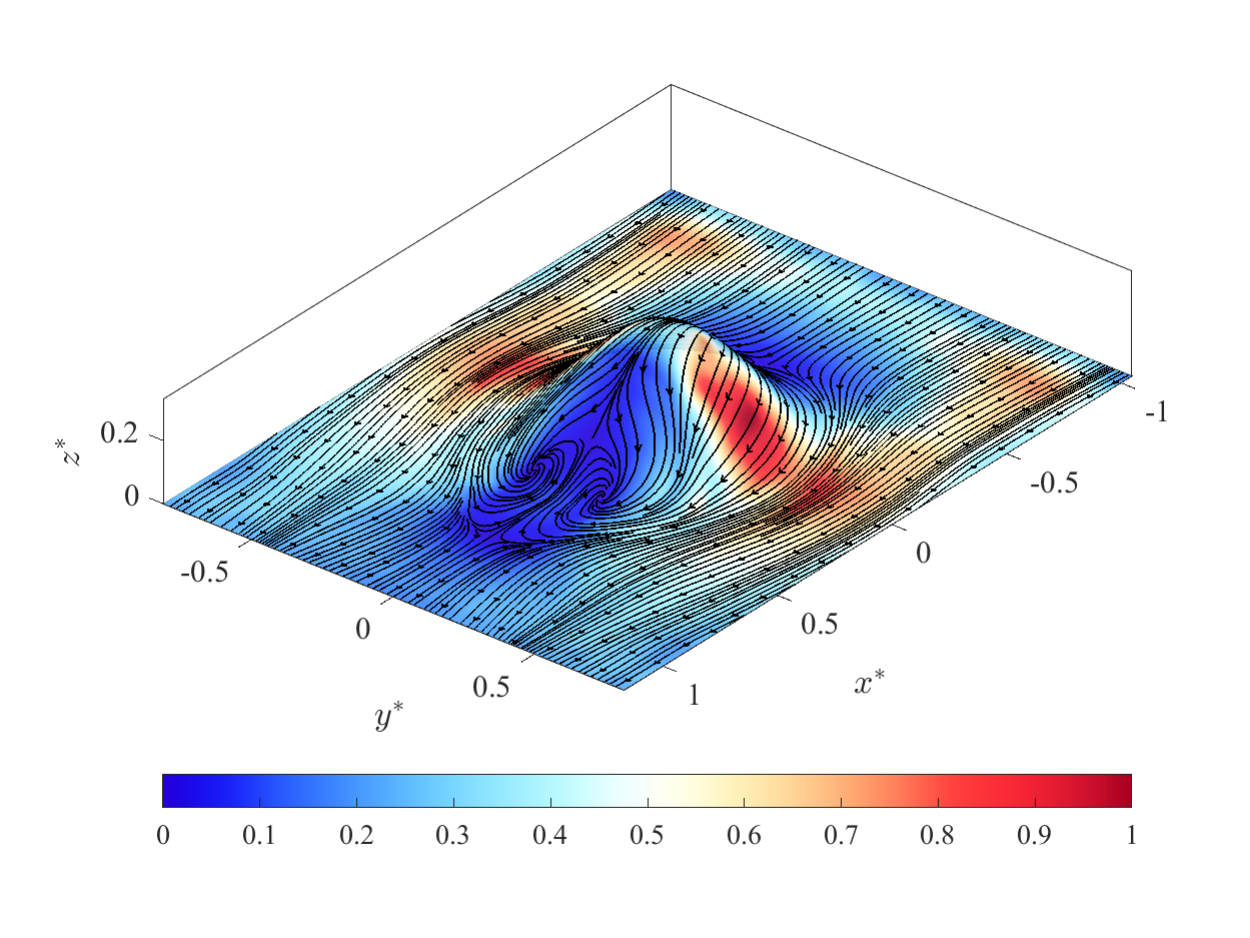}
			\label{tau_and_taunorm_3D}
		\end{minipage}%
	}%
	\subfigure[$(\bm{\tau},p_{w})$]{
		\begin{minipage}[t]{0.5\linewidth}
			\centering
			\includegraphics[width=1.0\columnwidth,trim={2cm 9.6cm 2cm 10.2cm},clip]{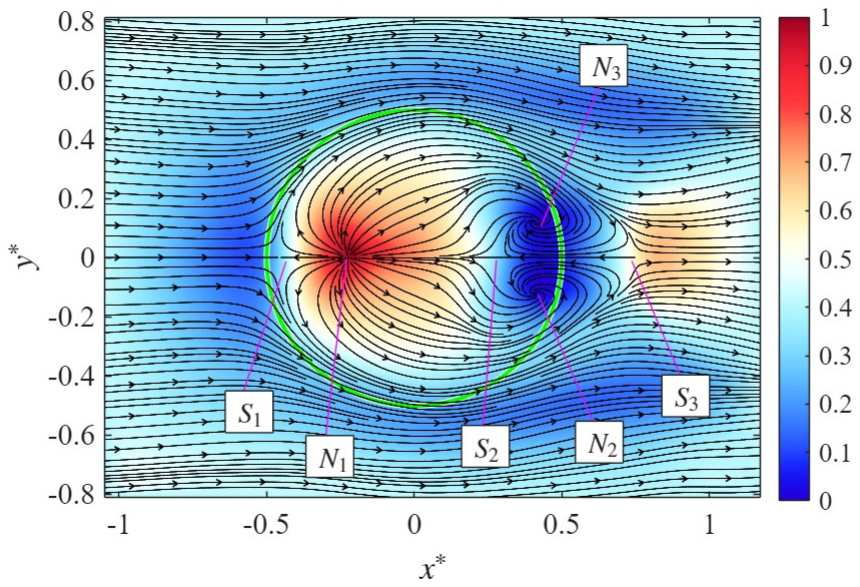}
			\label{tau_and_pw_2D}
		\end{minipage}%
	}%
	
		\subfigure[$(\bm{\tau},\bm{\omega}_{w})$]{
		\begin{minipage}[t]{0.5\linewidth}
			\centering
			\includegraphics[width=1.0\columnwidth,trim={0cm 0.2cm 0cm 0.2cm},clip]{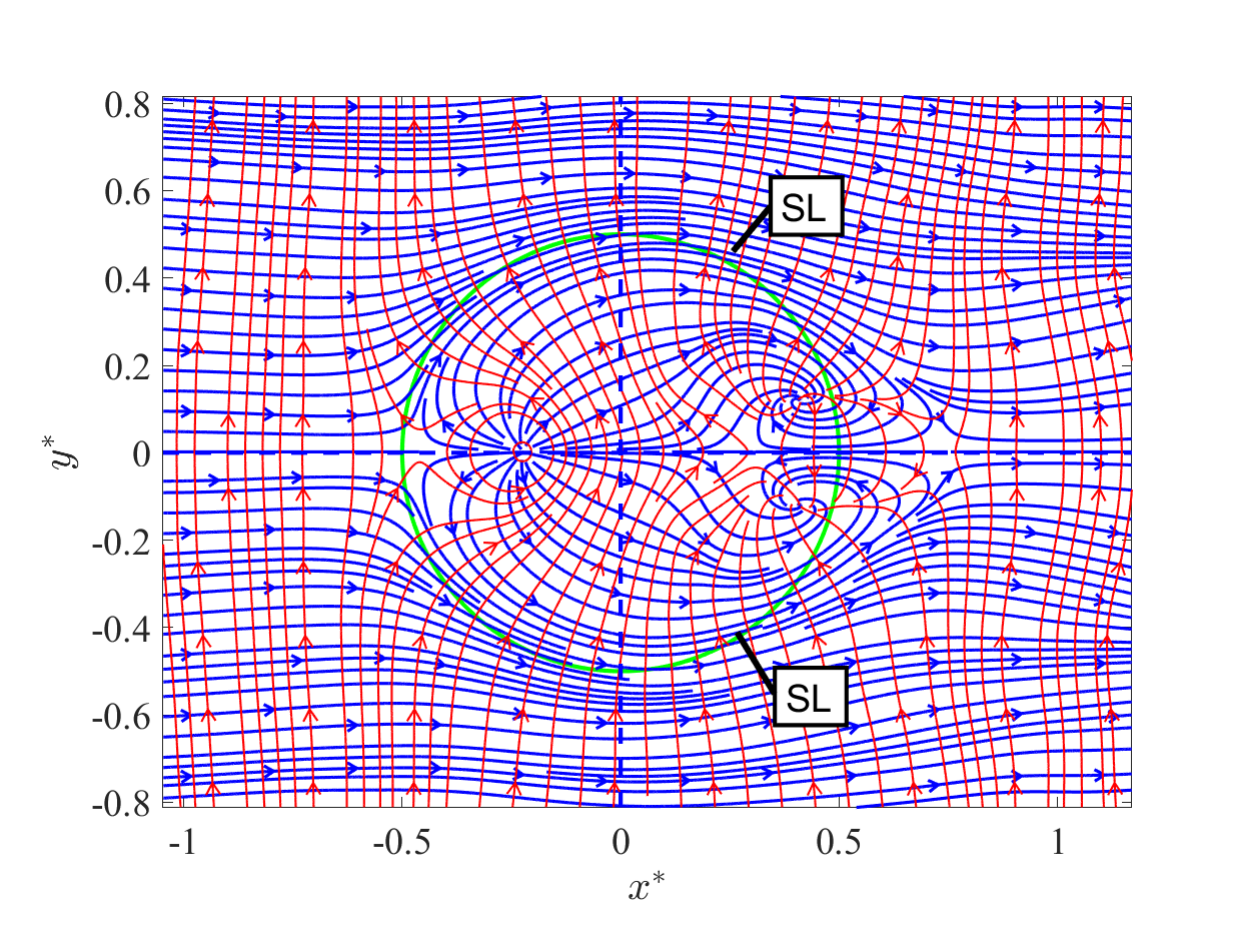}
			\label{tau_omega_lines}
		\end{minipage}%
	}%
	\caption{Skin-friction lines $(\bm{\tau}\text{-lines})$ superimposed on normalized background fields: (a) skin-friction magnitude $\lVert\bm{\tau}\rVert$ and (b) surface pressure $p_{w}$. The normalized fields are defined as $\bm{\tau}^{*}\equiv{\bm{\tau}}/{\max(\lVert\bm{\tau}\rVert)}$ and $p_{w}^{*}\equiv{(p_{w}-\min({p}_{w}))}/{(\max(p_{w})-\min(p_w))}$, respectively. (c) Orthogonal curvilinear net on $\bm{\Sigma}_{w}$ formed by the integral curves of the $\bm{\tau}$ and $\bm{\omega}_{w}$ fields, where ``SL'' denotes separation lines. The green circle indicates the edge of the hill on the bottom wall.} 
	\label{skin_friction_surface_pressure}
\end{figure}

\section{Experimental results and analysis}\label{Results and analysis}
\subsection{Features of skin friction and surface pressure}
Skin friction $(\bm{\tau})$ and surface pressure $(p_{w})$ are considered as the surface footprints of near-wall flow structures~\citep{Chen2019Relations,Chen2021features}.
Figure~\ref{tau_and_taunorm_3D} shows a 3D view of the $\bm{\tau}$-lines overlaid on the normalized $\lVert\bm{\tau}\rVert$ field, whereas the elementary $(\bm{\tau},p_{w})$ pair is illustrated in figure~\ref{tau_and_pw_2D}. When $(\bm{\tau},p_{w})$ is known, the instantaneous near-wall velocity and pressure fields can be reconstructed in a small neighborhood of the wall using the Taylor-series expansion solution of the NS equations~\citep{Zhang1985,Wu2000POF,Bewley2004,Liu2018AIAA,Chen2021turbulence,Chen2022Nearwall}. The number of critical points of the $\bm{\tau}$-field must satisfy the topological constraint imposed by the Poincar\'{e}--Bendixson (P--B) index formula~\citep{Hunt1978,LiuTS2019PAS}. For the present case, where the $\bm{\tau}$-lines in the far field are generally aligned with the mainstream direction, the P--B formula reduces to a simple form $\#{N}=\#{S}$ for the critical points enclosed by a sufficiently large closed boundary on the surface, where $\#{N}$ and $\#{S}$ denote the numbers of nodes and saddles, respectively. This indicates that nodes and saddles in the $\bm{\tau}$-field must occur or disappear in pairs. As seen in figure~\ref{tau_and_pw_2D}, there are three nodes $(N_{1,2,3})$ and three saddles $(S_{1,2,3})$, which indicates topological consistency. The surface pressure $p_{w}$ displays alternating positive and negative centers along the streamwise direction (figure~\ref{tau_and_pw_2D}).
The favorable pressure gradient accelerates the attached flow over the crest and around the flanks, producing the peaks in the $\lVert\bm{\tau}\rVert$ field (figure~\ref{tau_and_taunorm_3D}). The observed $\lVert\bm{\tau}\rVert$ pattern qualitatively agrees with the skin-friction coefficient derived from the wind-tunnel fringe imaging skin friction (FISF) tests $(Re_{h}=5\times10^5)$~\citep{Bell2012} and from the luminescent oil-film flow-tagging skin-friction measurements $(Re_{h}=1.12\times10^5)$~\citep{Husen2018}. Owing to the present Reynolds number $(Re_{h}=6.08\times10^4)$ being one order of magnitude lower than that of~\citet{Bell2012}, flow separation is both attenuated and delayed on the leeward side. As displayed in figure~\ref{tau_omega_lines}, owing to the relation $\bm{\tau}=\mu\bm{\omega}_{w}\times\bm{n}_{w}$, skin friction $\bm{\tau}$ and surface vorticity $\bm{\omega}_{w}$ form an orthogonal pair $(\bm{\tau},\bm{\omega}_{w})$ on the hill surface, which can alternatively be used to describe flow separation. The topology of the $\bm{\tau}$-field enables a physical description of local flow separation and attachment~\citep{Surana2006,Chong2012,Chen2021features}.  A separation line (SL) is defined as a skin-friction line to which all neighboring skin-friction lines converge asymptotically. Two distinct separation lines are associated with the necklace vortex (also termed horseshoe vortex)~\citep{Baker1979laminar} encircling the base of the hill. It is noted that the $(\bm{\tau},\bm{\omega}_{w})$ pair is a spatiotemporal accumulation effect of the BVF~\citep{Wu2000POF}. Foci in a skin-friction field are usually associated with tornado-like vortices, where the surrounding skin-friction lines spiral inward~\citep{Dellery2001}.
\begin{figure}[t]
	\centering
	\subfigure[$(\bm{\tau},\lVert\bm{\sigma}_{\pi}^{R}\rVert)$]{
		\begin{minipage}[t]{0.5\linewidth}
			\centering
			\includegraphics[width=1.0\columnwidth,trim={2cm 9.7cm 2cm 10.3cm},clip]{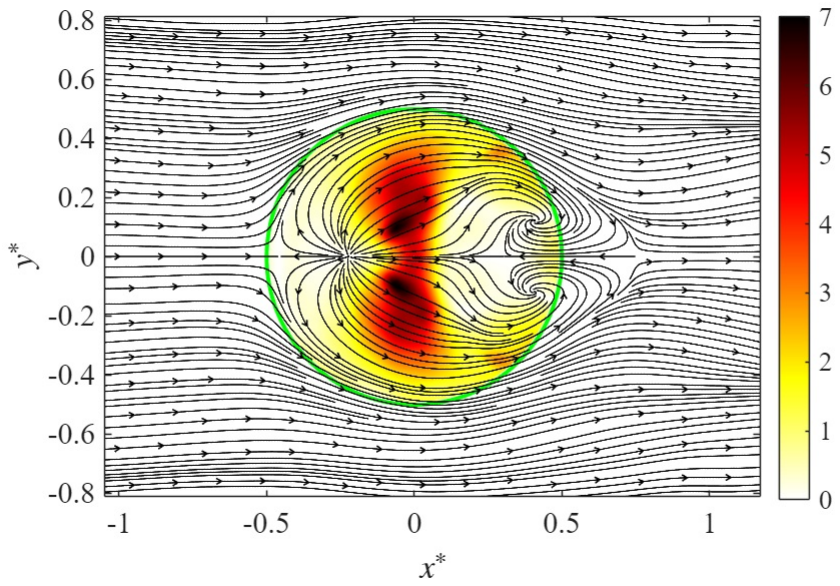}
			\label{sigma_pi_R_norm}
		\end{minipage}%
	}%
	\subfigure[$(\bm{\tau},\lVert\bm{\sigma}_{\pi}^{S}\rVert)$]{
		\begin{minipage}[t]{0.5\linewidth}
			\centering
			\includegraphics[width=1.0\columnwidth,trim={2cm 9.7cm 2cm 10.2cm},clip]{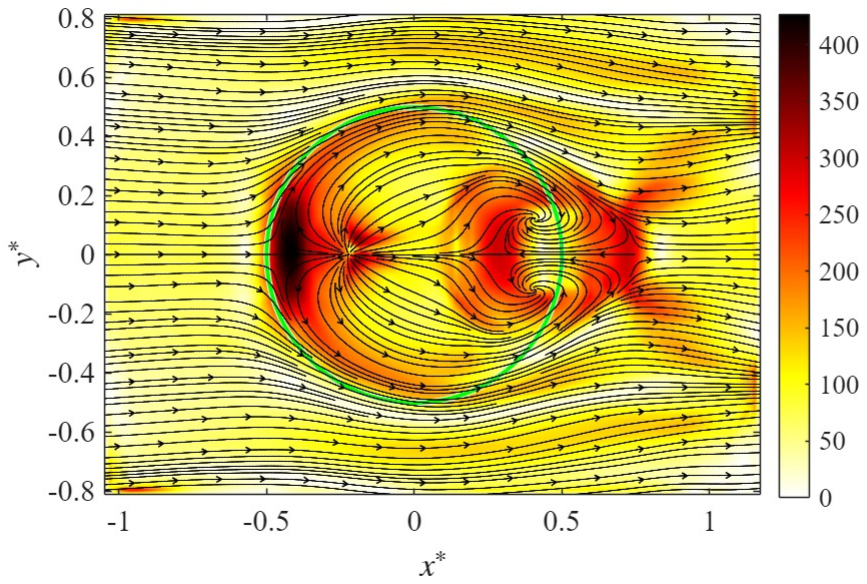}
			\label{sigma_pi_S_norm}
		\end{minipage}%
	}%
	
	\subfigure[$(\bm{\tau},\lVert\bm{\sigma}_{K}\rVert)$]{
		\begin{minipage}[t]{0.5\linewidth}
			\centering
			\includegraphics[width=1.0\columnwidth,trim={2cm 9.7cm 2cm 10.2cm},clip]{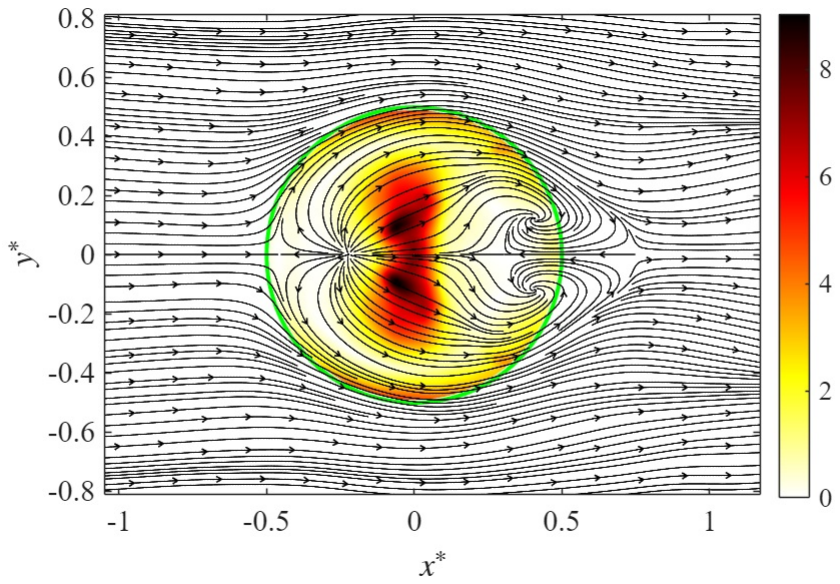}
			\label{sigma_K_norm}
		\end{minipage}%
	}%
	\subfigure[$(\bm{\tau},\lVert\bm{\sigma}_{L}\rVert)$]{
		\begin{minipage}[t]{0.5\linewidth}
			\centering
			\includegraphics[width=1.0\columnwidth,trim={2cm 9.7cm 2cm 10.2cm},clip]{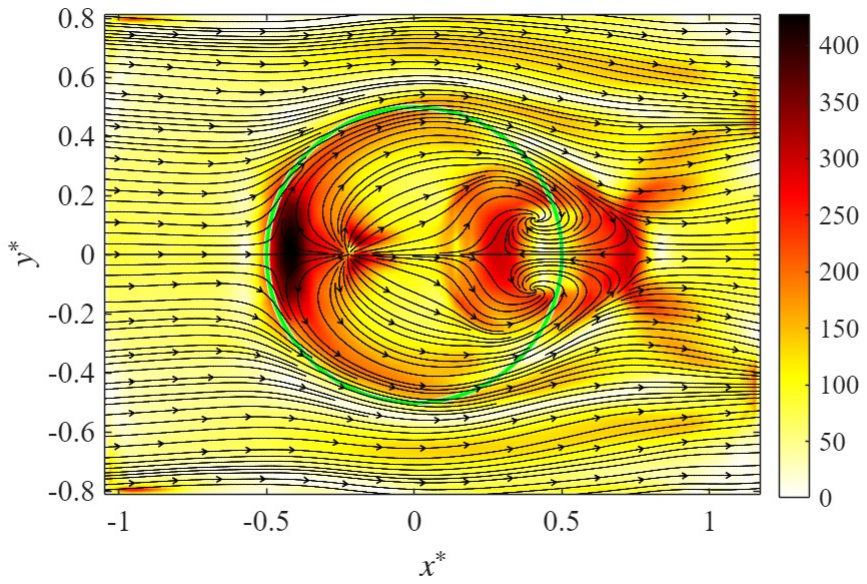}
			\label{sigma_Lyman_norm}
		\end{minipage}%
	}%
	\caption{Skin-friction lines $(\bm{\tau}\text{-lines})$ superimposed on normalized background fields of (a) $\lVert\bm{\sigma}_{\pi}^{R}\rVert$, (b) $\lVert\bm{\sigma}_{\pi}^{S}\rVert$, (c) $\lVert\bm{\sigma}_{K}\rVert$, and (d) $\lVert\bm{\sigma}_{L}\rVert$. The green circle indicates the edge of the hill on the bottom wall.} 
	\label{tangential_BVF}
\end{figure}
\begin{figure}[h!]
	\centering
	\subfigure[$(\bm{\tau},-2\nu\widehat{\lambda}_{z}\lbrack\omega_{\theta}\rbrack_{w})$]{
		\begin{minipage}[t]{0.5\linewidth}
			\centering
			\includegraphics[width=1.0\columnwidth,trim={0.45cm 0.1cm 0.45cm 0.5cm},clip]{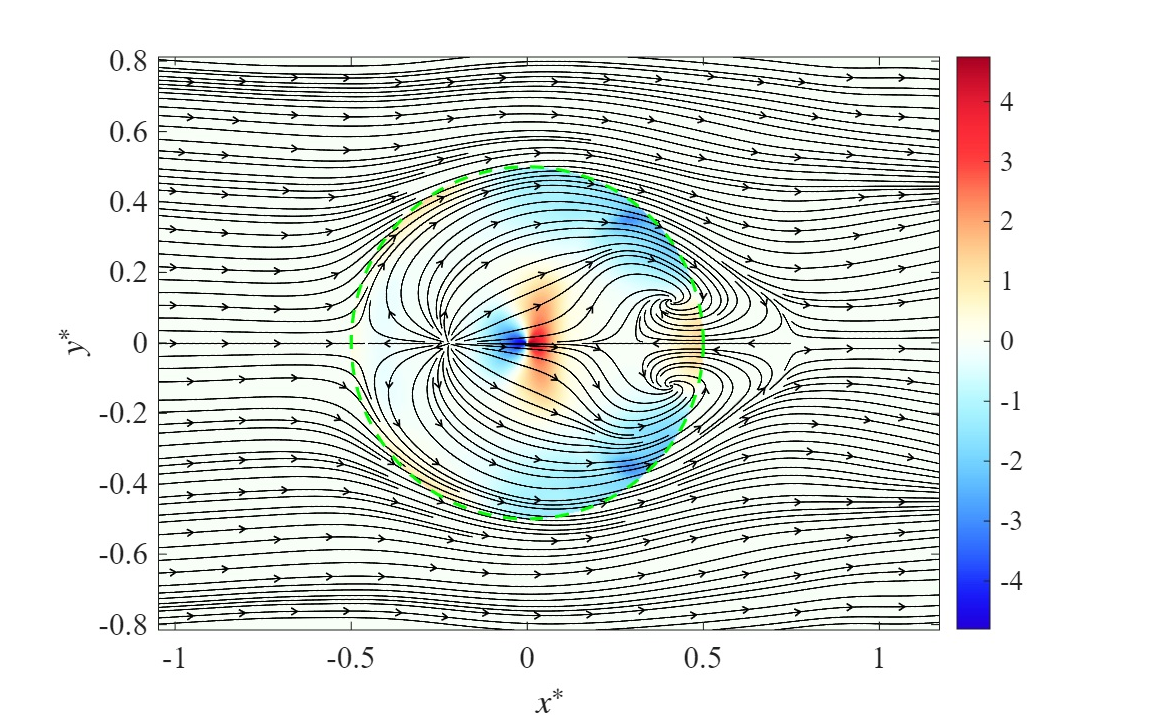}
			\label{sigma_R_theta}
		\end{minipage}%
	}%
	\subfigure[$(\bm{\tau},-2\nu\widehat{\lambda}_{\theta}\lbrack\omega_{z}\rbrack_{w})$]{
		\begin{minipage}[t]{0.5\linewidth}
			\centering
			\includegraphics[width=1.0\columnwidth,trim={0.45cm 0.1cm 0.45cm 0.5cm},clip]{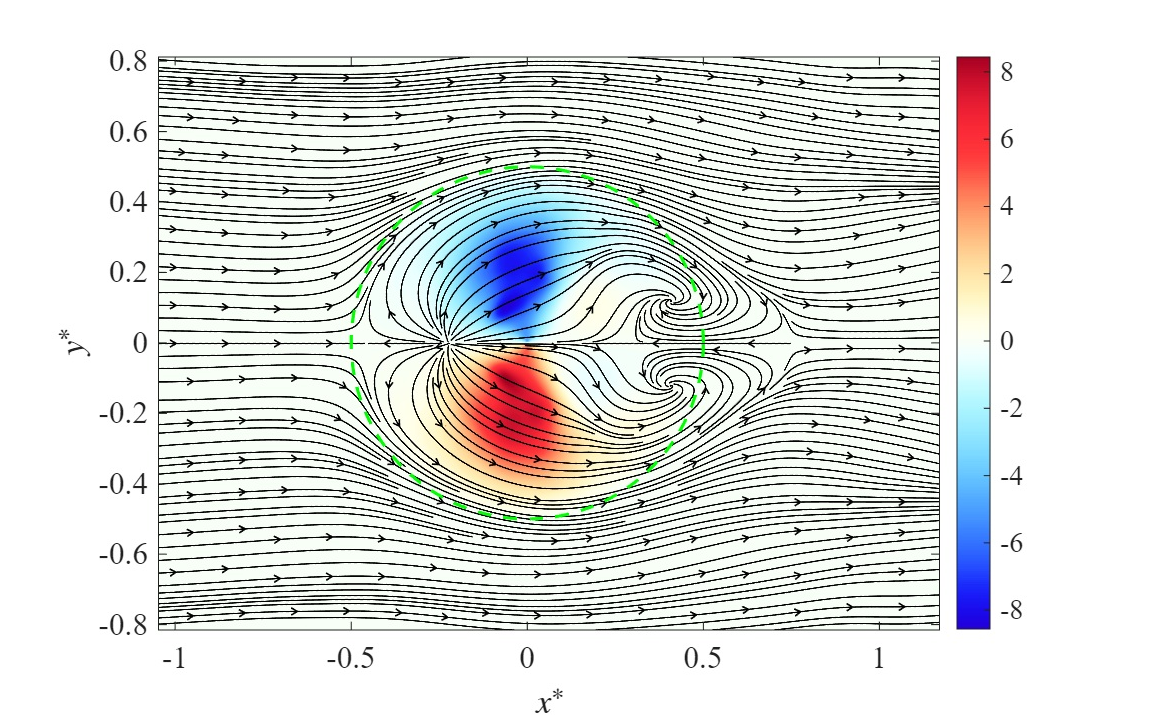}
			\label{sigma_R_z}
		\end{minipage}%
	}%
	
		\subfigure[$(\bm{\sigma}_{\pi}^{R},-2\nu\widehat{\lambda}_{z}\lbrack\omega_{\theta}\rbrack_{w})$]{
		\begin{minipage}[t]{0.5\linewidth}
			\centering
			\includegraphics[width=1.0\columnwidth,trim={0.45cm 0.1cm 0.45cm 0.5cm},clip]{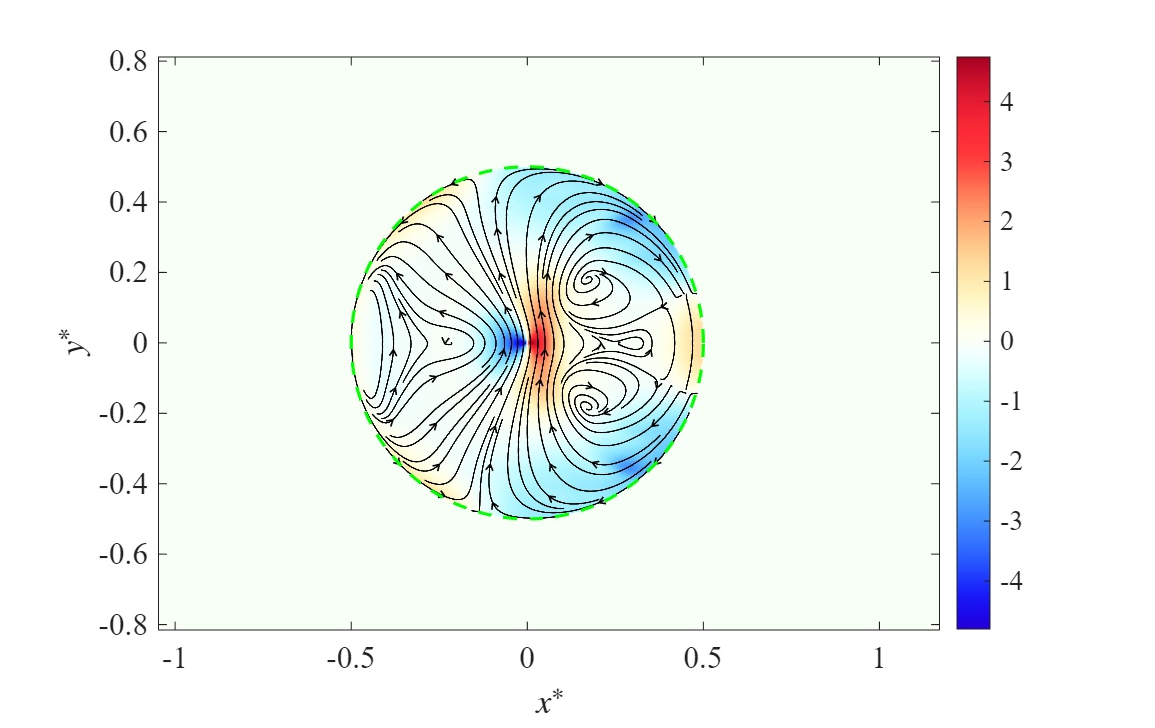}
			\label{sigma_R_theta1}
		\end{minipage}%
	}%
	\subfigure[$(\bm{\sigma}_{\pi}^{R},-2\nu\widehat{\lambda}_{\theta}\lbrack\omega_{z}\rbrack_{w})$]{
		\begin{minipage}[t]{0.5\linewidth}
			\centering
			\includegraphics[width=1.0\columnwidth,trim={0.45cm 0.1cm 0.45cm 0.5cm},clip]{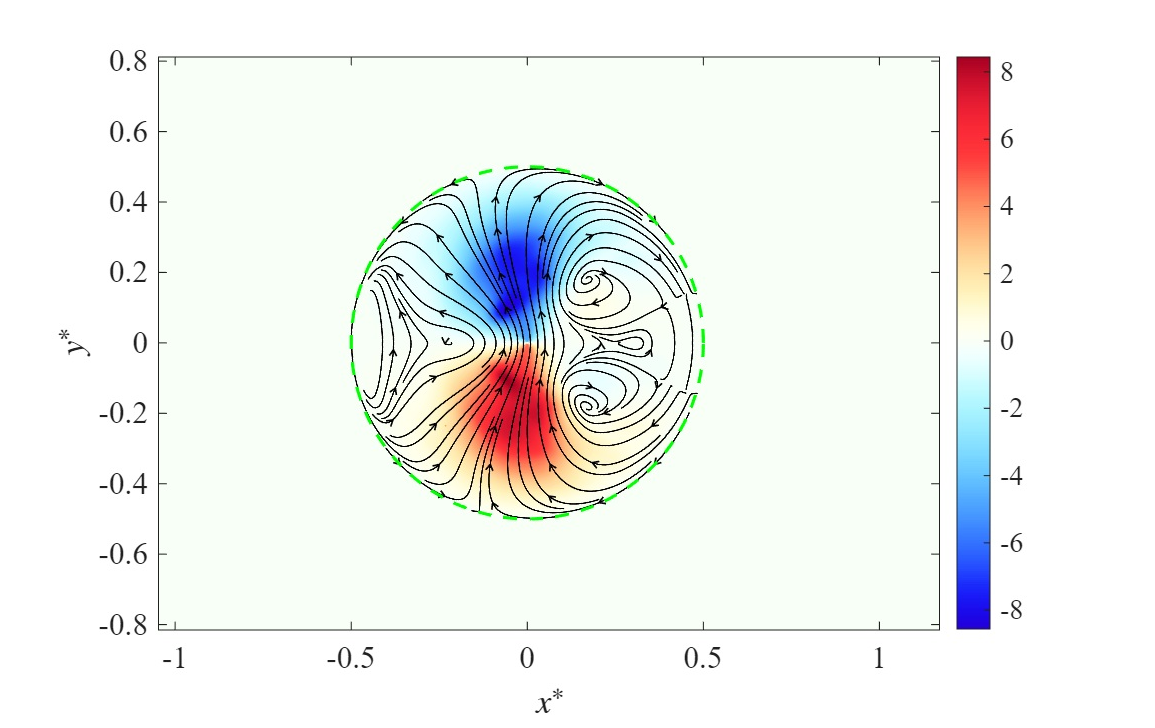}
			\label{sigma_R_z1}
		\end{minipage}%
	}%
	\caption{Skin-friction lines $(\bm{\tau}\text{-lines})$ and $\bm{\sigma}_{\pi}^R$--lines superimposed on the normalized components of $\bm{\sigma}_{\pi}^R$: (a) $(\bm{\tau},-2\nu\widehat{\lambda}_{z}[\omega_{\theta}]_{w})$, (b) $(\bm{\tau},-2\nu\widehat{\lambda}_{\theta}[\omega_{z}]_{w})$, (c) $(\bm{\sigma}_{\pi}^R,-2\nu\widehat{\lambda}_{z}[\omega_{\theta}]_{w})$, and (d) $(\bm{\sigma}_{\pi}^R,-2\nu\widehat{\lambda}_{\theta}[\omega_{z}]_{w})$. The green circle indicates the edge of the hill on the bottom wall.} 
	\label{sigma_pi_R_components}
\end{figure}
\begin{figure}[t]
	\centering
	\subfigure[$(\bm{\tau},\lVert\bm{\sigma}_{\pi}^{S}\rVert/\lVert\bm{\sigma}_{\pi}^{R}\rVert)$]{
		\begin{minipage}[t]{0.5\linewidth}
			\centering
			\includegraphics[width=1.0\columnwidth,trim={0cm 0.3cm 0cm 1.5cm},clip]{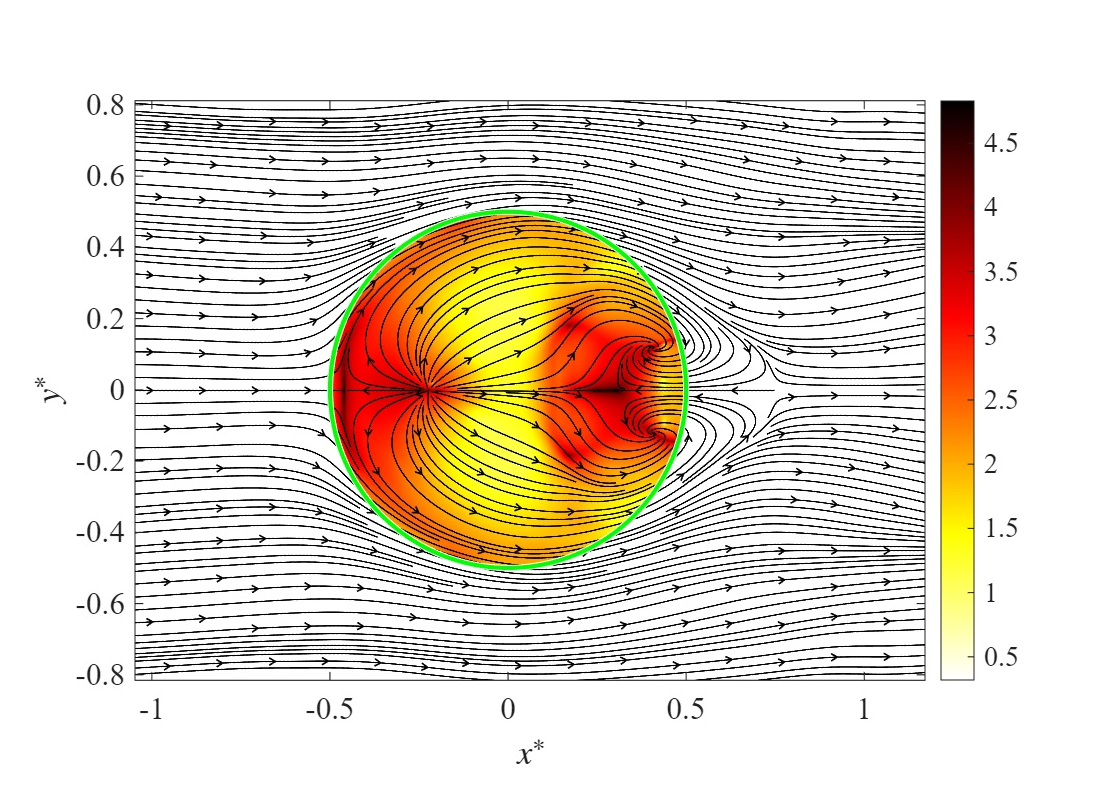}
			\label{ratio_sigmaS_sigmaR}
		\end{minipage}%
	}%
	\subfigure[$(\bm{\tau},\lVert\bm{\sigma}_{L}\rVert/\lVert\bm{\sigma}_{K}\rVert)$]{
		\begin{minipage}[t]{0.5\linewidth}
			\centering
			\includegraphics[width=1.0\columnwidth,trim={0cm 0.3cm 0cm 1.5cm},clip]{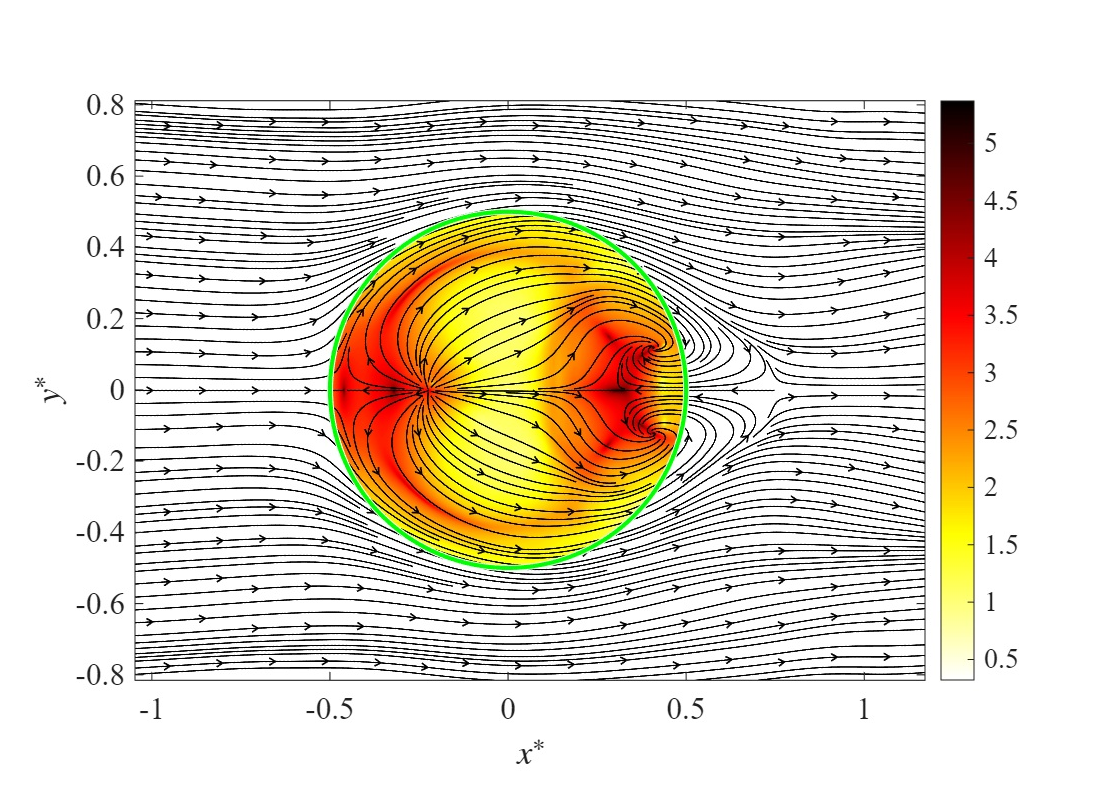}
			\label{ratio_sigmaL_sigmaK}
		\end{minipage}%
	}%
	\caption{Skin-friction lines $(\bm{\tau}\text{-lines})$ superimposed on the base-10 logarithm ratios of (a) $\lVert\bm{\sigma}_{\pi}^{S}\rVert/\lVert\bm{\sigma}_{\pi}^{R}\rVert$ and (b) $\lVert\bm{\sigma}_{L}\rVert/\lVert\bm{\sigma}_{K}\rVert$. The green circle indicates the edge of the hill on the bottom wall.} 
	\label{ratios}
\end{figure}
\begin{figure}[t]
	\centering
	\subfigure[]{
		\begin{minipage}[t]{0.5\linewidth}
			\centering
			\includegraphics[width=1.0\columnwidth,trim={5cm 0cm 5cm 0cm},clip]{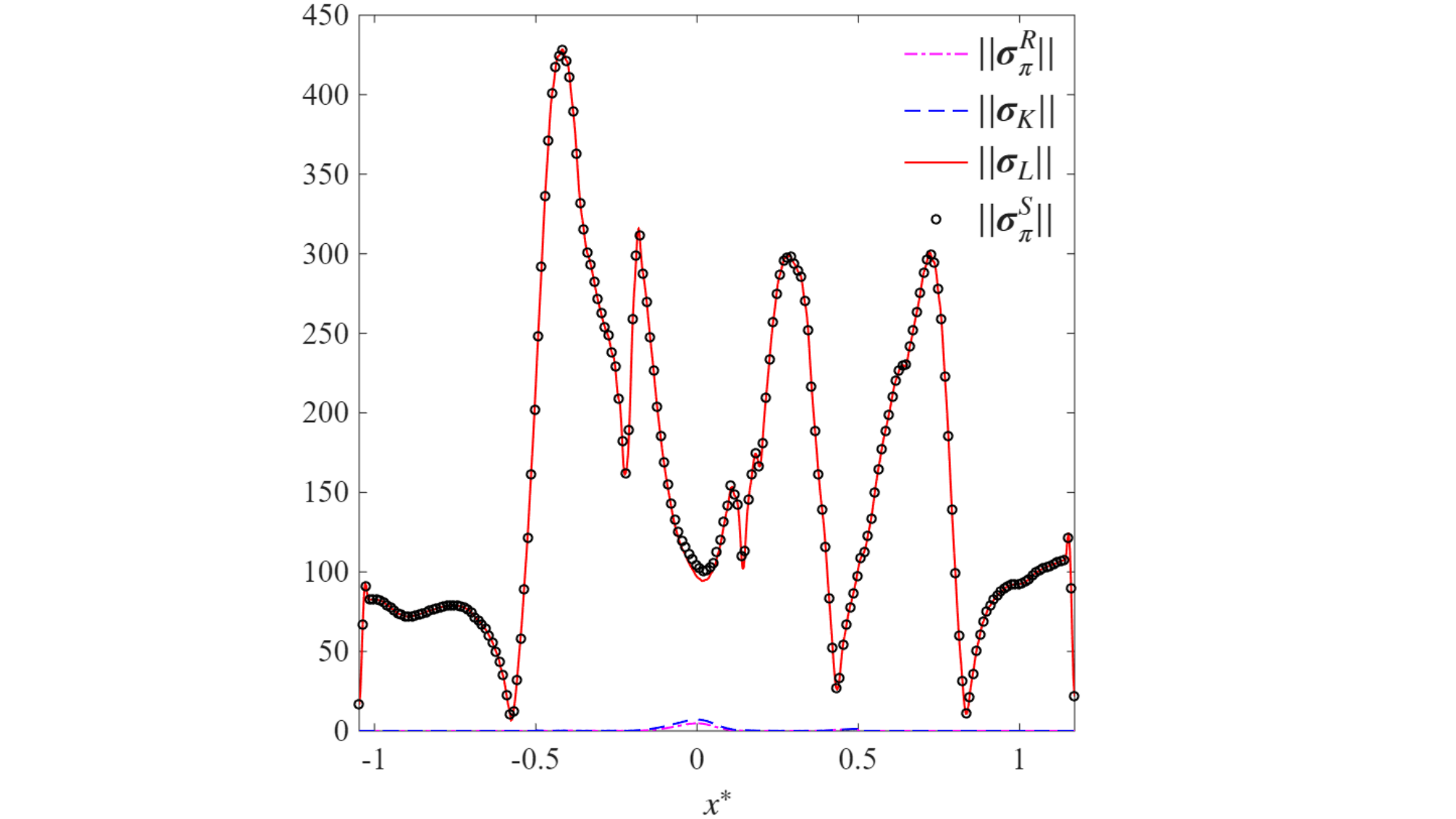}
			\label{comparison1}
		\end{minipage}%
	}%
	\subfigure[]{
		\begin{minipage}[t]{0.5\linewidth}
			\centering
			\includegraphics[width=1.0\columnwidth,trim={5cm 0cm 5cm 0cm},clip]{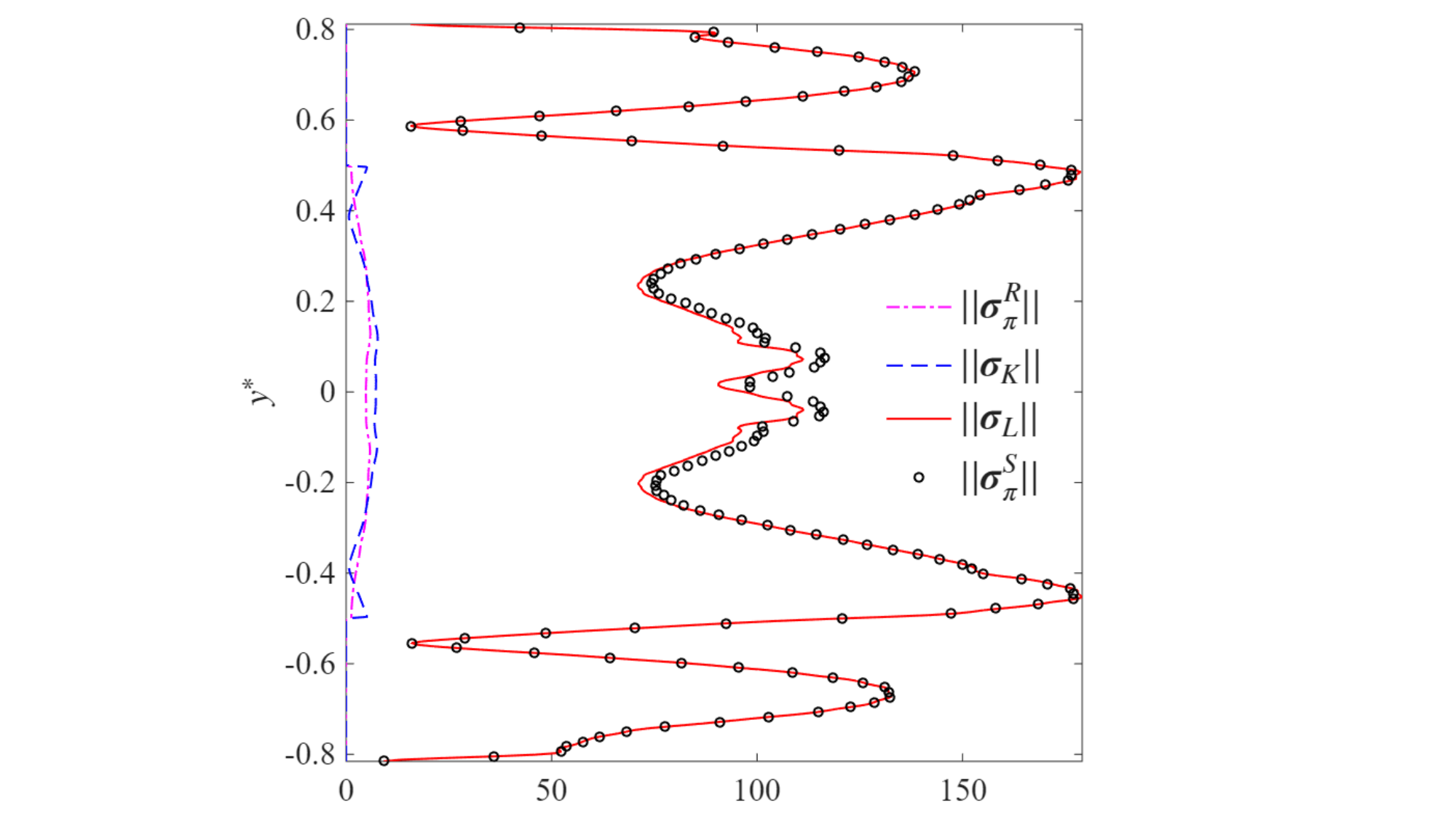}
			\label{comparison2}
		\end{minipage}%
	}%
	\caption{Comparison of normalized profiles of $\lVert\bm{\sigma}_{\pi}^{R}\rVert$, $\lVert\bm{\sigma}_{K}\rVert$, $\lVert\bm{\sigma}_{L}\rVert$ and $\lVert\bm{\sigma}_{\pi}^{S}\rVert$ along (a) $y=0$ and (b) $x=0$ over the hill surface.} 
	\label{line_compare}
\end{figure}
\begin{figure}[t]
	\centering
	\includegraphics[width=0.6\columnwidth,trim={0cm 0.65cm 0cm 1.5cm},clip]{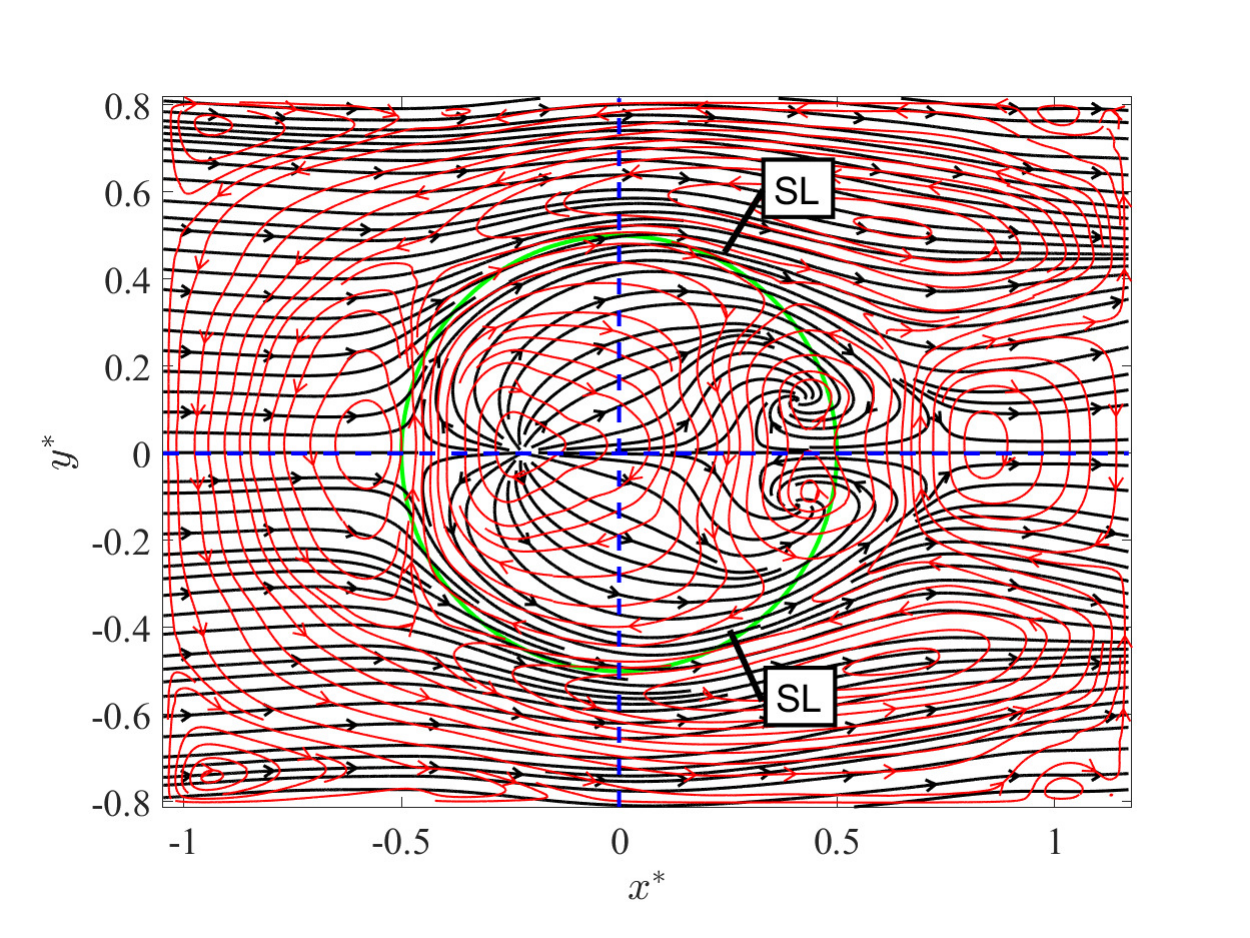}
	\caption{Integral lines of the skin friction $\bm{\tau}$ (black) and the Lyman flux $\bm{\sigma}_{L}$ (red) on the hill surface $\bm{\Sigma}_{w}$. ``SL'' indicates the separation lines. The green circle marks the edge of the hill on the bottom wall.} 
	\label{pair_tau_sigmap_2D}
\end{figure}

\subsection{Quadruple decomposition of the boundary vorticity flux on the hill surface}
\begin{figure}[t]
	\centering
	\subfigure[$(\bm{\tau},{\sigma}_{n}^{R})$]{
		\begin{minipage}[t]{0.5\linewidth}
			\centering
			\includegraphics[width=1.0\columnwidth,trim={2cm 9.75cm 2cm 10.2cm},clip]{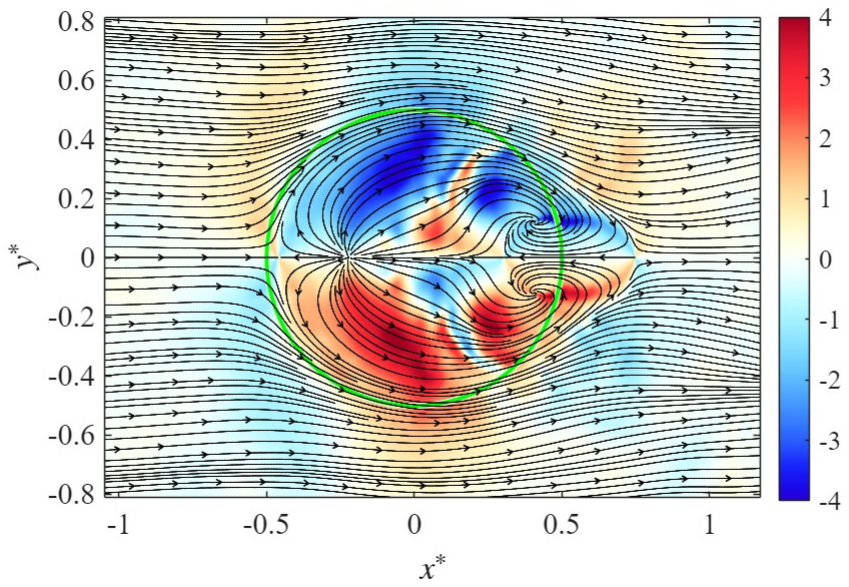}
			\label{sigma_n_R}
		\end{minipage}%
	}%
	\subfigure[$(\bm{\tau},{\sigma}_{n}^{S})$]{
		\begin{minipage}[t]{0.5\linewidth}
			\centering
			\includegraphics[width=1.0\columnwidth,trim={2cm 9.75cm 2cm 10.2cm},clip]{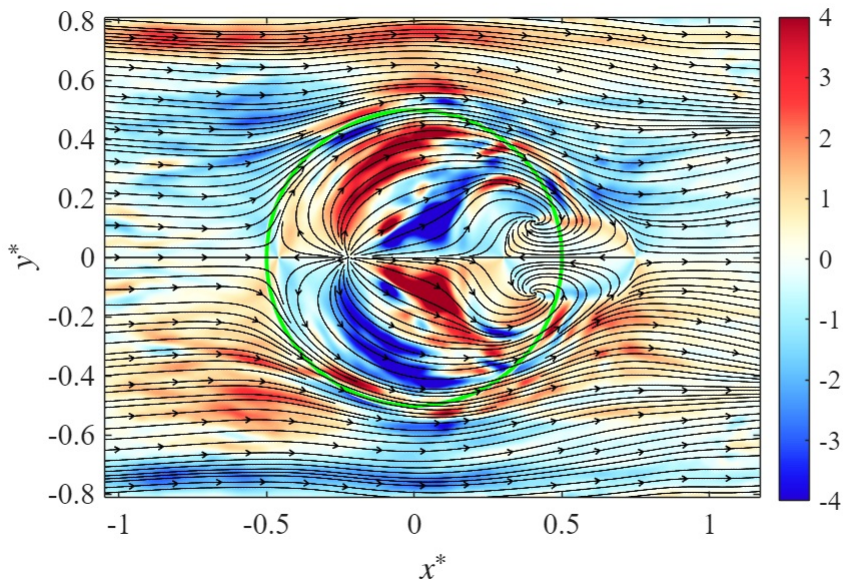}
			\label{sigma_n_S}
		\end{minipage}%
	}%
	
	\subfigure[$(\bm{\tau},{\sigma}_{n})$]{
		\begin{minipage}[t]{0.5\linewidth}
			\centering
			\includegraphics[width=1.0\columnwidth,trim={2cm 9.75cm 2cm 10.2cm},clip]{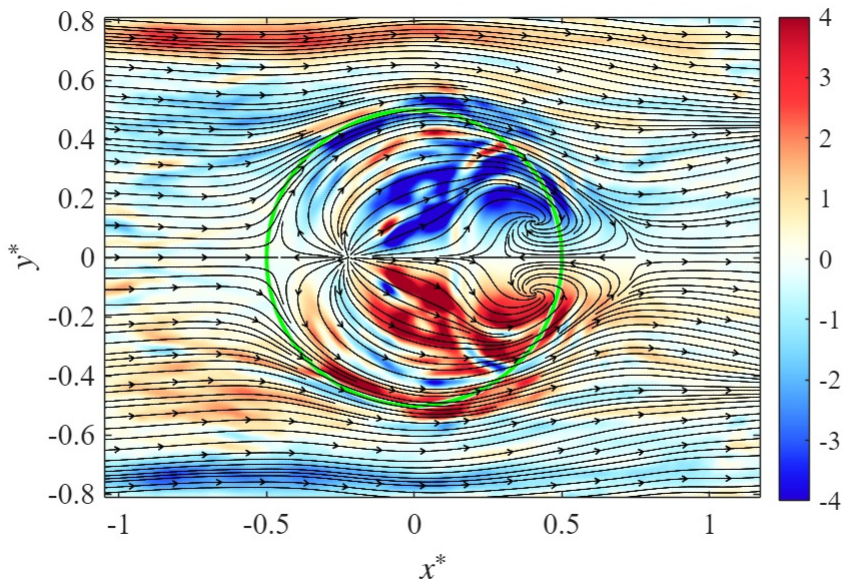}
			\label{sigma_n_total}
		\end{minipage}%
	}%
	\caption{Skin-friction lines $(\bm{\tau}\text{-lines})$ superimposed on normalized background fields of (a) $\sigma_{n}^{R}$, (b) $\sigma_{n}^{S}$, and (c) $\sigma_{n}$. The green circle indicates the edge of the hill on the bottom wall.} 
	\label{wall_normal_BVF}
\end{figure}
Using~\eqref{L17b} and~\eqref{L17c} for the decomposition of the tangential BVF $\bm{\sigma}_{\pi}$, we demonstrate the spatial distributions of $\lVert\bm{\sigma}_{\pi}^{R}\rVert$ (figure~\ref{sigma_pi_R_norm}), $\lVert\bm{\sigma}_{\pi}^{S}\rVert$ (figure~\ref{sigma_pi_S_norm}), $\lVert\bm{\sigma}_{K}\rVert$ (figure~\ref{sigma_K_norm}), and $\lVert\bm{\sigma}_{L}\rVert$ (figure~\ref{sigma_Lyman_norm}), where $\lVert\bm{\cdot}\rVert$ denotes the Frobenius norm. Under the surface orthonormal basis $\left\{\widehat{\bm{e}}_{1},\widehat{\bm{e}}_{2},\widehat{\bm{e}}_{3}\right\}=\left\{\widehat{\bm{e}}_{\theta},\widehat{\bm{e}}_{z},\widehat{\bm{e}}_{\zeta}\right\}$ (where $\widehat{\bm{e}}_{\zeta}=\bm{n}_{w}$), the curvature tensor pair $(\bm{K}_{w},\breve{\bm{K}}_{w})$ is expressed as $\bm{K}_{w}=\widehat{\lambda}_{\theta}\widehat{\bm{e}}_{\theta}+\widehat{\lambda}_{z}\widehat{\bm{e}}_{z}$ and $\breve{\bm{K}}_{w}=\widehat{\lambda}_{z}\widehat{\bm{e}}_{\theta}+\widehat{\lambda}_{\theta}\widehat{\bm{e}}_{z}$, where $\widehat{\lambda}_{\theta}$ and $\widehat{\lambda}_{z}$ are two principal curvatures of $\bm{\Sigma}_{w}$. Consequently,  $\bm{\sigma}_{\pi}^{R}$ is expanded as
\begin{equation}
\bm{\sigma}_{\pi}^{R}=-2\nu\widehat{\lambda}_{z}[\omega_{\theta}]_{w}\widehat{\bm{e}}_{\theta}-2\nu\widehat{\lambda}_{\theta}[\omega_{z}]_{w}\widehat{\bm{e}}_{z},
\end{equation}
which arises from the coupling between the boundary spin mode $(\bm{\omega}_{w}=[\bm{s}_{\Sigma}]_{w})$ and the conjugate curvature tensor $\breve{\bm{K}}_{w}$. Since $[\bm{R}_{\Sigma}]_{w}=\bm{0}$ always holds, this ascending mechanism can generate the rigid-rotation mode slightly away from the wall for $\bm{\sigma}_{\pi}^{R}\neq\bm{0}$, which could facilitate or suppress the transformation of the separated shear layer into an axial vortex.
Based on the experimental results, we find that the crest and flank regions are dominated by $-2\nu\widehat{\lambda}_{z}[\omega_{\theta}]_{w}$ (figures~\ref{sigma_R_theta} and~\ref{sigma_R_theta1}) and $-2\nu\widehat{\lambda}_{\theta}[\omega_{z}]_{w}$ (figures~\ref{sigma_R_z} and ~\ref{sigma_R_z1}), respectively. The former domination stems primarily from the high magnitude of $\widehat{\lambda}_{z}$, whereas the latter arises from high magnitude of $[\omega_{z}]_{w}$ (strong shearing effect). The distribution of $\bm{\sigma}_{\pi}^{R}$ tends to weaken spanwise-oriented axial vortices (dominated by the rigid-rotation mode) within the near-wall region over a Gaussian bump. The BVF constituent $\bm{\sigma}_{K}$ in~\eqref{sigma_K} comprises the two ascending mechanisms responsible for generating the spin mode, which is expanded as
\begin{equation}
	\bm{\sigma}_{K}=\nu(2\widehat{\lambda}_{z}+\widehat{\lambda}_{\theta})[\omega_{\theta}]_{w}\widehat{\bm{e}}_{\theta}+\nu(2\widehat{\lambda}_{\theta}+\widehat{\lambda}_{z})[\omega_{z}]_{w}\widehat{\bm{e}}_{z}.
\end{equation}
As shown in figure~\ref{sigma_K_norm}, $\bm{\sigma}_{K}$ exhibits a pattern similar to that of $\bm{\sigma}_{\pi}^{R}$, with differences attributed to $\widehat{\lambda}_{\theta}[\omega_{\theta}]_{w}$ and $\widehat{\lambda}_{z}[\omega_{z}]_{w}$ from $\bm{K}_{w}\bm{\cdot}\bm{\omega}_{w}$. The Lyman flux $\bm{\sigma}_{L} (=\bm{\sigma}_{p})$ (figure~\ref{sigma_Lyman_norm}) dominates the boundary spin creation rate measured by $\bm{\sigma}_{\pi}^{S}$, which differs from $\bm{\sigma}_{K}$ (figure~\ref{sigma_K_norm}) by approximately 1.68 orders of magnitude in terms of the base-10 logarithm of the ratio ${\rm max}(\lVert\bm{\sigma}_{L}\rVert)/{\rm max}(\lVert\bm{\sigma}_{K}\lVert)$, where ${\rm max}$ denotes the maximum Frobenius norm in the considered region. Using Prandtl's attached boundary--layer theory, it follows that $\lVert\bm{\sigma}_{L}\rVert/\lVert\bm{\sigma}_{K}\lVert\sim\mathcal{O}(Re_{h}^{1/2})\sim\mathcal{O}(100)$, which is essentially consistent with the present results. Since the attached--boundary--layer scaling cannot be uniform over a separated leeward side, figures~\ref{ratio_sigmaS_sigmaR} and~\ref{ratio_sigmaL_sigmaK} display the distributions of the base-10 logarithm of the local ratios $\lVert\bm{\sigma}_{\pi}^{S}\rVert/\lVert\bm{\sigma}_{\pi}^{R}\rVert$ and $\lVert\bm{\sigma}_{L}\rVert/\lVert\bm{\sigma}_{K}\rVert$ in the region $r<3a$, which fall within the range about $0.5-5$. The patterns of these two ratios closely resemble those of $\lVert\bm{\sigma}_{\pi}^{S}\rVert$ and $\lVert\bm{\sigma}_{L}\rVert$, which further supports the dominance of the surface pressure gradient in boundary spin creation at such a relatively high Reynolds number. This conclusion is reinforced by comparing the magnitudes of the BVF constituents along $y=0$ (figure~\ref{comparison1}) and $x=0$ (figure~\ref{comparison2}) over the hill. The magnitudes of $\bm{\sigma}_{L}$ and $\bm{\sigma}_{\pi}^{S}$ are very close to each other and are obviously higher than those of $\bm{\sigma}_{\pi}^{R}$ and $\bm{\sigma}_{K}$. As the newly generated vorticity within the separated shear layer interacts with preexisting near-wall flow structures, the pressure distribution across the entire domain (including the surface pressure and its gradient) is altered, thereby initiating a subsequent round of boundary vorticity generation. This closed-loop coupling mechanism was clearly elucidated by the classical triple-deck theory~\citep{Smith1986}. Moreover, figure~\ref{pair_tau_sigmap_2D} shows that the Lyman flux $\bm{\sigma}_{L}$ tends to align with a separation line in the separation zone, which is consistent with the sufficient criterion for separation watch (i.e., $\lVert\bm{\sigma}_{\pi}\times\bm{\tau}\rVert/\lVert\bm{\sigma}_{\pi}\rVert\lVert\bm{\tau}\rVert=\mathcal{O}({\rm Re}^{-1/8})$) used to identify the onset of boundary layer separation~\citep{Wu2000POF}. Equivalently, the surface pressure gradient $\widehat{\bm{\nabla}}_{\pi}\widetilde{P}_{w}$ tends to be perpendicular to a separation line~\citep{LiuTS2016}.
\begin{figure}[t]
	\centering	\subfigure[$(\bm{\tau},\lVert\bm{\omega}_{w}\rVert)$]{
		\begin{minipage}[t]{0.5\linewidth}
			\centering
			\includegraphics[width=1.0\columnwidth,trim={0cm 0.5cm 0cm 1.5cm},clip]{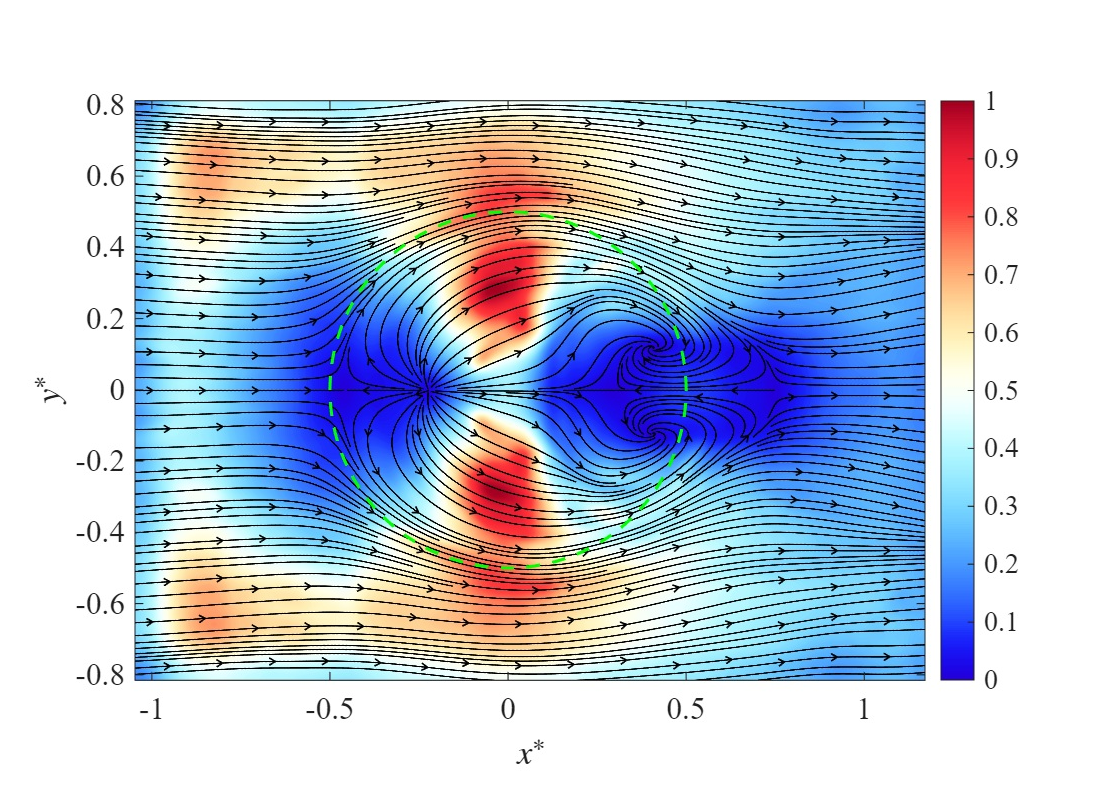}
			\label{tau_and_omeganorm}
		\end{minipage}%
	}%
	\subfigure[$(\bm{\omega}_{w},\lVert\bm{\omega}_{w}\rVert)$]{
		\begin{minipage}[t]{0.5\linewidth}
			\centering
			\includegraphics[width=1.0\columnwidth,trim={0cm 0.5cm 0cm 1.5cm},clip]{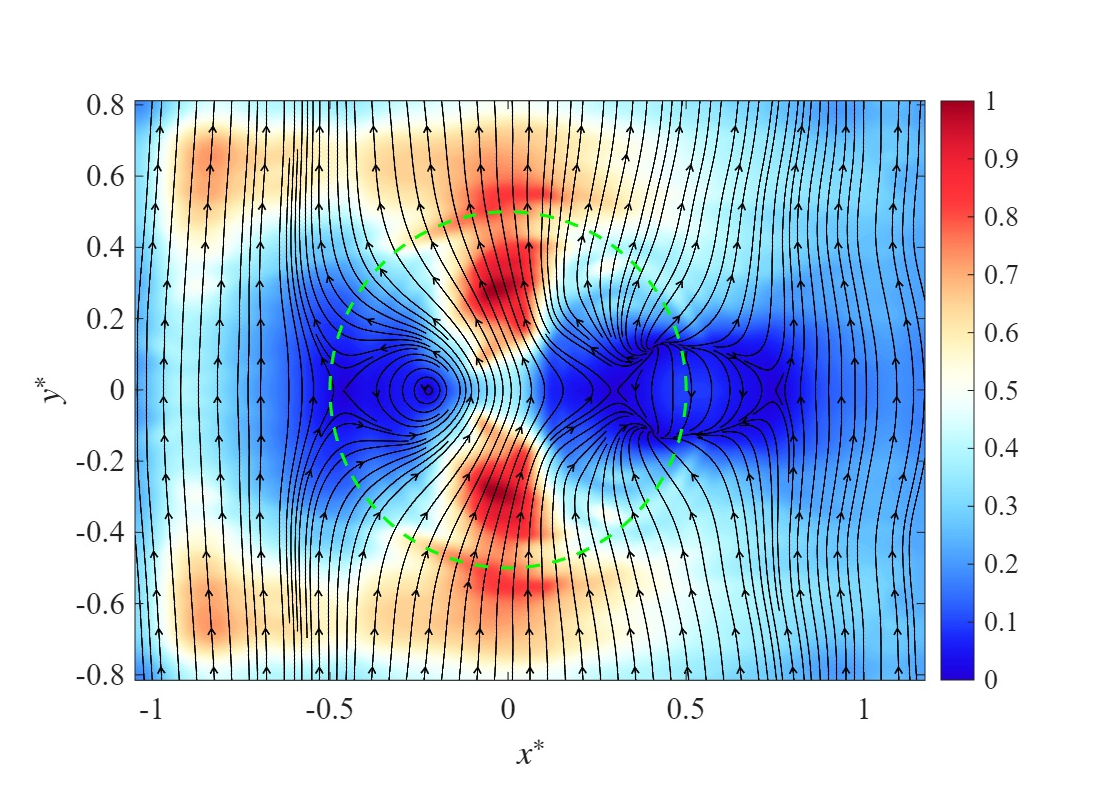}
			\label{omega_and_omeganorm}
		\end{minipage}%
	}%
	
	\subfigure[$(\bm{\tau},\kappa_{g,\xi})$]{
		\begin{minipage}[t]{0.5\linewidth}
			\centering
			\includegraphics[width=1.0\columnwidth,trim={0cm 0.5cm 0cm 1.5cm},clip]{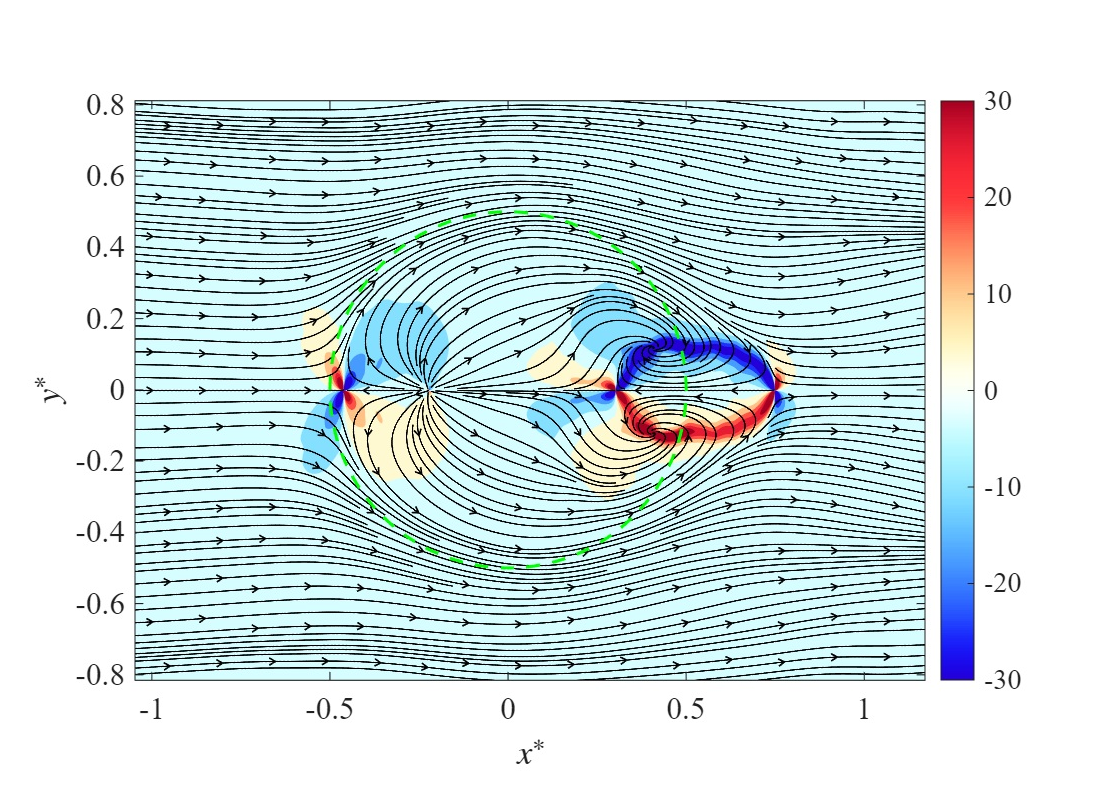}
			\label{kappa_gtau_and_tau_lines}
		\end{minipage}%
	}%
		\subfigure[$(\bm{\omega}_{w},\kappa_{g,\xi})$]{
		\begin{minipage}[t]{0.5\linewidth}
			\centering
			\includegraphics[width=1.0\columnwidth,trim={0cm 0.5cm 0cm 1.5cm},clip]{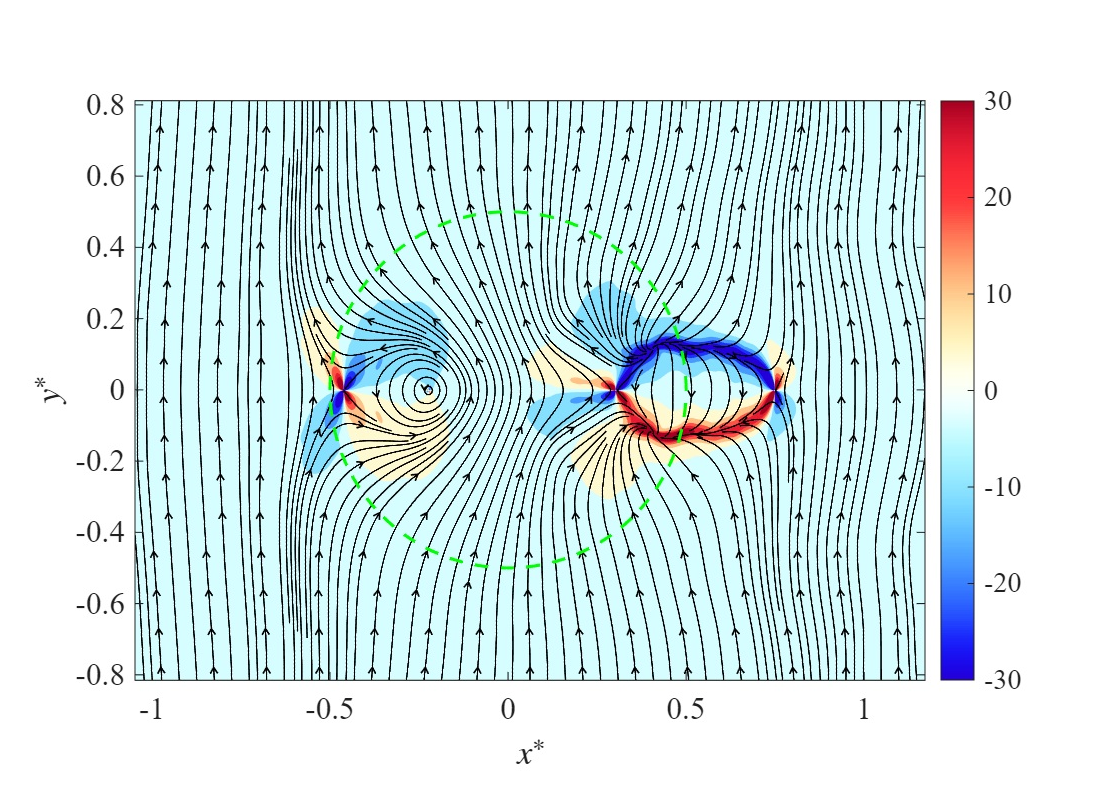}
			\label{kappa_gtau_and_omega_lines}
		\end{minipage}%
	}%
	
		\subfigure[$(\bm{\tau},-\nu\partial_{\omega}\lVert\bm{\omega}_{w}\rVert)$]{
		\begin{minipage}[t]{0.5\linewidth}
			\centering
			\includegraphics[width=1.0\columnwidth,trim={0.3cm 0.2cm 0.3cm 0.6cm},clip]{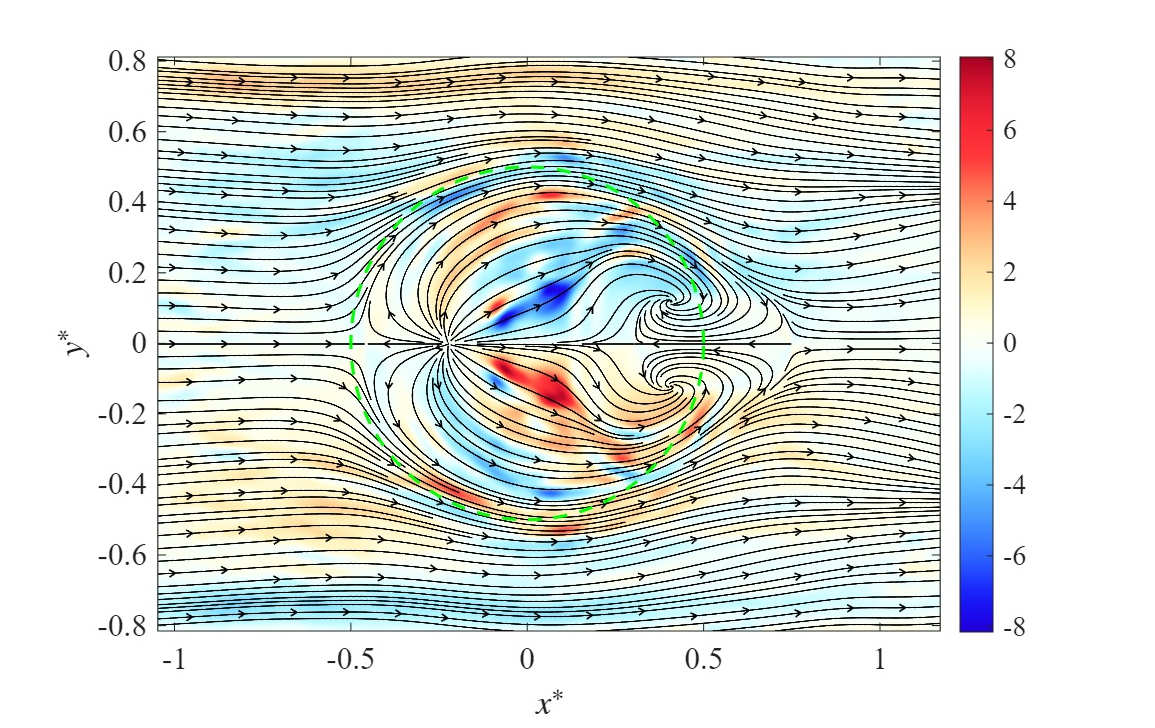}
			\label{tau_and_mdd}
		\end{minipage}%
	}%
		\subfigure[$(\bm{\omega}_{w},-\nu\partial_{\omega}\lVert\bm{\omega}_{w}\rVert)$]{
		\begin{minipage}[t]{0.5\linewidth}
			\centering
			\includegraphics[width=1.0\columnwidth,trim={0.3cm 0.2cm 0.3cm 0.6cm},clip]{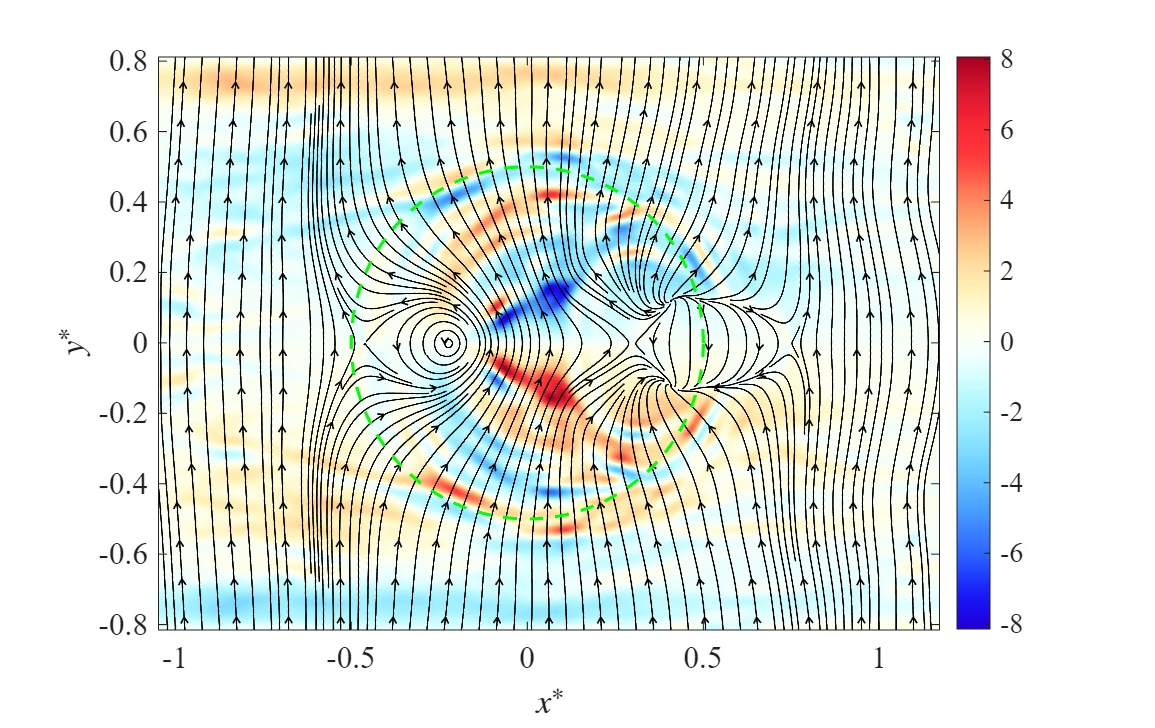}
			\label{omega_and_mdd}
		\end{minipage}%
	}%
	\caption{Skin-friction lines ($\bm{\tau}$-lines) and surface vorticity lines ($\bm{\omega}_{w}$-lines) superimposed on normalized fields of (a,b) $\lVert\bm{\omega}_{w}\rVert$, (c,d) $\kappa_{g,\xi}$, and (e,f) $-\nu\partial_{\omega}\lVert\bm{\omega}_{w}\rVert$. The green circle marks the edge of the hill on the bottom wall.}
	\label{kappa_gtau_omega}
\end{figure}

The wall-normal BVF \(\bm{\sigma}_{n}\) is decomposed according to \eqref{L23a} and \eqref{L23b}. The normalized distributions of \(\sigma_{n}^{R}\) (figure~\ref{sigma_n_R}), \(\sigma_{n}^{S}\) (figure~\ref{sigma_n_S}), and \(\sigma_{n}\) (figure~\ref{sigma_n_total}) are each overlaid with skin-friction lines (\(\bm{\tau}\)-lines). 
In figure~\ref{sigma_n_R}, high-magnitude regions of $\sigma_{n}^{R}$ appear on the hill flanks owing to the high vorticity magnitude (figures~\ref{tau_and_omeganorm} and~\ref{omega_and_omeganorm}), as well as near the downstream spiraling nodes $(N_{2}~\text{and}~N_{3})$ owing to the elevated $\kappa_{g,\tau}(\equiv\kappa_{g,\xi})$ of the spiraling $\bm{\tau}$-lines (figures~\ref{kappa_gtau_and_tau_lines} and~\ref{kappa_gtau_and_omega_lines}). Owing to the influence of $-\nu\partial_{\omega}\lVert\bm{\omega}_{w}\rVert$ (figures~\ref{tau_and_mdd} and~\ref{omega_and_mdd}), the distribution of $\sigma_{n}^{S}$ differs markedly from $\sigma_{n}^{R}$ (figure~\ref{sigma_n_S}). Note that the geodesic curvature \(\kappa_{g,\xi}\) and the vorticity derivative \(\partial_{\omega}\lVert\bm{\omega}_{w}\rVert\) are odd functions of the coordinate \(y\). Consequently, by \eqref{jj123}, all three quantities must exhibit antisymmetric patterns about \(y=0\). The sum of $\sigma_{n}^{R}$ and $\sigma_{n}^{S}$ yields the wall-normal BVF $\sigma_{n}$ in figure~\ref{sigma_n_total}. It is seen that the high-magnitude regions of $\sigma_{n}$ are contributed by both $\sigma_{n}^{R}$ and $\sigma_{n}^{S}$ at the flanks of the hill and by $\sigma_{n}^{R}$ around the downstream nodes ($N_{2}$ and $N_{3}$), respectively. The latter is responsible for the arch vortex formed by the rolled-up separated shear layers on the leeward side of the mountain.
\begin{figure}[t]
	\centering
	\subfigure[$F_{\Omega}=\bm{\xi}\bm{\cdot}\widehat{\bm{\nabla}}_{\pi}\widetilde{P}_{w}+\nu\bm{\omega}_{w}\bm{\cdot}\bm{K}_{w}\bm{\cdot}\bm{\omega}_{w}$]{
		\begin{minipage}[t]{0.5\linewidth}
			\centering
			\includegraphics[width=1.0\columnwidth,trim={0.7cm 1.5cm 0.7cm 1.4cm},clip]{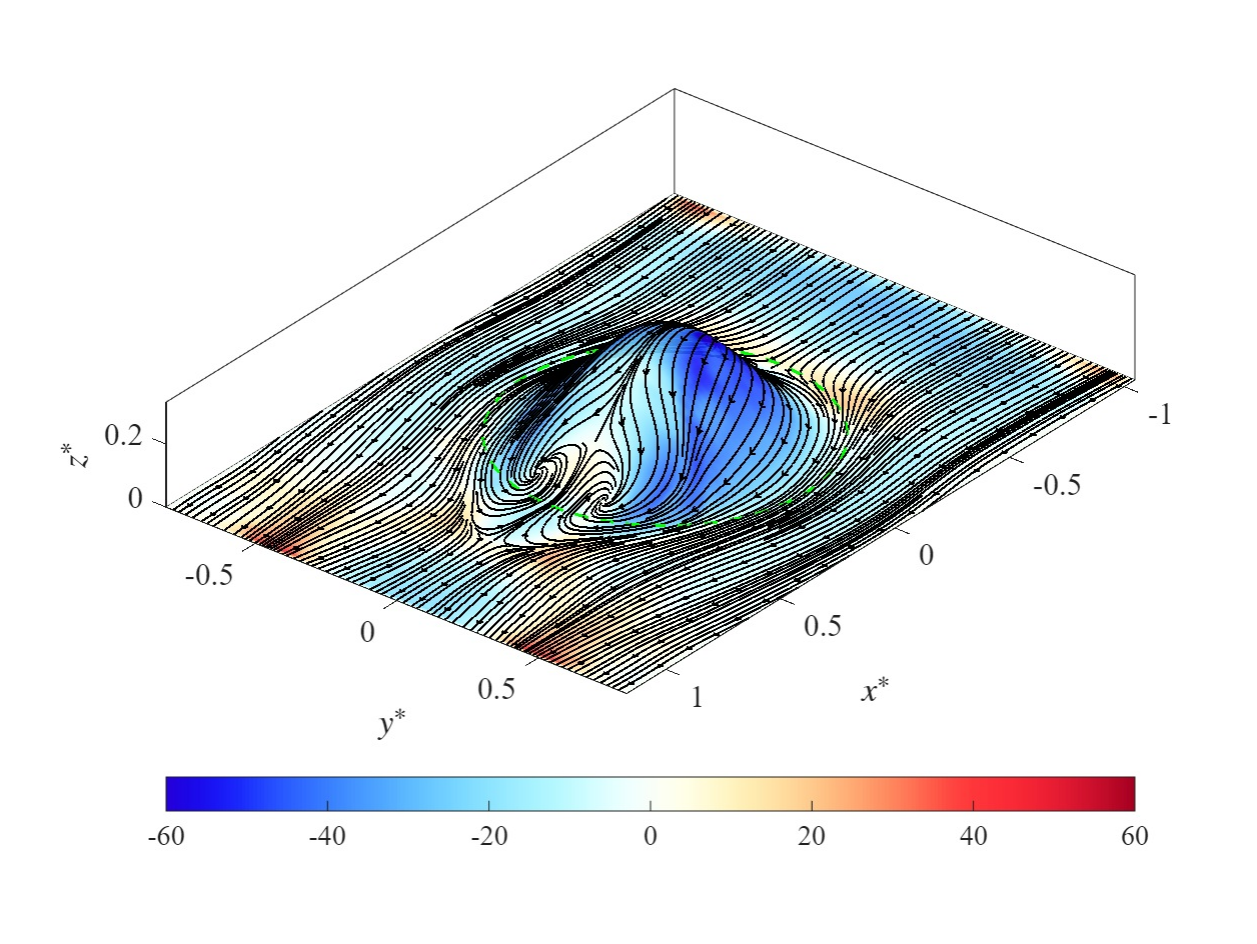}
			\label{FOmega_fig}
		\end{minipage}%
	}%
	\subfigure[$F_{RS}=-2\nu\bm{\xi}\bm{\cdot}\bm{K}_{w}\bm{\cdot}\bm{\xi}$]{
		\begin{minipage}[t]{0.5\linewidth}
			\centering
			\includegraphics[width=1.0\columnwidth,trim={0.7cm 1.5cm 0.7cm 1.4cm},clip]{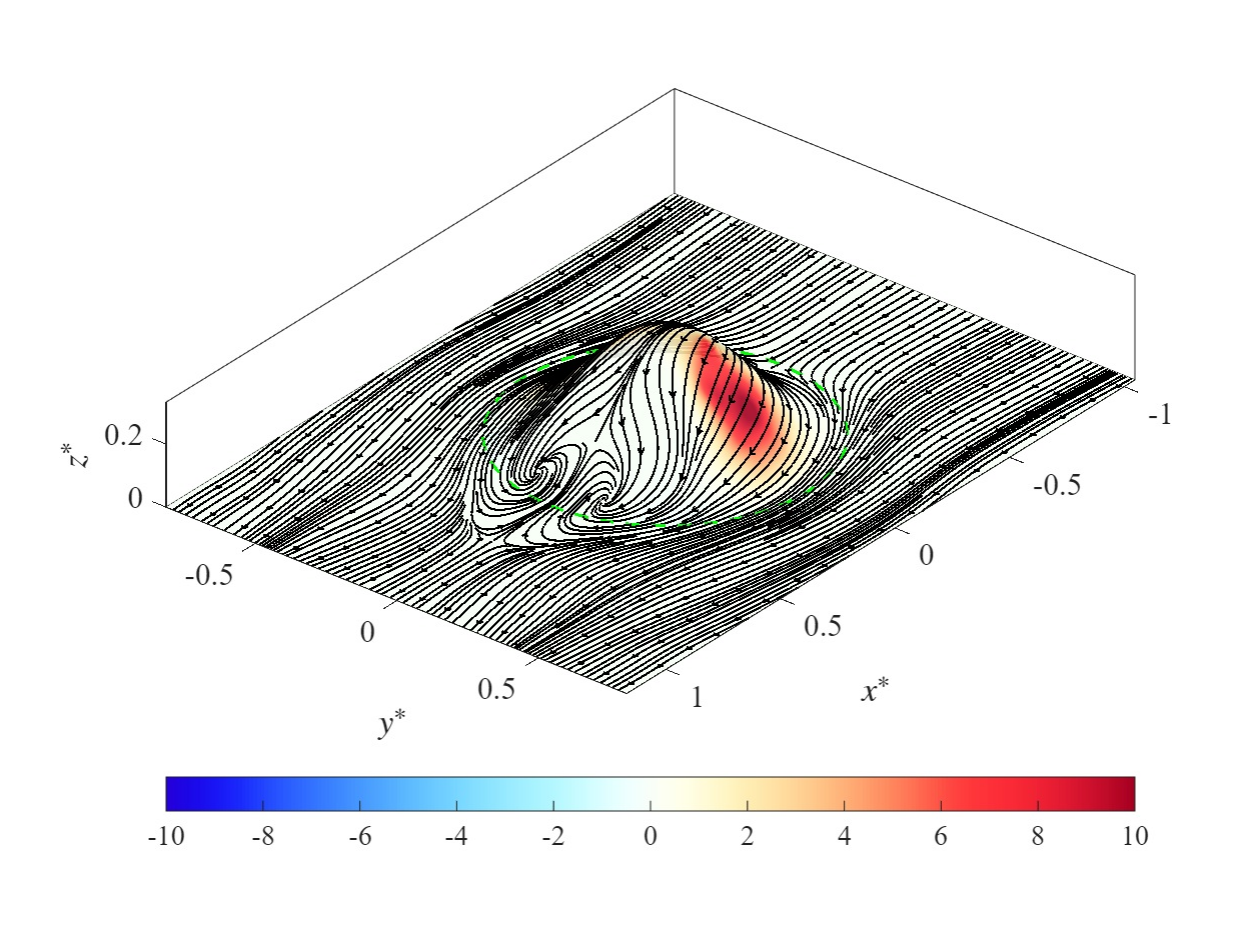}
			\label{FRS_fig}
		\end{minipage}%
	}%
	
	\subfigure[$F_{SS}=2\nu\bm{\xi}\bm{\cdot}\bm{K}_{w}\bm{\cdot}\bm{\xi}+\nu\bm{\omega}_{w}\bm{\cdot}\bm{K}_{w}\bm{\cdot}\bm{\omega}_{w}
	+\bm{\xi}\bm{\cdot}\widehat{\bm{\nabla}}_{\pi}\widetilde{P}_{w}$]{
		\begin{minipage}[t]{0.5\linewidth}
			\centering
			\includegraphics[width=1.0\columnwidth,trim={0.7cm 1.5cm 0.7cm 1.4cm},clip]{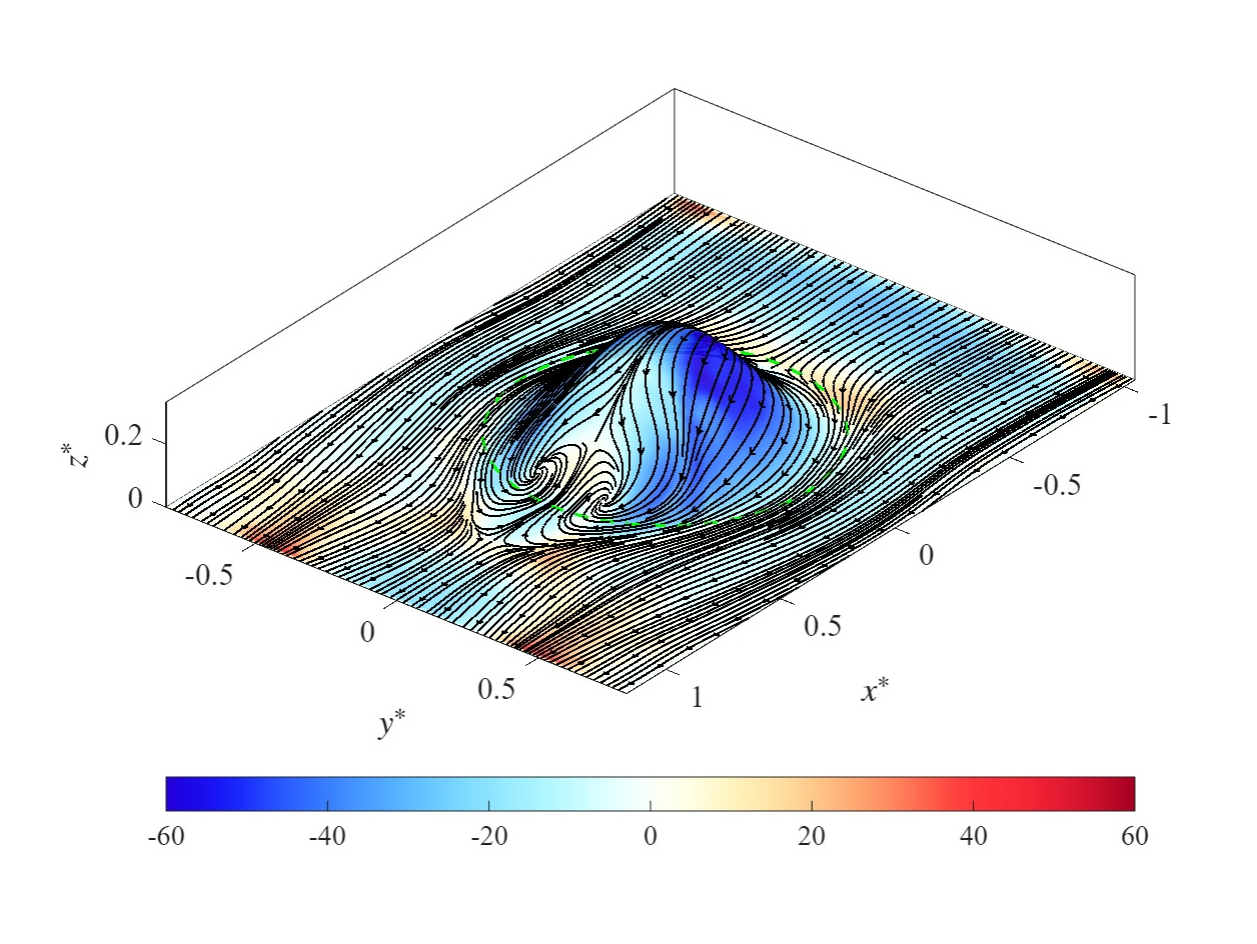}
			\label{FSS_fig}
		\end{minipage}%
	}%
	\subfigure[$2\nu\bm{\xi}\bm{\cdot}\bm{K}_{w}\bm{\cdot}\bm{\xi}$]{
		\begin{minipage}[t]{0.5\linewidth}
			\centering
			\includegraphics[width=1.0\columnwidth,trim={0.7cm 1.5cm 0.7cm 1.4cm},clip]{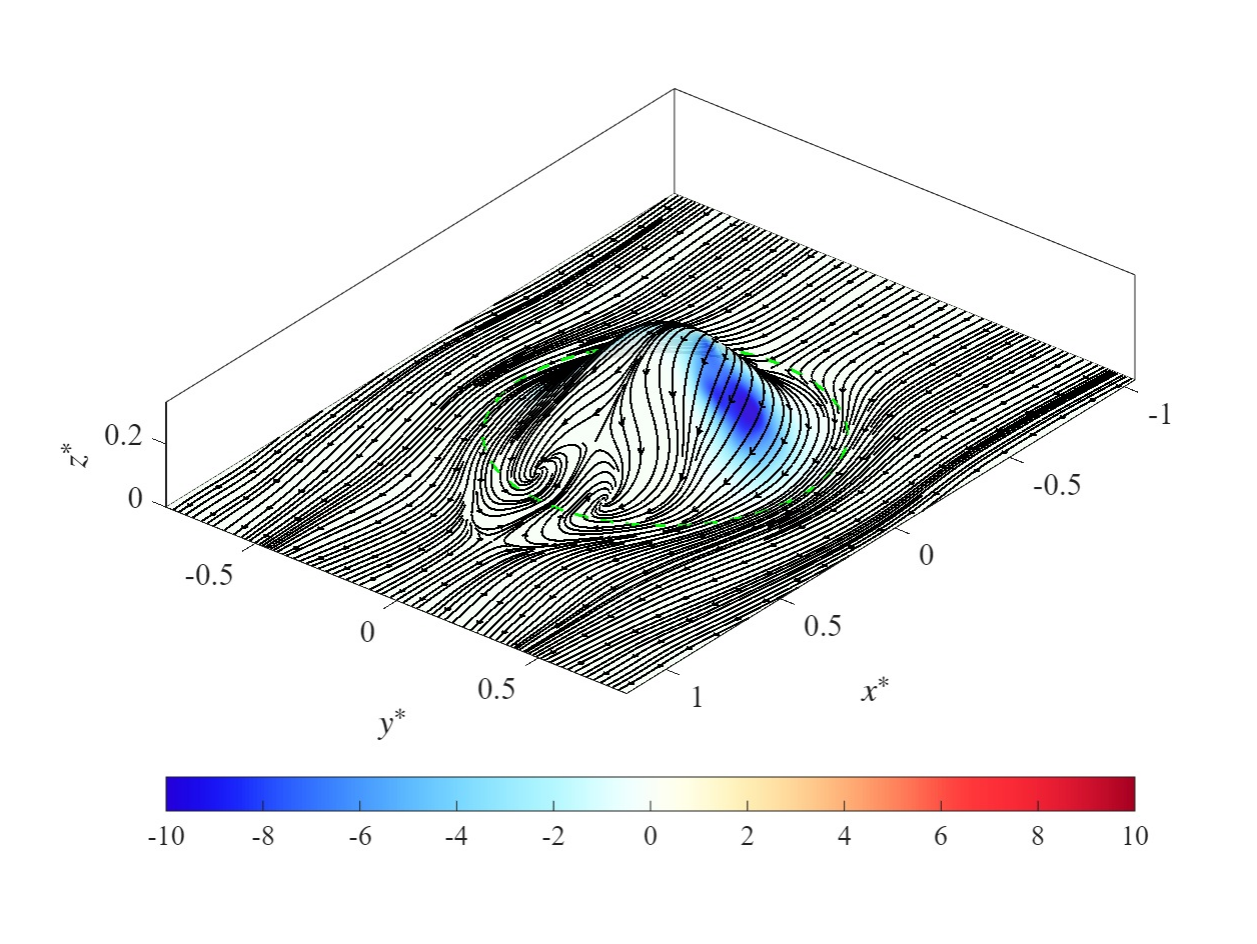}
			\label{FSS1_fig}
		\end{minipage}%
	}%
	
	\subfigure[$\nu\bm{\omega}_{w}\bm{\cdot}\bm{K}_{w}\bm{\cdot}\bm{\omega}_{w}$]{
		\begin{minipage}[t]{0.5\linewidth}
			\centering
			\includegraphics[width=1.0\columnwidth,trim={0.7cm 1.5cm 0.7cm 1.4cm},clip]{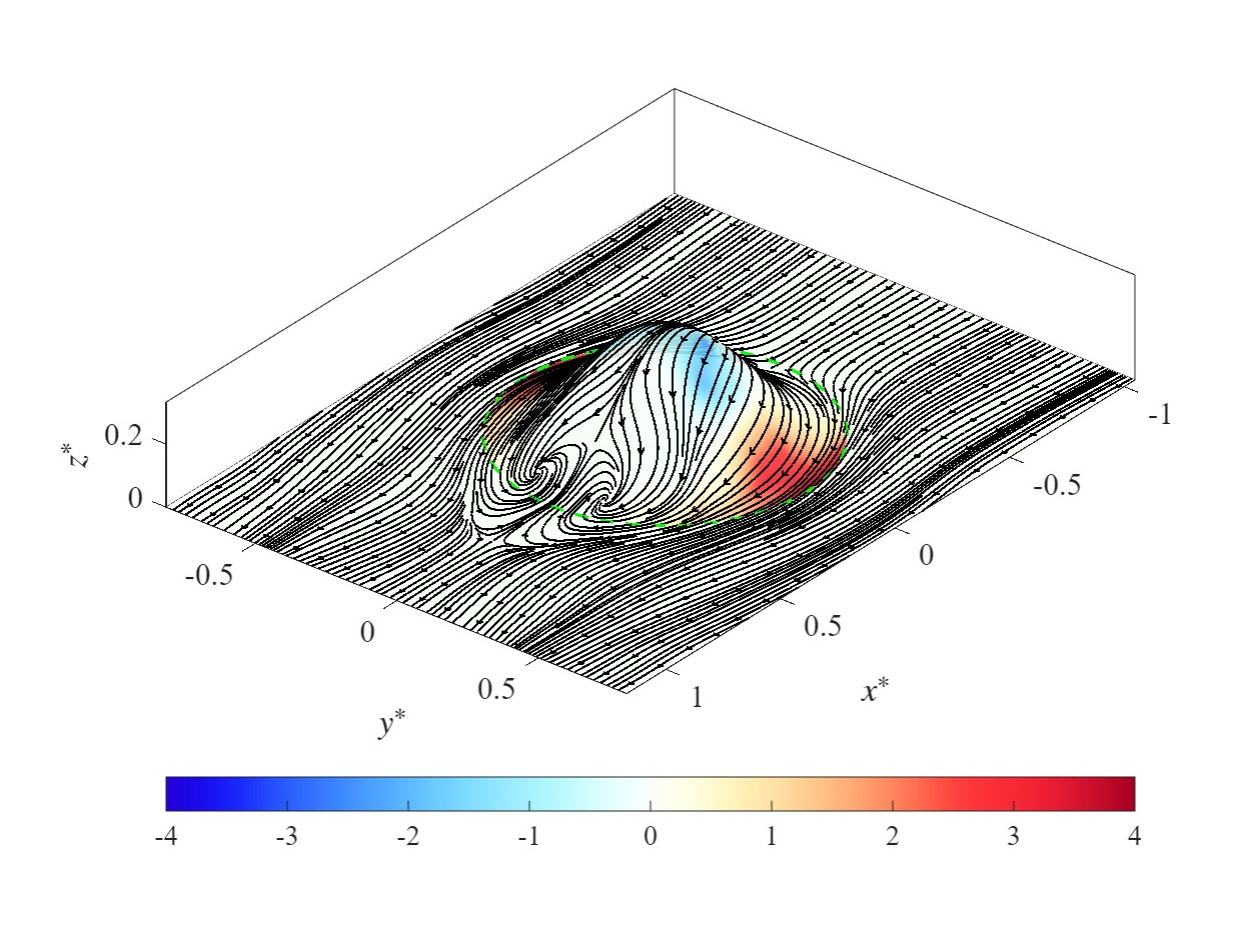}
			\label{FSS2_fig}
		\end{minipage}%
	}%
	\subfigure[$\bm{\xi}\bm{\cdot}\widehat{\bm{\nabla}}_{\pi}\widetilde{P}_{w}$]{
		\begin{minipage}[t]{0.5\linewidth}
			\centering
			\includegraphics[width=1.0\columnwidth,trim={0.7cm 1.5cm 0.7cm 1.4cm},clip]{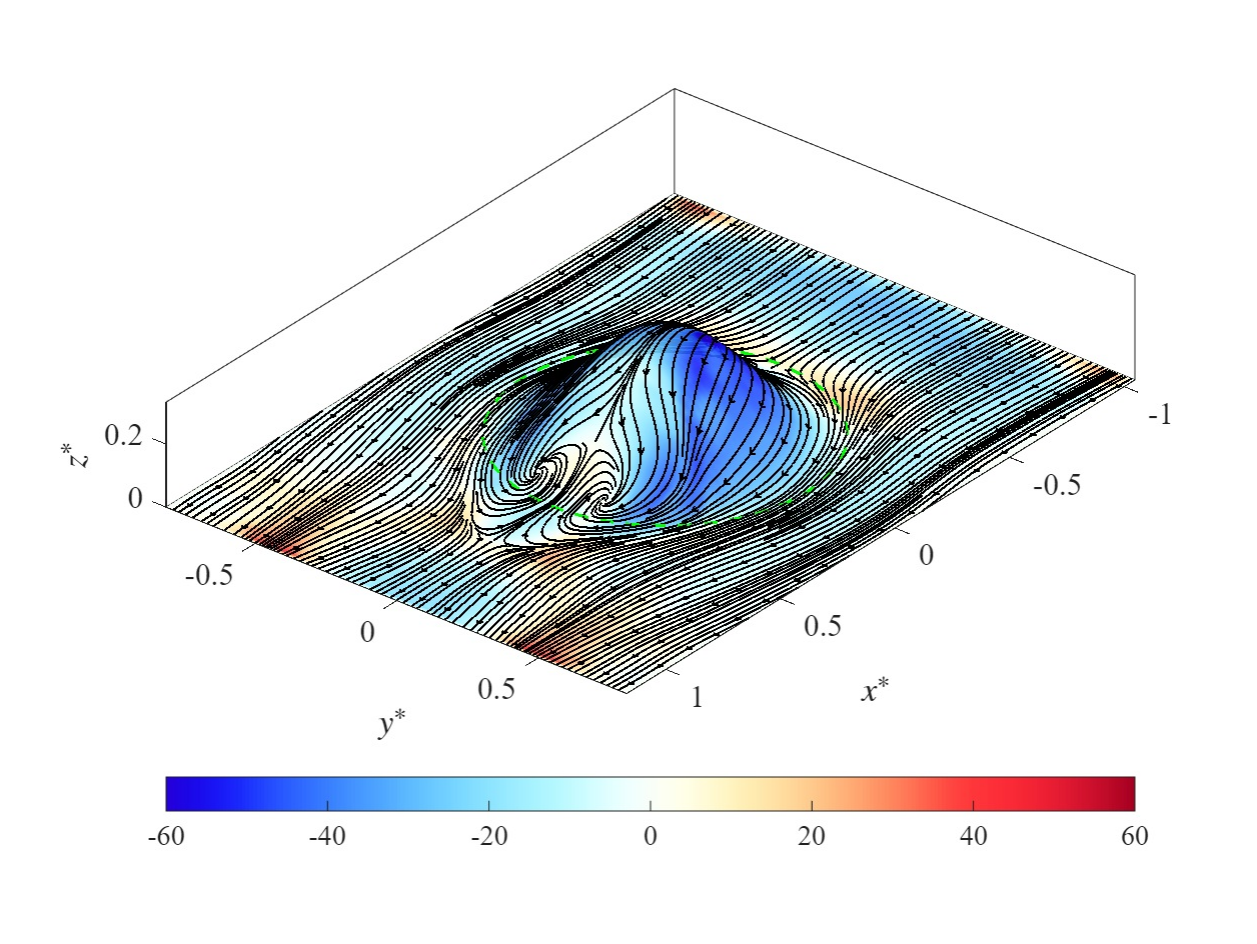}
			\label{FSS3_fig}
		\end{minipage}%
	}%
	\caption{Skin-friction lines ($\bm{\tau}$-lines) superimposed on normalized fields of (a) $F_{\Omega}$, (b) $F_{RS}$, (c) $F_{SS}$, (d) $2\nu\bm{\xi}\bm{\cdot}\bm{K}_{w}\bm{\cdot}\bm{\xi}$, (e) $\nu\bm{\omega}_{w}\bm{\cdot}\bm{K}_{w}\bm{\cdot}\bm{\omega}_{w}$, and (f) $\bm{\xi}\bm{\cdot}\widehat{\bm{\nabla}}_{\pi}\widetilde{P}_{w}$. The green circle indicates the edge of the hill on the bottom wall.} 
	\label{hill_BEF_constituents}
\end{figure}
\begin{figure}[t]
	\centering
	\subfigure[]{
		\begin{minipage}[t]{0.5\linewidth}
			\centering
			\includegraphics[width=1.0\columnwidth,trim={5cm 0cm 5cm 0cm},clip]{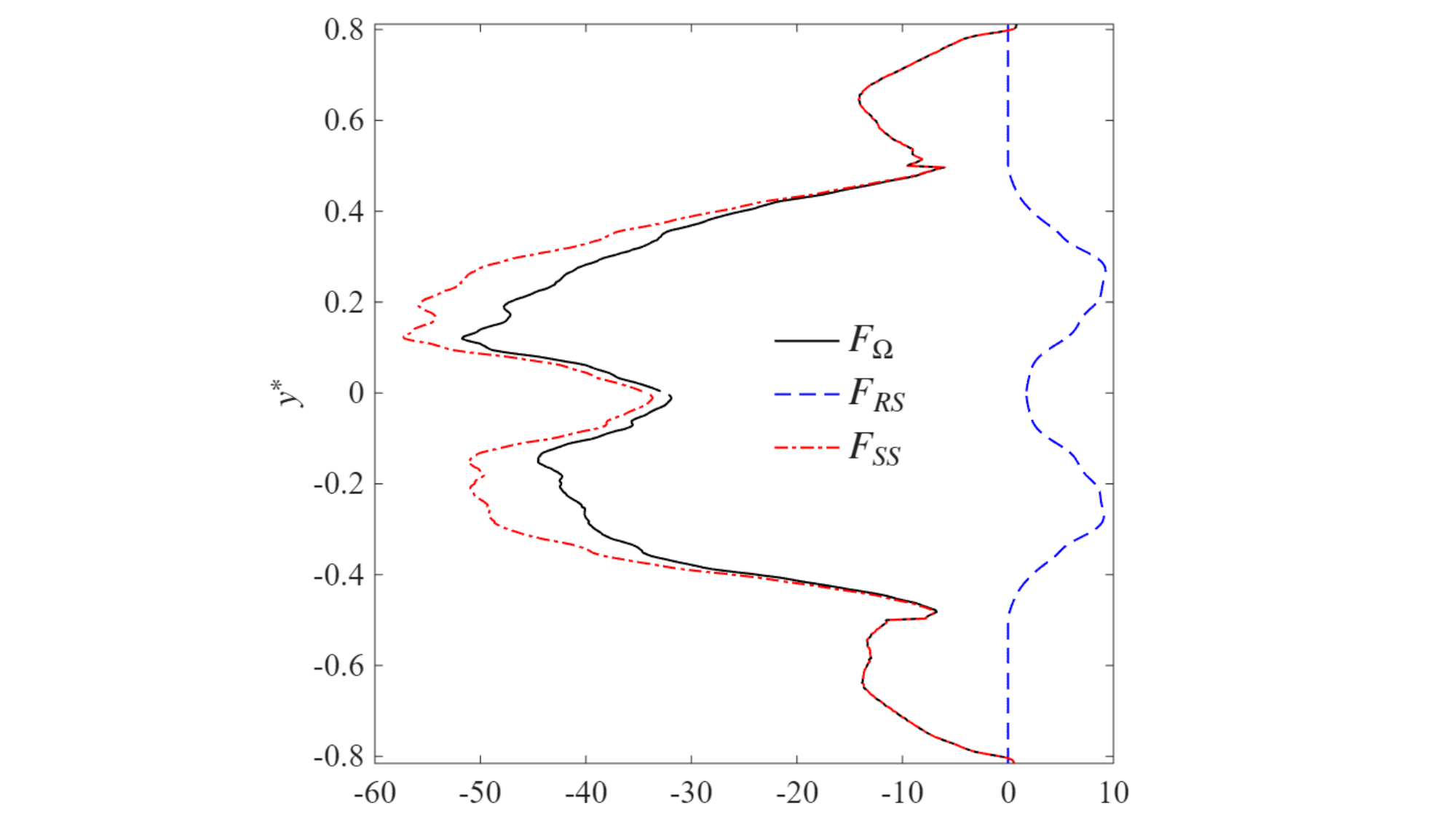}
			\label{comparisonBEF1}
		\end{minipage}%
	}%
	\subfigure[]{
		\begin{minipage}[t]{0.5\linewidth}
			\centering
			\includegraphics[width=1.0\columnwidth,trim={5cm 0cm 5cm 0cm},clip]{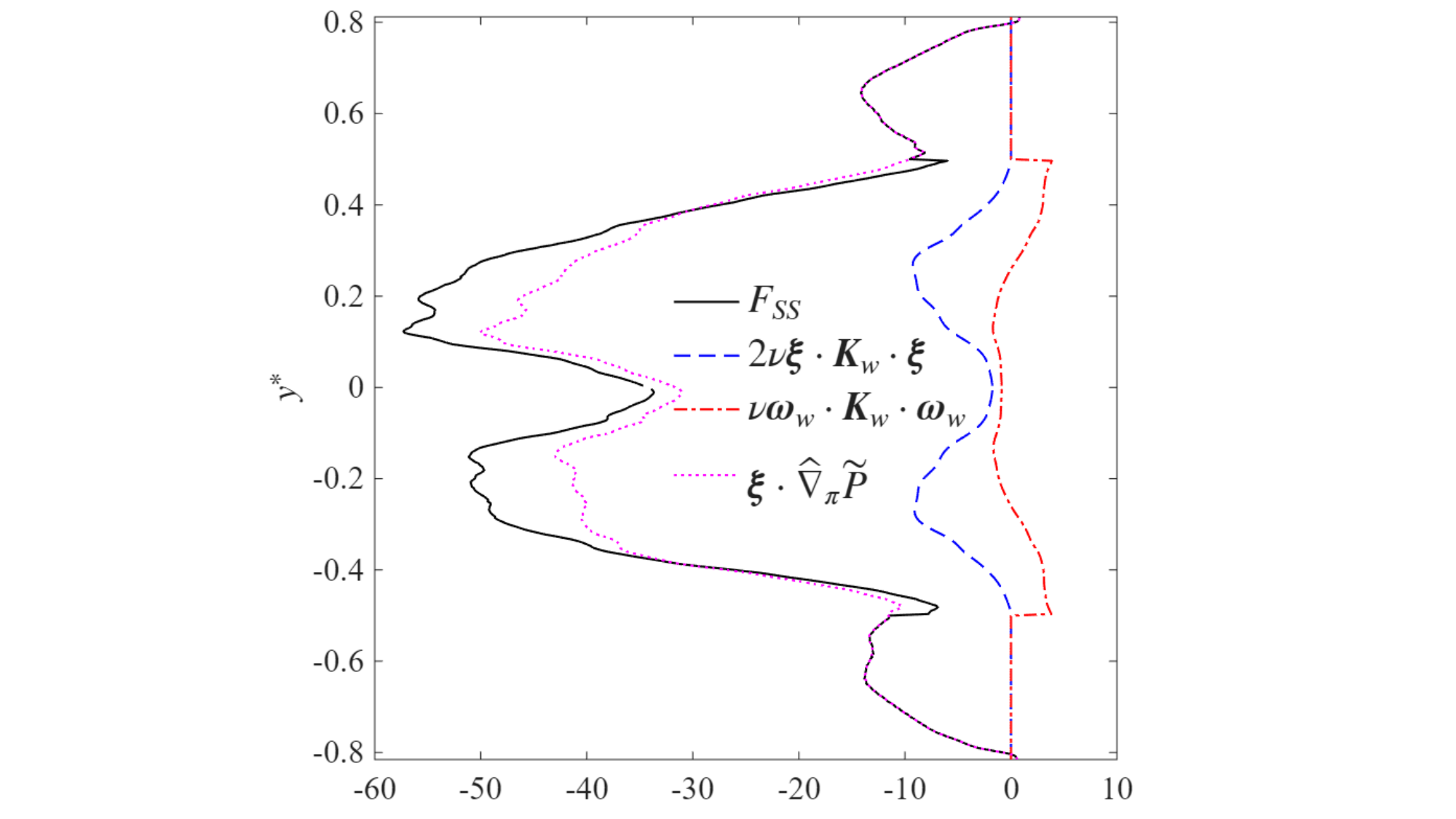}
			\label{comparisonBEF2}
		\end{minipage}%
	}%
	\caption{Comparison of normalized profiles of (a) $(F_{\Omega},F_{RS},F_{SS})$, and (b) $F_{SS}$ together with its three constituents along the $x=0$ curve across the surface.} 
	\label{line_compare2}
\end{figure}

\section{Decomposition of the boundary enstrophy flux}\label{BEF_decomp}
Enstrophy, defined as half the square of the vorticity magnitude $(\Omega\equiv\omega^2/2)$, measures the local intensity of the rotation rate of a fluid volume element irrespective of direction. On the local surface $\bm{\Sigma}_{\zeta}$, using~\eqref{ZZ1}, the enstrophy is decomposed as
\begin{subequations}\label{enstophy1ab}
	\begin{gather}
		\Omega = \Omega_{RR}+\Omega_{SS}+\Omega_{RS}+\Omega_{n},\label{enstrophy1a} \\
		\Omega_{RR} \equiv \frac{1}{2}\lVert\bm{R}_{\Sigma}\rVert^{2},~
		\Omega_{SS}\equiv\frac{1}{2}\lVert\bm{S}_{\Sigma}\rVert^{2},~
		\Omega_{RS}\equiv\bm{R}_{\Sigma}\bm{\cdot}\bm{S}_{\Sigma},~
		\Omega_{n}\equiv\frac{1}{2}\omega_{n}^2,\label{enstrophy1b}
	\end{gather}
\end{subequations}
where $\Omega_{RS}$ is the enstrophy constituent arising from orbital--spin coupling; $\Omega_{R}$ and $\Omega_{S}$ denote the enstrophy constituents associated with the surface-element-based rigid-rotation and spin modes, respectively; and $\Omega_{n}$ arises solely from the surface-normal vorticity component.
The boundary enstrophy flux (BEF) quantifies the rate of enstrophy generation on a no-slip wall per unit area~\citep{Wu1993,LiuTS2016,Chen2019Relations}, which is defined as
\begin{equation}\label{FOmega1}
F_{\Omega}\equiv\nu\left[\frac{\partial\Omega}{\partial n}\right]_{w}=\bm{\omega}_{w}\bm{\cdot}\bm{\sigma}.
\end{equation}
It indicates the boundary sources and sinks of the enstrophy field and is notably linked to topological features such as isolated critical points and separation/attachment lines in a $\bm{\tau}$-field~\citep{Chen2021features}. Enstrophy can be diffused across a boundary only if both vorticity and its flux are nonzero on the boundary~\citep{Wu1995}.   Following~\eqref{enstrophy1a} and~\eqref{enstrophy1b}, the BEF $F_{\Omega}$ in~\eqref{FOmega1} is decomposed as
\begin{subequations}\label{FOmega12345}
	\begin{gather}
		F_{\Omega}=F_{RR}+F_{SS}+F_{RS}+F_{n}, \label{mm1}\\
		F_{RR}\equiv\nu\left[\frac{\partial\Omega_{RR}}{\partial n}\right]_{w}=\left[\bm{R}_{\Sigma}\right]_{w}\bm{\cdot}\bm{\sigma}_{\pi}^{R}, \label{mm2}\\
		F_{SS}\equiv\nu\left[\frac{\partial\Omega_{SS}}{\partial n}\right]_{w}=\left[\bm{S}_{\Sigma}\right]_{w}\bm{\cdot}\bm{\sigma}_{\pi}^{S}, \label{mm3}\\
		F_{RS}\equiv\nu\left[\frac{\partial\Omega_{RS}}{\partial n}\right]_{w}=\left[\bm{R}_{\Sigma}\right]_{w}\bm{\cdot}\bm{\sigma}_{\pi}^{S}+\left[\bm{S}_{\Sigma}\right]_{w}\bm{\cdot}\bm{\sigma}_{\pi}^{R}, \label{mm4}\\
		F_{n}\equiv\nu\left[\frac{\partial\Omega_{n}}{\partial n}\right]_{w}=\left[\omega_{n}\right]_{w}\sigma_{n}.\label{mm5}
	\end{gather}
\end{subequations}
On a stationary wall, the velocity adherence condition $\bm{u}_{w}=\bm{0}$ implies $[\bm{R}_{\Sigma}]_{w}=\bm{0}$, $\bm{\omega}_{w}=[\bm{s}_{\Sigma}]_{w}$, and $\bm{\omega}_{n}=\bm{0}$. Therefore, $F_{RR}=F_{n}=0$, and~\eqref{FOmega12345} simplifies to
\begin{subequations}\label{FOmega12345new}
	\begin{gather}
		F_{\Omega}=F_{SS}+F_{RS}=\bm{\omega}_{w}\bm{\cdot}\bm{\sigma}_{\pi}, \label{mm1new}\\
		F_{SS}=\bm{\omega}_{w}\bm{\cdot}\bm{\sigma}_{\pi}^{S},~
		F_{RS}=\bm{\omega}_{w}\bm{\cdot}\bm{\sigma}_{\pi}^{R}. \label{mm4new}
	\end{gather}
\end{subequations}
By using~\eqref{L17b} and~\eqref{L17c}, $F_{RS}$ and $F_{SS}$ are explicitly expressed as
\begin{subequations}\label{A1ab}
	\begin{gather}
		F_{RS}=-2\nu\bm{\xi}\bm{\cdot}\widehat{\bm{\nabla}}_{\pi}\vartheta_{w}-2\nu\bm{\xi}\bm{\cdot}\bm{K}_{w}\bm{\cdot}\bm{\xi}, \label{A1a} \\
		F_{SS}=2\nu\bm{\xi}\bm{\cdot}\widehat{\bm{\nabla}}_{\pi}\vartheta_{w}+2\nu\bm{\xi}\bm{\cdot}\bm{K}_{w}\bm{\cdot}\bm{\xi}+\nu\bm{\omega}_{w}\bm{\cdot}\bm{K}_{w}\bm{\cdot}\bm{\omega}_{w}
		+\bm{\xi}\bm{\cdot}\widehat{\bm{\nabla}}_{\pi}\widetilde{P}_{w}, \label{A1b}
	\end{gather}
\end{subequations}
where we have used the identity
\begin{equation}\label{ID1}
	\bm{\xi}\bm{\cdot}\bm{K}_{w}\bm{\cdot}\bm{\xi}=\bm{\omega}_{w}\bm{\cdot}\breve{\bm{K}}_{w}\bm{\cdot}\bm{\omega}_{w}.
\end{equation}
Summing~\eqref{A1a} and~\eqref{A1b} yields the BEF $F_{\Omega}$ as
\begin{equation}\label{A2}
	F_{\Omega}=\bm{\xi}\bm{\cdot}\widehat{\bm{\nabla}}_{\pi}\widetilde{P}_{w}+\nu\bm{\omega}_{w}\bm{\cdot}\bm{K}_{w}\bm{\cdot}\bm{\omega}_{w}.
\end{equation}

Equations~\eqref{A1a},~\eqref{A1b} and~\eqref{A2} reveal all the physical mechanisms of enstrophy creation on a no-slip stationary wall. Both $F_{RS}$ and $F_{SS}$ receive contributions from the coupling between skin friction and the surface dilatation gradient. Notably, the two quadratic terms, $\bm{\xi}\bm{\cdot}\bm{K}_{w}\bm{\cdot}\bm{\xi}$ and $\bm{\omega}_{w}\bm{\cdot}\bm{K}_{w}\bm{\cdot}\bm{\omega}_{w}$, appear in~\eqref{A1a} and~\eqref{A1b}, yet only the latter contributes to $F_{\Omega}$ in~\eqref{A2}. The longitudinal--transverse coupling term $\bm{\xi}\bm{\cdot}\widehat{\bm{\nabla}}_{\pi}\widetilde{P}_{w}$ contributes solely to the boundary generation of the spin mode, arising from the intrinsic coupling between skin friction and the surface pressure gradient. Figure~\ref{hill_BEF_constituents} shows the normalized contour maps overlaid with skin-friction lines for $F_{\Omega}$ (figure~\ref{FOmega_fig}), $F_{RS}$ (figure~\ref{FRS_fig}), $F_{SS}$ (figure~\ref{FSS_fig}), $2\nu\bm{\xi}\bm{\cdot}\bm{K}_{w}\bm{\cdot}\bm{\xi}$ (figure~\ref{FSS1_fig}), $\nu\bm{\omega}_{w}\bm{\cdot}\bm{K}_{w}\bm{\cdot}\bm{\omega}_{w}$ (figure~\ref{FSS2_fig}), and $\bm{\xi}\bm{\cdot}\widehat{\bm{\nabla}}_{\pi}\widetilde{P}_{w}$ (figure~\ref{FSS3_fig}). In figures~\ref{comparisonBEF1} and~\ref{comparisonBEF2}, the various BEF constituents are compared along 
$x=0$ curve across the hill surface. From figure~\ref{FOmega_fig}, we observe that $F_{\Omega}$ is predominantly negative on the hill surface, indicating that the wall-generated enstrophy is diffused into the fluid interior. As shown in figure~\ref{comparisonBEF1}, $F_{\Omega}$ (figure~\ref{FOmega_fig}) is dominated by the negative $F_{SS}$ (figure~\ref{FSS_fig}) and attenuated by the positive $F_{RS}$ (figure~\ref{FRS_fig}), implying that orbital--spin coupling term $F_{RS}$ suppresses the boundary generation of the spin mode. As illustrated in figure~\ref{comparisonBEF2}, $F_{SS}$ (figure~\ref{FSS_fig}) is dominated by $2\nu\bm{\xi}\bm{\cdot}\bm{K}_{w}\bm{\cdot}\bm{\xi}$ and $\bm{\xi}\bm{\cdot}\widehat{\bm{\nabla}}_{\pi}\widetilde{P}_{w}$ and is modulated by $\nu\bm{\omega}_{w}\bm{\cdot}\bm{K}_{w}\bm{\cdot}\bm{\omega}_{w}$. Under the surface orthonormal basis $\left\{\widehat{\bm{e}}_{1},\widehat{\bm{e}}_{2},\widehat{\bm{e}}_{3}\right\}=\left\{\widehat{\bm{e}}_{\theta},\widehat{\bm{e}}_{z},\widehat{\bm{e}}_{\zeta}\right\}$ (where $\widehat{\bm{e}}_{3}=\widehat{\bm{e}}_{\zeta}=\bm{n}_{w}$), we have
\begin{subequations}
	\begin{gather}
		\bm{\xi}\bm{\cdot}\bm{K}_{w}\bm{\cdot}\bm{\xi}=\widehat{\lambda}_{z}[\omega_{\theta}]_{w}^2+\widehat{\lambda}_{\theta}[\omega_{z}]_{w}^2, \\
		\bm{\omega}_{w}\bm{\cdot}\bm{K}_{w}\bm{\cdot}\bm{\omega}_{w}=\widehat{\lambda}_{\theta}[\omega_{\theta}]_{w}^2+\widehat{\lambda}_{z}[\omega_{z}]_{w}^2.
	\end{gather}
\end{subequations}
On either flank of the mountain away from the crest, $[\omega_{z}]_{w}\gg[\omega_{\theta}]_{w}$, so that $\bm{\xi}\bm{\cdot}\bm{K}_{w}\bm{\cdot}\bm{\xi}\approx\widehat{\lambda}_{\theta}[\omega_{z}]_{w}^2$ and $\bm{\omega}_{w}\bm{\cdot}\bm{K}_{w}\bm{\cdot}\bm{\omega}_{w}\approx\widehat{\lambda}_{z}[\omega_{z}]_{w}^2$. Since $\widehat{\lambda}_{\theta}<0$ for $r\in(0,3a)$, it follows that $\bm{\xi}\bm{\cdot}\bm{K}_{w}\bm{\cdot}\bm{\xi}<0$ (or $F_{RS}>0$) at the flanks, consistent with figures~\ref{FRS_fig} and~\ref{FSS1_fig}. Since $\widehat{\lambda}_{z}$ varies from positive to negative along the $z$-direction within $r\in(0,3a)$ (i.e., from the hill foot to the hill crest), so does $\bm{\omega}_{w}\bm{\cdot}\bm{K}_{w}\bm{\cdot}\bm{\omega}_{w}$, which is consistent with figure~\ref{FSS2_fig}.

\section{Conclusions}\label{Conclusions and discussions} 
\begin{enumerate}
	\item Using a differential-geometric approach, we propose a quadruple decomposition of the boundary vorticity flux (BVF) within the framework of boundary vorticity dynamics:
	\begin{equation*}
	\bm{\sigma}=\bm{\sigma}_{\pi}^{R}+\bm{\sigma}_{\pi}^{S}+\bm{\sigma}_{n}^{R}+\bm{\sigma}_{n}^{S}.
	\end{equation*}
	Beyond the single-vorticity-based boundary vorticity dynamics established by Lighthill and Wu, the present work provides general mathematical measures for the boundary sources of the rigid-rotation and spin modes inherent in vorticity.
	\item By using the surface-element-based decomposition for the tangential vorticity $(\bm{\omega}_{\pi}={\bm{R}}_{\Sigma}+{\bm{S}}_{\Sigma})$ on the material surface instantaneously coinciding with the local surface $\bm{\Sigma}_{\zeta}$ (see~\eqref{SEB}), the tangential BVF is decomposed as $\bm{\sigma}_{\pi}=\bm{\sigma}_{\pi}^{R}+\bm{\sigma}_{\pi}^{S}$ (see~\eqref{L17ab}), which reveals the boundary creation rates (i.e., the boundary sources) of the rigid rotation mode $\bm{R}_{\Sigma}$ and the spin mode $\bm{S}_{\Sigma}$, respectively. Both vorticity modes are found to be generated by the surface dilatation gradient (i.e., $\pm2\nu\bm{n}_{w}\times\widehat{\bm{\nabla}}_{\pi}\vartheta_{w}$) and the vorticity-curvature coupling mechanisms (i.e., $-2\nu\breve{\bm{K}}_{w}\bm{\cdot}\bm{\omega}_{w}$ for $\bm{\sigma}_{\pi}^{R}$, while $2\nu\breve{\bm{K}}_{w}\bm{\cdot}\bm{\omega}_{w}$ and $\nu\bm{K}_{w}\bm{\cdot}\bm{\omega}_{w}$ for $\bm{\sigma}_{\pi}^{S}$), whereas the surface pressure gradient (i.e., the Lyman flux $\bm{\sigma}_{L}$ in this case) contributes solely to the generation of $\bm{S}_{\Sigma}$. 
	\item By utilizing the streamline-based decomposition of the wall-normal vorticity $(\bm{\omega}_{n}={\bm{R}}_{L}+{\bm{S}}_{L})$ constructed from the parallel velocity field $\bm{u}_{\pi}$ and the Levi-Civita connection $\bm{\nabla}_{C}$ on $\bm{\Sigma}_{\zeta}$ (see~\eqref{q8abff}), the wall-normal BVF is decomposed as $\bm{\sigma}_{n}=\bm{\sigma}_{n}^{R}+\bm{\sigma}_{n}^{S}$ (see~\eqref{L230ab}), which accounts for the boundary creation rates (i.e., the boundary sources) of the rigid rotation mode $\bm{R}_{L}$ and the spin mode $\bm{S}_{L}$, respectively. 
	\item The applicability of the proposed quadruple BVF decomposition is demonstrated using global skin-friction and surface-pressure measurements in a low-speed wind tunnel, revealing subtle surface flow patterns of the BVF and BEF constituents that serve as footprints of near-wall coherent structures. Notably, in the tangential direction, the Lyman flux $\bm{\sigma}_{L}$ dominates the boundary spin generation by the surface pressure gradient, which is significantly higher than the vorticity-curvature interaction terms (i.e., $\bm{\sigma}_{\pi}^{R}$ and $\bm{\sigma}_{K}$). In the wall-normal direction, $\bm{\sigma}_{n}^{R}$ dominates $\bm{\sigma}_{n}$ around the two downstream nodes due to high geodesic curvature of the skin-friction lines, which is responsible for the arch vortex formed by the rolled-up separated shear layers on the leeward side of the mountain.
    \item We extend the decomposition of the BVF to the boundary enstrophy flux (BEF) $(F=F_{RR}+F_{SS}+F_{RS}+F_{n})$, which reveals the boundary sources of various enstrophy constituents on a no-slip stationary wall (see~\eqref{FOmega12345new}~--~\eqref{A2}). Note that the terms $F_{RR}$ and $F_{n}$ vanish because of $\bm{u}_{w}=\bm{0}$. The total BEF $F_{\Omega}$ is predominantly negative on the hill surface, indicating that the wall-generated enstrophy is diffused into the fluid interior. $F_{\Omega}$ is dominated by $F_{SS}$ and attenuated by $F_{RS}$. The two quadratic terms, $\bm{\xi}\bm{\cdot}\bm{K}_{w}\bm{\cdot}\bm{\xi}~(=\bm{\omega}_{w}\bm{\cdot}\breve{\bm{K}}_{w}\bm{\cdot}\bm{\omega}_{w})$ and $\bm{\omega}_{w}\bm{\cdot}\bm{K}_{w}\bm{\cdot}\bm{\omega}_{w}$, describe the intrinsic couplings between $(\bm{\xi},\bm{\omega}_{w})$ and $(\bm{K}_{w},\breve{\bm{K}}_{w})$ in $F_{RS}$ and $F_{SS}$, while $\bm{\omega}_{w}\bm{\cdot}\bm{K}_{w}\bm{\cdot}\bm{\omega}_{w}$ contributes solely to $F_{\Omega}$. $F_{SS}$ is dominated by the first and second dominant terms ($\bm{\xi}\bm{\cdot}\widehat{\bm{\nabla}}_{\pi}\widetilde{P}_{w}$ and $2\nu\bm{\xi}\bm{\cdot}\bm{K}_{w}\bm{\cdot}\bm{\xi}$), which is modulated by $\nu\bm{\omega}_{w}\bm{\cdot}\bm{K}_{w}\bm{\cdot}\bm{\omega}_{w}$.
	\item Future works could incorporate an arbitrarily moving
	and continuously deforming boundary into the present framework, thereby extending the application of the proposed theory from stationary no-slip walls to fluid--structure interaction problems such as flapping wings, compliant coatings, and deforming membranes. A second direction is to extend the application to wall-bounded turbulent flows, where the proposed decompositions of BVF and BEF can help clarify how rigid-rotation and spin modes affect the formation of near-wall coherent structures.
\end{enumerate}

\section*{Declarations}
\begin{itemize}
	\item \textbf{Funding} 	This work was funded by the National Natural Science
	Foundation of China (Grant No. 12402262).
	\item \textbf{Conflict of interest} The authors have no conflicts to disclose.
	\item \textbf{Data availability} 	Data will be made available on request.
	\item \textbf{Author contributions}	Tao Chen was involved in conceptualization, formal analysis, funding acquisition, investigation, visualization, and methodology. He also contributed to the original draft preparation, review and editing of the manuscript. Tianshu Liu designed the experiment and acquired the experimental data for analysis.
	\item \textbf{Author ORCID} Tao Chen https://orcid.org/0000-0001-6838-204X; Tianshu Liu https://orcid.org/0000-0001-6297-1660.
\end{itemize}

\backmatter


\begin{appendices}
\section{A novel representation on the geodesic curvature of a skin friction line}\label{geodesic}
Based on classical differential geometry~\citep{Chen2017DG}, the geodesic curvature of a skin-friction line (i.e., a $\bm{\tau}$-line) is defined as
\begin{equation}\label{kappa1}
	\kappa_{g,\xi}\equiv\frac{\partial\bm{e}_{\tau}}{\partial s_{\tau}}\bm{\cdot}\bm{e}_{\omega},
\end{equation}
where ${s}_{\tau}$ is the arc-length parameter of a $\bm{\tau}$-line.
This form, however, lacks an explicit representation of rotation along a $\bm{\tau}$-line. For this reason, it may be inconvenient for mechanical analysis. By virtue of the Levi-Civita operator $\widehat{\bm{\nabla}}_{C}$ and noting the identity
\begin{equation}\label{kappa2}
	\widehat{\bm{\nabla}}_{\pi}\bm{e}_{\tau}=\widehat{\bm{\nabla}}_{C}\bm{e}_{\tau}+(\bm{K}_{w}\bm{\cdot}\bm{e}_{\tau})\bm{n}_{w},
\end{equation}
\eqref{kappa1} can be equivalently written as
\begin{equation}\label{kappa3}
	\kappa_{g,\xi}=\bm{e}_{\tau}\bm{\cdot}\widehat{\bm{\nabla}}_{\pi}\bm{e}_{\tau}\bm{\cdot}\bm{e}_{\omega}=\bm{e}_{\tau}\bm{\cdot}\widehat{\bm{\nabla}}_{C}\bm{e}_{\tau}\bm{\cdot}\bm{e}_{\omega}.
\end{equation}
The first equality in~\eqref{kappa3} can be equivalently rewritten as
\begin{eqnarray}\label{kappa4}
	\kappa_{g,\xi}&=&\bm{e}_{\tau}\bm{\cdot}(\widehat{\bm{\nabla}}_{\pi}\bm{e}_{\tau}-\bm{e}_{\tau}\widehat{\bm{\nabla}}_{\pi}+\bm{e}_{\tau}\widehat{\bm{\nabla}}_{\pi})\bm{\cdot}\bm{e}_{\omega}\nonumber\\
	&=&\bm{e}_{\tau}\bm{\cdot}(\widehat{\bm{\nabla}}_{\pi}\bm{e}_{\tau}-\bm{e}_{\tau}\widehat{\bm{\nabla}}_{\pi})\bm{\cdot}\bm{e}_{\omega}\nonumber\\
	&=&2\bm{e}_{\tau}\bm{\cdot}\mathscr{A}(\widehat{\bm{\nabla}}_{\pi}\bm{e}_{\tau})\bm{\cdot}\bm{e}_{\omega}\nonumber\\
	&=&(\widehat{\bm{\nabla}}_{\pi}\times\bm{e}_{\tau})\times\bm{e}_{\tau}\bm{\cdot}\bm{e}_{\omega}\nonumber\\
	&=&(\widehat{\bm{\nabla}}_{\pi}\times\bm{e}_{\tau})\cdot(\bm{e}_{\tau}\bm{\times}\bm{e}_{\omega})\nonumber\\
	&=&\bm{n}_{w}\bm{\cdot}(\widehat{\bm{\nabla}}_{\pi}\times\bm{e}_{\tau}),
\end{eqnarray}
where $\mathscr{A}$ is the antisymmetrization operator; $\bm{e}_{\tau}\widehat{\bm{\nabla}}_{\pi}=\widehat{\bm{\nabla}}_{\pi}\bm{e}_{\tau}^{\mathrm{T}}$, with ``$\mathrm{T}$'' denoting the transpose. This indicates that $\kappa_{g,\xi}$ is equal to the wall-normal component of the surface curl of $\bm{e}_{\tau}$.
From~\eqref{kappa2}, the curl of $\bm{e}_{\tau}$ is obtained as
\begin{equation}\label{hp1}
	\widehat{\bm{\nabla}}_{\pi}\times\bm{e}_{\tau}=\widehat{\bm{\nabla}}_{C}\times\bm{e}_{\tau}+(\bm{K}_{w}\bm{\cdot}\bm{e}_{\tau})\times\bm{n}_{w}.
\end{equation}
Therefore, an equivalent expression of~\eqref{kappa4} is found as
\begin{equation}\label{hp2}
	\kappa_{g,\xi}=\bm{n}_{w}\bm{\cdot}(\widehat{\bm{\nabla}}_{C}\times\bm{e}_{\tau}).
\end{equation}
Since $\bm{e}_{\tau}=\bm{e}_{\omega}\times\bm{n}_{w}$, it follows from~\eqref{kappa4} that
\begin{eqnarray}\label{hp3}
\kappa_{g,\xi}&=&\bm{n}_{w}\bm{\cdot}\widehat{\bm{\nabla}}_{\pi}\times(\bm{e}_{\omega}\times\bm{n}_{w})\nonumber\\
&=&\bm{n}_{w}\bm{\cdot}\big(\big(\widehat{\bm{\nabla}}_{\pi}\bm{\cdot}\bm{n}_{w}\big)\bm{e}_{\omega}
-\big(\widehat{\bm{\nabla}}_{\pi}\bm{\cdot}\bm{e}_{\omega}\big)\bm{n}_{w}
+\bm{n}_{w}\bm{\cdot}\widehat{\bm{\nabla}}_{\pi}\bm{e}_{\omega}
-\bm{e}_{\omega}\bm{\cdot}\widehat{\bm{\nabla}}_{\pi}\bm{n}_{w}\big)\nonumber\\
&=&-\widehat{\bm{\nabla}}_{\pi}\bm{\cdot}\bm{e}_{\omega},
\end{eqnarray}
This indicates that $\kappa_{g,\xi}$ is equal to the negative divergence of 
$\bm{e}_{\omega}$. Equation~\eqref{hp3} can be alternatively derived from~\eqref{kappa1} as follows:
\begin{eqnarray}
\kappa_{g,\xi}&=&\frac{\partial\bm{e}_{\tau}}{\partial s_{\tau}}\bm{\cdot}\bm{e}_{\omega}\nonumber\\
&=&\frac{\partial}{\partial s_{\tau}}\left(\bm{e}_{\tau}\bm{\cdot}\bm{e}_{\omega}\right)-\bm{e}_{\tau}\bm{\cdot}\frac{\partial\bm{e}_{\omega}}{\partial s_{\tau}}\nonumber\\
&=&-\bm{e}_{\tau}\bm{\cdot}\frac{\partial\bm{e}_{\omega}}{\partial s_{\tau}}\nonumber\\
&=&-\bm{e}_{\tau}\bm{\cdot}\frac{\partial\bm{e}_{\omega}}{\partial s_{\tau}}-\bm{e}_{\omega}\bm{\cdot}\frac{\partial\bm{e}_{\omega}}{\partial s_{\omega}}\nonumber\\
&=&-\widehat{\bm{\nabla}}_{\pi}\bm{\cdot}\bm{e}_{\omega}.
\end{eqnarray}

\section{Near-wall Taylor-series expansion and wall-normal BVF decomposition}\label{Near-wall Taylor-series expansion and the decomposition of the wall-normal BVF}
By applying the Taylor-series expansion to the NS equations and using the velocity adherence condition, we obtain the formal solution of the velocity field in a small neighborhood of the wall~\citep{Chen2022Nearwall}:
\begin{subequations}
	\begin{gather}
		\bm{u}_{\pi}=\bm{\xi}\zeta+\left(\frac{1}{2\nu}\widehat{\bm{\nabla}}_{\pi}\breve{P}_{w}+\frac{1}{2}{\rm tr}(\bm{K}_{w})\bm{\xi}\right)\zeta^2+\bm{\mathcal{O}}(\zeta^3), \\
		u_{\zeta}=\vartheta_{w}\zeta-\frac{1}{2}\left(\widehat{\bm{\nabla}}_{\pi}\bm{\cdot}\bm{\xi}-{\rm tr}(\bm{K}_{w})\vartheta_{w}-\left[\frac{\partial\vartheta}{\partial n}\right]_{w}\right)\zeta^2+\mathcal{O}(\zeta^3),
	\end{gather}
\end{subequations}
where the pressure $\breve{P}$ is defined by
\begin{equation}
\breve{P}\equiv\frac{p-(\mu_{\vartheta}-\mu)\vartheta}{\rho}=\frac{p-(\mu_{b}+\frac{1}{3}\mu)\vartheta}{\rho}=\widetilde{P}+\nu\vartheta.
\end{equation}
On the local surface $\bm{\Sigma}_{\zeta}$, we introduce the right-handed orthogonal frame $(\bm{t},\bm{t}_{\perp},\bm{n}_{\Sigma})$ satisfying $\bm{t}_{\perp}\equiv\bm{n}_{\Sigma}\times\bm{t}$ for $\bm{u}_{\pi}\neq\bm{0}$ (\S\ref{Decomposition of the wall-normal BVF}). In general, as $\zeta\rightarrow 0^+$, one can define the orthogonal pair $(\bm{t}_{w},[\bm{t}_{\perp}]_{w})$ as
\begin{equation}
\bm{t}_{w}\equiv\lim\limits_{\zeta\rightarrow0^+}\bm{t}=\lim\limits_{\zeta\rightarrow0^+}\frac{\bm{u}_{\pi}}{\lVert\bm{u}_{\pi}\rVert},~[\bm{t}_{\perp}]_{w}\equiv\lim\limits_{\zeta\rightarrow0^+}\bm{t}_{\perp}=\bm{n}_{w}\times\bm{t}_{w}.
\end{equation}
Consequently, we obtain
\begin{itemize}
	\item At a regular point where $\bm{\omega}_{w}\neq\bm{0}$, it holds that 
	\begin{subequations}
	\begin{equation}\label{A41}
	\bm{t}_{w}=\frac{\bm{\xi}}{\lVert\bm{\xi}\rVert}=\bm{e}_{\xi},~[\bm{t}_{\perp}]_{w}=\bm{n}_{w}\times\bm{e}_{\xi}=\bm{e}_{\omega};
	\end{equation}
	\item At a critical point where $\bm{\omega}_{w}=\bm{0}$ and $\widehat{\bm{\nabla}}_{\pi}\breve{P}_{w}\neq\bm{0}$, it holds that 
	\begin{equation}\label{A42}
		\bm{t}_{w}=\frac{\widehat{\bm{\nabla}}_{\pi}\breve{P}_{w}}{\lVert\widehat{\bm{\nabla}}_{\pi}\breve{P}_{w}\rVert},~[\bm{t}_{\perp}]_{w}=\bm{n}_{w}\times\widehat{\bm{\nabla}}_{\pi}\breve{P}_{w}.
	\end{equation}
	\end{subequations}
\end{itemize}
By utilizing $\bm{u}_{w}=\bm{0}$ and the identity
\begin{equation}
\bm{\nabla}_{\pi}\bm{u}_{\pi}=\bm{\nabla}_{C}\bm{u}_{\pi}+(\bm{K}\bm{\cdot}\bm{u}_{\pi})\bm{n}_{\Sigma},
\end{equation}
the wall-normal derivative of $\bm{\nabla}_{C}\bm{u}_{\pi}$ evaluates to
\begin{eqnarray}
\left[\frac{\partial}{\partial n}\bm{\nabla}_{C}\bm{u}_{\pi}\right]_{w}&=&\widehat{\bm{\nabla}}_{\pi}\left[\frac{\partial\bm{u}_{\pi}}{\partial n}\right]_{w}-\left(\bm{K}_{w}\bm{\cdot}\left[\frac{\partial\bm{u}_{\pi}}{\partial n}\right]_{w}\right)\bm{n}_{w}\nonumber\\
&=&\widehat{\bm{\nabla}}_{\pi}\bm{\xi}
-\left(\bm{K}_{w}\bm{\cdot}\bm{\xi}\right)\nonumber\\
&=&\widehat{\bm{\nabla}}_{C}\bm{\xi}.
\end{eqnarray}
We claim that $\bm{\xi}=\bm{0}$ at a critical point does not imply that \(\bm{\xi}\) vanishes in the vicinity of this point.
Note that~\eqref{q80ff} and~\eqref{q8aff} can be equivalently written as
\begin{subequations}
	\begin{gather}
		\bm{R}_{L}(\bm{t})=2(\bm{t}\bm{\cdot}\bm{\nabla}_{C}\bm{u}_{\pi}\bm{\cdot}\bm{t}_{\perp})\bm{n}_{\Sigma}, \\
		\bm{S}_{L}(\bm{t})=-2(\bm{t}\bm{\cdot}\mathscr{S}[\bm{\nabla}_{C}\bm{u}_{\pi}]\bm{\cdot}\bm{t}_{\perp})\bm{n}_{\Sigma}.
	\end{gather}
\end{subequations}
The two wall-normal BVF constituents are well defined via the following limits:
\begin{subequations}
	\begin{gather}
		\bm{\sigma}_{n}^{R}\equiv\nu\left[\frac{\partial\bm{R}_{L}}{\partial n}\right]_{w}\equiv\lim\limits_{\zeta\rightarrow 0^+}\frac{\bm{R}_{L}-[\bm{R}_{L}]_{w}}{\zeta}=\lim\limits_{\zeta\rightarrow 0^+}\frac{\bm{R}_{L}}{\zeta}, \\
		\bm{\sigma}_{n}^{S}\equiv\nu\left[\frac{\partial\bm{S}_{L}}{\partial n}\right]_{w}\equiv\lim\limits_{\zeta\rightarrow 0^+}\frac{\bm{S}_{L}-[\bm{S}_{L}]_{w}}{\zeta}=\lim\limits_{\zeta\rightarrow 0^+}\frac{\bm{S}_{L}}{\zeta},
	\end{gather}
\end{subequations}
which are evaluated as
\begin{subequations}
	\begin{gather}
		\bm{\sigma}_{n}^{R}=2\nu(\bm{t}_{w}\bm{\cdot}\widehat{\bm{\nabla}}_{C}\bm{\xi}\bm{\cdot}[\bm{t}_{\perp}]_{w})\bm{n}_{w}=2\nu\bm{t}_{w}\times(\bm{t}_{w}\bm{\cdot}\widehat{\bm{\nabla}}_{C}\bm{\xi}), \\
		\bm{\sigma}_{n}^{S}=-2\nu(\bm{t}_{w}\bm{\cdot}\mathscr{S}[\widehat{\bm{\nabla}}_{C}\bm{\xi}]\bm{\cdot}[\bm{t}_{\perp}]_{w})\bm{n}_{w}=-2\nu\bm{t}_{w}\times(\bm{t}_{w}\bm{\cdot}\mathscr{S}[\widehat{\bm{\nabla}}_{C}\bm{\xi}]),
	\end{gather}
\end{subequations}
where $(\bm{t}_{w},[\bm{t}_{\perp}]_{w})$ is given by~\eqref{A41} at a regular point, and by~\eqref{A42} at a critical point of the skin-friction field.

\section{Geometry analysis of the hill model}\label{geometry}
The hill surface $\bm{\Sigma}_{w}$ can be parametrized by the curvilinear coordinate system $(\theta,z)$, where $z$ is the height coordinate and $\theta$ is the polar angle on the cross section at a fixed height $z$. The physical space is described by the Cartesian coordinate system $(X,Y,Z)$ with the orthonormal basis $(\bm{i}_{X},\bm{i}_{Y},\bm{i}_{Z})$, where the origin $O$ is placed at the location $(X,Y,Z)=(0,0,0)$. This smooth mapping admits the explicit form
\begin{eqnarray}
&&\bm{x}=(\theta,z)\in\mathscr{D}_{\bm{x}}\subset\mathbb{R}^2\mapsto\nonumber\\ &&\bm{\Sigma}_{w}(\theta,z)=\begin{bmatrix} 
	\bm{i}_{X} & \bm{i}_{Y} & \bm{i}_{Z}
\end{bmatrix}	\begin{bmatrix}
	X \\
	Y \\
	Z
\end{bmatrix}(\theta,z)
=
\begin{bmatrix} 
	\bm{i}_{X} & \bm{i}_{Y} & \bm{i}_{Z}
\end{bmatrix}
\begin{bmatrix}
	X_{c}+r(z)\cos\theta\\
	Y_{c}+r(z)\sin\theta\\
	z
\end{bmatrix}\in\mathbb{R}^{3},
\end{eqnarray}
where $(X_{c},Y_{c})$ corresponds to the axis of rotational symmetry $(r=0)$. The maximum height of the hill is $2\lambda a$, whereas the base diameter is $6a$. Thus, the height-to-base-diameter ratio is $\lambda/3$. The domain of definition is given by $\mathscr{D}_{\bm{x}}\equiv(0,2\pi)\times(0,2\lambda a)$. Here, $r=r(z)$ is the inverse function of the height function
\begin{equation}
	z(r)=\lambda a\cos\left(\frac{\pi r}{3a}\right)+\lambda a.
\end{equation}
The first-order derivative of $z(r)$ with respect to $r$ is denoted by
\begin{equation}
\dot{z}(r)\equiv\frac{dz}{dr}=z_{r}=-\frac{\lambda\pi}{3}\sin\left(\frac{\pi r}{3a}\right).
\end{equation}
The first- and second-order derivatives of $r(z)$ with respect to $z$ take the form
\begin{subequations}
	\begin{gather}
		\dot{r}(z)\equiv\frac{dr}{dz}=r_{z}=-\frac{3}{\lambda\pi} \csc\left(\frac{\pi r}{3a}\right), \\
		\ddot{r}(z)\equiv\frac{d^2r}{dz^2}=r_{zz}= -\frac{3}{\lambda^2 \pi a} \csc^2\left(\frac{\pi r}{3a}\right) \cot\left(\frac{\pi r}{3a}\right).
	\end{gather}
\end{subequations}
At a point $P(\bm{x})\in\bm{\Sigma}_{w}$, the local covariant basis vectors $\left\{\widehat{\bm{g}}_{\theta},\widehat{\bm{g}}_{z}\right\}$ spanning the tangent plane $T_{P}\bm{\Sigma}_{w}$ are expressed as
\begin{subequations}
	\begin{gather}
		\widehat{\bm{g}}_{\theta}(\theta,z)\equiv\frac{\partial\bm{\Sigma}_{w}}{\partial\theta}(\theta,z)
		=\begin{bmatrix} 
			\bm{i}_{X} & \bm{i}_{Y} & \bm{i}_{Z}
		\end{bmatrix}
		\begin{bmatrix}
			-r(z)\sin\theta\\
			r(z)\cos\theta\\
			0
		\end{bmatrix}, \\
		\widehat{\bm{g}}_{z}(\theta,z)\equiv\frac{\partial\bm{\Sigma}_{w}}{\partial z}(\theta,z)
		=\begin{bmatrix} 
			\bm{i}_{X} & \bm{i}_{Y} & \bm{i}_{Z}
		\end{bmatrix}
		\begin{bmatrix}
			\dot{r}(z)\cos\theta\\
			\dot{r}(z)\sin\theta\\
			1
		\end{bmatrix},
	\end{gather}
\end{subequations}
which represent the tangent vectors to the coordinate curves. The unit normal vector to $\bm{\Sigma}_{w}$ is given by
\begin{equation}\label{nw}
	\bm{n}_{w}\equiv\frac{\widehat{\bm{g}}_{\theta}\times\widehat{\bm{g}}_{z}}{\lVert\widehat{\bm{g}}_{\theta}\times\widehat{\bm{g}}_{z}\rVert}	=	\begin{bmatrix} 
		\bm{i}_{X} & \bm{i}_{Y} & \bm{i}_{Z}
	\end{bmatrix}\frac{1}{\sqrt{1+\dot{r}^2(z)}}\begin{bmatrix}
		\cos\theta\\
		\sin\theta\\
		-\dot{r}(z)
	\end{bmatrix}.
\end{equation}
Upon normalizing the surface covariant basis, we obtain an orthonormal triad $\left\{\widehat{\bm{e}}_{1},\widehat{\bm{e}}_{2},\widehat{\bm{e}}_{3}\right\}=\left\{\widehat{\bm{e}}_{\theta},\widehat{\bm{e}}_{z},\bm{n}_{w}\right\}$ attached to the surface, where
\begin{subequations}
	\begin{gather}
		\widehat{\bm{e}}_{\theta}\equiv\frac{\widehat{\bm{g}}_{\theta}}{\lVert\widehat{\bm{g}}_{\theta}\rVert}
		=\begin{bmatrix} 
			\bm{i}_{X} & \bm{i}_{Y} & \bm{i}_{Z}
		\end{bmatrix}
		\begin{bmatrix}
			-\sin\theta\\
			\cos\theta\\
			0
		\end{bmatrix},\label{etheta} \\
		\widehat{\bm{e}}_{z}\equiv\frac{\widehat{\bm{g}}_{z}}{\lVert\widehat{\bm{g}}_{z}\rVert}
		=\begin{bmatrix} 
			\bm{i}_{X} & \bm{i}_{Y} & \bm{i}_{Z}
		\end{bmatrix}\frac{1}{\sqrt{1+\dot{r}^2(z)}}\begin{bmatrix}
			\dot{r}(z)\cos\theta\\
			\dot{r}(z)\sin\theta\\
			1
		\end{bmatrix}.\label{ez}
	\end{gather}
\end{subequations}
\begin{subequations}
The surface metric tensor is expanded as
\begin{equation}
	\bm{G}_{w}=\widehat{g}_{\alpha\beta}\widehat{\bm{g}}^{\alpha}\widehat{\bm{g}}^{\beta}=\widehat{g}_{\theta\theta}\widehat{\bm{g}}^{\theta}\widehat{\bm{g}}^{\theta}+\widehat{g}_{zz}\widehat{\bm{g}}^{z}\widehat{\bm{g}}^{z},
\end{equation}
where the covariant component matrix is evaluated as 
\begin{equation}
	\left[\widehat{g}_{\alpha\beta} \right] = 
	\begin{bmatrix} 
		\widehat{g}_{\theta\theta} & 0 \\ 
		0 & \widehat{g}_{zz} 
	\end{bmatrix} = 
	\begin{bmatrix} 
		r^2(z) & 0 \\ 
		0 & 1 + \dot{r}^2(z) 
	\end{bmatrix}.
\end{equation}
\end{subequations}
The surface curvature tensor pair $(\bm{K}_{w},\breve{\bm{K}}_{w})$ is evaluated as
\begin{equation}
	\bm{K}_{w}=\widehat{\lambda}_{\theta}\widehat{\bm{e}}_{\theta}\widehat{\bm{e}}_{\theta}+\widehat{\lambda}_{z}\widehat{\bm{e}}_{z}\widehat{\bm{e}}_{z},~\breve{\bm{K}}_{w}=\widehat{\lambda}_{z}\widehat{\bm{e}}_{\theta}\widehat{\bm{e}}_{\theta}+\widehat{\lambda}_{\theta}\widehat{\bm{e}}_{z}\widehat{\bm{e}}_{z},
\end{equation}
where the principal curvatures $(\widehat{\lambda}_{\theta},\widehat{\lambda}_{z})$ are given by
\begin{subequations}
	\begin{gather}
		\widehat{\lambda}_\theta = -\frac{1}{r\left[1 + \dot{r}^2(z)\right]^{\frac{1}{2}}} = -\frac{\lambda\pi}{3r}\sin\left(\frac{\pi r}{3a}\right)\left[1 + \frac{\lambda^2\pi^2}{9}\sin^2\left(\frac{\pi r}{3a}\right)\right]^{-\frac{1}{2}},\\
		\widehat{\lambda}_z = \frac{\ddot{r}(z)}{\left[1 + \dot{r}^2(z)\right]^{\frac{3}{2}}} = -\frac{\lambda\pi^2}{9a}\cos\left(\frac{\pi r}{3a}\right)\left[1 + \frac{\lambda^2\pi^2}{9}\sin^2\left(\frac{\pi r}{3a}\right)\right]^{-\frac{3}{2}}.
	\end{gather}
\end{subequations}
Both $\widehat{\lambda}_{\theta}$ and $\widehat{\lambda}_{z}$ converge to the same limit $-\lambda\pi^2/(9a)$ when approaching the peak of the hill surface (i.e., $r\rightarrow 0^+$). The normalized maps of \(\widehat{\lambda}_{\theta}\) and \(\widehat{\lambda}_{z}\) are shown in figures C1(a) and C1(b), respectively.
\begin{figure}[t]
	\centering
		\subfigure[$(\bm{\tau},\widehat{\lambda}_{\theta})$]{
		\begin{minipage}[t]{0.5\linewidth}
			\centering
			\includegraphics[width=1.0\columnwidth,trim={0.7cm 1.5cm 0.7cm 1.4cm},clip]{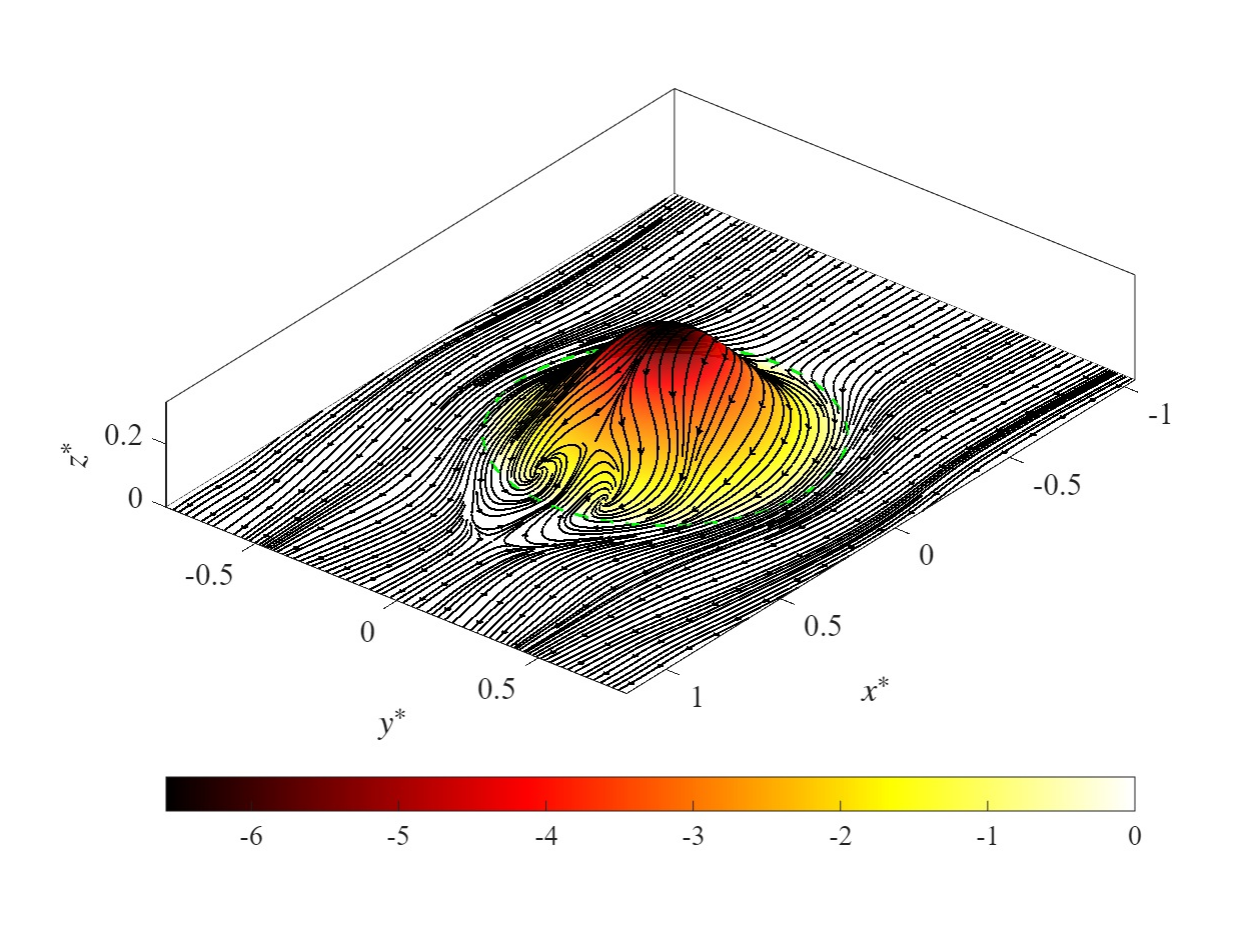}
			\label{Lambda_tt}
		\end{minipage}%
	}%
	\subfigure[$(\bm{\tau},\widehat{\lambda}_{z})$]{
		\begin{minipage}[t]{0.5\linewidth}
			\centering
			\includegraphics[width=1.0\columnwidth,trim={0.7cm 1.5cm 0.7cm 1.4cm},clip]{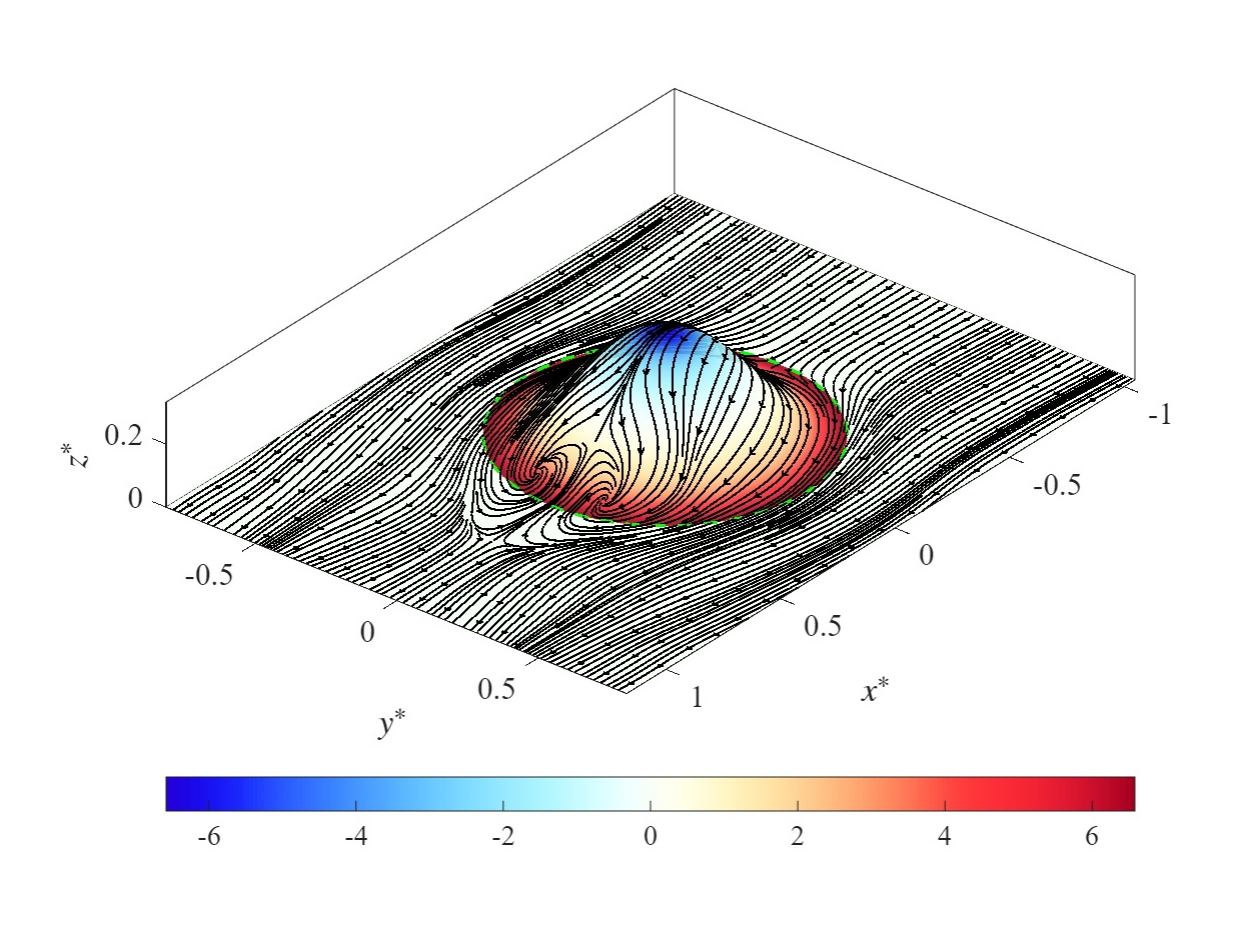}
			\label{Lambda_zz}
		\end{minipage}%
	}%
	\caption{Skin-friction lines ($\bm{\tau}$-lines) superimposed on normalized principal curvatures: (a) $\widehat{\lambda}_{\theta}$ and (b) $\widehat{\lambda}_{z}$. The green circle marks the edge of the hill on the bottom wall.}
		\label{hill_curvatures}
	\end{figure}

\section{Transformation of vorticity components between different bases}\label{Transformation of vorticity components}
The vorticity at a point $P(\bm{x})\in\bm{\Sigma}_{w}$ may be expressed in either the Cartesian basis $\left\{\bm{i}_{X},\bm{i}_{Y},\bm{i}_{Z}\right\}$ or the surface-attached orthonormal triad $\left\{\widehat{\bm{e}}_{1},\widehat{\bm{e}}_{2},\widehat{\bm{e}}_{3}\right\}=\left\{\widehat{\bm{e}}_{\theta},\widehat{\bm{e}}_{z},\widehat{\bm{e}}_{\zeta}\right\}$, with $\widehat{\bm{e}}_{\zeta}=\bm{n}_{w}$ denoting the wall-normal direction:
\begin{equation}
\bm{\omega}
=\begin{bmatrix} 
	\bm{i}_{X} & \bm{i}_{Y} & \bm{i}_{Z}
\end{bmatrix}
\begin{bmatrix}
	\omega_{X}\\
	\omega_{Y}\\
	\omega_{Z}
\end{bmatrix}
=\begin{bmatrix} 
\widehat{\bm{e}}_{\theta} & \widehat{\bm{e}}_{z} & \widehat{\bm{e}}_{\zeta}
\end{bmatrix}
\begin{bmatrix}
	\omega_{\theta}\\
	\omega_{z}\\
	\omega_{\zeta}
\end{bmatrix}.
\end{equation}
For brevity, the subscript ``$w$'' is omitted throughout this appendix.
Using~\eqref{nw},~\eqref{etheta} and~\eqref{ez}, the relation between the vorticity components in the two bases is obtained as
\begin{subequations}
	\begin{gather}
		\begin{bmatrix}
			\omega_X \\
			\omega_Y \\
			\omega_Z
		\end{bmatrix}
		=
		\begin{bmatrix}
			-\sin\theta & \dfrac{r_z\cos\theta}{\sqrt{1+r_z^2}} & \dfrac{\cos\theta}{\sqrt{1+r_z^2}} \\[2mm]
			\cos\theta & \dfrac{r_z\sin\theta}{\sqrt{1+r_z^2}} & \dfrac{\sin\theta}{\sqrt{1+r_z^2}} \\[2mm]
			0 & \dfrac{1}{\sqrt{1+r_z^2}} & -\dfrac{r_z}{\sqrt{1+r_z^2}}
		\end{bmatrix}
		\begin{bmatrix}
			\omega_\theta \\
			\omega_z \\
			\omega_\zeta
		\end{bmatrix},\label{F2}\\[2mm]
		\begin{bmatrix}
			\omega_\theta \\
			\omega_z \\
			\omega_\zeta
		\end{bmatrix}
		=
		\begin{bmatrix}
			-\sin\theta & \cos\theta & 0 \\[2mm]
			\dfrac{r_z \cos\theta}{\sqrt{1 + r_z^2}} & \dfrac{r_z \sin\theta}{\sqrt{1 + r_z^2}} & \dfrac{1}{\sqrt{1 + r_z^2}} \\[2mm]
			\dfrac{\cos\theta}{\sqrt{1 + r_z^2}} & \dfrac{\sin\theta}{\sqrt{1 + r_z^2}} & -\dfrac{r_z}{\sqrt{1 + r_z^2}}
		\end{bmatrix}
		\begin{bmatrix}
			\omega_X \\
			\omega_Y \\
			\omega_Z
		\end{bmatrix},\label{F3}
	\end{gather}
\end{subequations}
where the orientation angle $\theta$ is determined from
\begin{subequations}
	\begin{gather}
		\sin\theta=\frac{Y-Y_{c}}{\sqrt{(X-X_c)^2+(Y-Y_c)^2}},\\
		\cos\theta=\frac{X-X_{c}}{\sqrt{(X-X_c)^2+(Y-Y_c)^2}}.
	\end{gather}
\end{subequations}
The last component of~\eqref{F3} gives 
\begin{equation}\label{C4_condition}
	\omega_{\zeta}=\dfrac{\cos\theta}{\sqrt{1 + r_z^2}}\omega_{X}+\dfrac{\sin\theta}{\sqrt{1 + r_z^2}}\omega_{Y}
-\dfrac{r_z}{\sqrt{1 + r_z^2}}\omega_{Z}.
\end{equation}
Since the wall-normal vorticity component vanishes on $\bm{\Sigma}_{w}$ (i.e., $\omega_{n}=\omega_{\zeta}=0$),~\eqref{C4_condition} implies that
\begin{equation}\label{C5eq}
	\omega_{Z}=\frac{\cos\theta\omega_{X}+\sin\theta\omega_{Y}}{r_{z}}=z_{r}\left(\cos\theta\omega_{X}+\sin\theta\omega_{Y}\right).
\end{equation}
Substituting~\eqref{C5eq} into the second component of~\eqref{F3} yields
\begin{equation}\label{C6eq}
	\omega_{z}=-\sqrt{1+{z}_{r}^2}\left(\cos\theta\omega_{X}+\sin\theta\omega_{Y}\right).
\end{equation}
Combining~\eqref{C5eq} and~\eqref{C6eq} gives
\begin{eqnarray}\label{C7eq}
\omega_{z}=\sqrt{1+r_z^2}\omega_{Z}=-\frac{\sqrt{1+z_{r}^{2}}}{z_{r}}\omega_{Z}.
\end{eqnarray}
From the first component of~\eqref{F3} together with~\eqref{C6eq}, we arrive at
\begin{eqnarray}\label{F8}
	\begin{bmatrix}
		\omega_\theta \\
		\omega_z
	\end{bmatrix}
	=
	\begin{bmatrix}
		-\sin\theta & \cos\theta \\
		-\sqrt{1+z_{r}^2}\cos\theta & -\sqrt{1+z_{r}^2}\sin\theta
	\end{bmatrix}
	\begin{bmatrix}
		\omega_X \\
		\omega_Y
	\end{bmatrix}.
\end{eqnarray}
Letting $z_{r}\rightarrow 0^{-}$ (i.e., $\lambda\rightarrow 0^+$),~\eqref{F8} reduces to
\begin{eqnarray}\label{F9}
	\begin{bmatrix}
		\omega_\theta \\
		\omega_z
	\end{bmatrix}
	=
	\begin{bmatrix}
		-\sin\theta & \cos\theta \\
		-\cos\theta & -\sin\theta
	\end{bmatrix}
	\begin{bmatrix}
		\omega_X \\
		\omega_Y
	\end{bmatrix}.
\end{eqnarray}

\section{Relationship between the present work and~\citet{Chen2025AMM}}\label{literature}
In this appendix, we discuss the relation between the BVF and BEF decompositions developed in the present work and those reported in~\citet{Chen2025AMM}. Although the two studies share apparent similarities in their final results, they are built upon entirely different premises rooted in distinct vorticity decompositions.
\subsection{Comparison of the boundary vorticity flux decompositions}
The starting point of~\citet{Chen2025AMM} is the DVD based on a regular streamline $\mathscr{C}\subset\mathbb{R}^3$:
\begin{subequations}
	\begin{gather}
		\bm{\omega}=\overline{\bm{R}}_{L}(\overline{\bm{t}})+\overline{\bm{S}}_{L}(\overline{\bm{t}}),\label{D1a} \\
		\overline{\bm{R}}_{L}(\overline{\bm{t}})=2\overline{\bm{t}}\times(\overline{\bm{t}}\bm{\cdot}\bm{A}),~\overline{\bm{S}}_{L}(\overline{\bm{t}})=(\bm{\omega}\bm{\cdot}\overline{\bm{t}})\overline{\bm{t}}-2\overline{\bm{t}}\times(\overline{\bm{t}}\bm{\cdot}\bm{D}),\label{D1b}
	\end{gather}
\end{subequations}
where $\overline{\bm{R}}_{L}$ is the orbital-rotation mode, and $\overline{\bm{S}}_{L}$ is the generalized spin mode; $\overline{\bm{t}}\equiv\bm{u}/\lVert\bm{u}\rVert$ denotes the unit tangent vector to $\mathscr{C}$, and $\bm{u}$ is the full velocity vector.
For an incompressible viscous flow interacting with a stationary no-slip wall, in the $\bm{\tau}$-frame $(\bm{e}_{\tau},\bm{e}_{\omega},\bm{n}_{w})$ (where $\bm{e}_{\tau}=\bm{e}_{\xi}$), the BVF is decomposed as~\citep{Chen2025AMM}
	\begin{equation}
	\overline{\bm{\sigma}}=\overline{\bm{\sigma}}^{R}+\overline{\bm{\sigma}}^{S},~\overline{\bm{\sigma}}^{R}\equiv\nu\left[\partial_{n}\overline{\bm{R}}_{L}\right]_{w},~\overline{\bm{\sigma}}^{S}\equiv\nu\left[\partial_{n}\overline{\bm{S}}_{L}\right]_{w}.
\end{equation}
\begin{itemize}
	\item Decomposition of the tangential BVF
\begin{subequations}\label{B3abc}
	\begin{gather}
		\overline{\bm{\sigma}}_{\pi}=\overline{\bm{\sigma}}_{\pi}^{R}+\overline{\bm{\sigma}}_{\pi}^{S},\label{B3a}\\
		\overline{\bm{\sigma}}_{\pi}^{R}=-2\nu b_{\xi\xi}\bm{\omega}_{w},\label{B3b}\\
		\overline{\bm{\sigma}}_{\pi}^{S}=\nu{b}_{\xi\omega}\bm{\xi}+\nu(2b_{\xi\xi}+b_{\omega\omega})\bm{\omega}_{w}+\bm{n}_{w}\times\widehat{\bm{\nabla}}_{\pi}\breve{P}_{w};\label{B3c}
	\end{gather}
\end{subequations}
	\item Decomposition of the wall-normal BVF
	\begin{subequations}\label{B4abc}
		\begin{gather}
			\overline{\bm{\sigma}}_{n}=\overline{\bm{\sigma}}_{n}^{R}+\overline{\bm{\sigma}}_{n}^{S},\\
			\overline{\bm{\sigma}}_{\pi}^{R}=2\nu\kappa_{g,\xi}\lVert\bm{\omega}_{w}\rVert\bm{n}_{w},\\
			\overline{\bm{\sigma}}_{\pi}^{S}=(-\nu\partial_{\omega}\lVert\bm{\omega}_{w}\rVert-\nu\kappa_{g,\xi}\lVert\bm{\omega}_{w}\rVert)\bm{n}_{w}.
		\end{gather}
	\end{subequations}
\end{itemize}
Note that in~\eqref{B3abc} and~\eqref{B4abc}, it holds that
\begin{subequations}
	\begin{gather}
		\bm{K}_{w}=b_{\xi\xi}\bm{e}_{\xi}\bm{e}_{\xi}+b_{\omega\omega}\bm{e}_{\omega}\bm{e}_{\omega}+b_{\xi\omega}\bm{e}_{\xi}\bm{e}_{\omega}+b_{\omega\xi}\bm{e}_{\omega}\bm{e}_{\xi},\\
		b_{\xi\xi}=\bm{e}_{\xi}\bm{\cdot}\bm{K}_{w}\bm{\cdot}\bm{e}_{\xi},\\
		b_{\omega\omega}=\bm{e}_{\omega}\bm{\cdot}\bm{K}_{w}\bm{\cdot}\bm{e}_{\omega},\\
		b_{\xi\omega}=\bm{e}_{\xi}\bm{\cdot}\bm{K}_{w}\bm{\cdot}\bm{e}_{\omega}=b_{\omega\xi}.
	\end{gather}
\end{subequations}

The starting point of the present work comprises two DVD formulations: one defined with respect to the material surface instantaneously coinciding with the local surface $\bm{\Sigma}_{\zeta}$ (i.e.,~\eqref{ZZ1} and~\eqref{ss10a}), and the other constructed from the parallel velocity field $\bm{u}_{\pi}$ (instead of the full velocity $\bm{u}$) and the Levi-Civita connection $\bm{\nabla}_{C}$ on $\bm{\Sigma}_{\zeta}$ (i.e.,~\eqref{q80ff} and~\eqref{q8aff}). In the $\bm{\tau}$-frame $(\bm{e}_{\xi},\bm{e}_{\omega},\bm{n}_{w})$, the matrix representations of the curvature tensor pair $(\bm{K}_{w},\breve{\bm{K}}_{w})$ are 
\begin{equation}
\bm{K}_{w}\mapsto 
\begin{bmatrix}
	b_{\xi\xi} & b_{\xi\omega}\\
	b_{\xi\omega} & b_{\omega\omega}
\end{bmatrix},~
\breve{\bm{K}}_{w}\mapsto 
\begin{bmatrix}
b_{\omega\omega} & -b_{\xi\omega}\\
	-b_{\xi\omega} & b_{\xi\xi}
\end{bmatrix}.
\end{equation}
For an incompressible viscous flow interacting with a stationary no-slip wall, we obtain
\begin{itemize}
	\item Decomposition of the tangential BVF~\eqref{L17ab}
	\begin{subequations}\label{newx1}
		\begin{gather}
			\bm{\sigma}_{\pi}=\bm{\sigma}_{\pi}^{R}+\bm{\sigma}_{\pi}^{S},~
			\bm{\sigma}_{\pi}^{R}\equiv\nu\left[\partial_{n}\bm{R}_{\Sigma}\right]_{w},~\bm{\sigma}_{\pi}^{S}\equiv\nu\left[\partial_{n}\bm{S}_{\Sigma}\right]_{w}, \label{newx1a} \\
			\bm{\sigma}_{\pi}^{R}
			=2\nu{b}_{\xi\omega}\bm{\xi}-	2\nu b_{\xi\xi}\bm{\omega}_{w}, \label{newx1b} \\
			\bm{\sigma}_{\pi}^{S}
			=-\nu{b}_{\xi\omega}\bm{\xi}+\nu(2b_{\xi\xi}+b_{\omega\omega})\bm{\omega}_{w}
			+\bm{n}_{w}\times\widehat{\bm{\nabla}}_{\pi}\widetilde{P}_{w}. \label{newx1c}
		\end{gather}
	\end{subequations}
	\item Decomposition of the wall-normal BVF~\eqref{L230ab}
\begin{subequations}\label{newy1}
	\begin{gather}
		\bm{\sigma}_{n}=\bm{\sigma}_{n}^{R}+\bm{\sigma}_{n}^{S},~\bm{\sigma}_{n}^{R}\equiv\nu\left[\partial_{n}\bm{R}_{L}\right]_{w},~\bm{\sigma}_{n}^{S}\equiv\nu\left[\partial_{n}\bm{S}_{L}\right]_{w},\label{newy1a}\\
		\bm{\sigma}_{n}^{R}=2\nu\kappa_{g,\xi}\lVert\bm{\omega}_{w}\rVert\bm{n}_{w},\label{newy1b}\\
		\bm{\sigma}_{n}^{S}=(-\nu\partial_{\omega}\lVert\bm{\omega}_{w}\rVert-\nu\kappa_{g,\xi}\lVert\bm{\omega}_{w}\rVert)\bm{n}_{w}.\label{newy1c}
	\end{gather}
\end{subequations}
\end{itemize}
 Comparing~\eqref{B3abc} and~\eqref{newx1} shows that
\begin{subequations}
	\begin{gather}
		\bm{\sigma}_{\pi}^{R}=2\nu{b}_{\xi\omega}\bm{\xi}+\overline{\bm{\sigma}}_{\pi}^{R},\label{D7a}\\
		\bm{\sigma}_{\pi}^{S}=-2\nu{b}_{\xi\omega}\bm{\xi}+\overline{\bm{\sigma}}_{\pi}^{S}.\label{D7b}
	\end{gather}
\end{subequations}
Comparing~\eqref{B4abc} and~\eqref{newy1} shows that
\begin{equation}
	\bm{\sigma}_{n}^{R}=\overline{\bm{\sigma}}_{n}^{R},~\bm{\sigma}_{n}^{S}=\overline{\bm{\sigma}}_{n}^{S}.
\end{equation}
\subsection{Comparison of the boundary enstrophy flux decompositions}
In~\citet{Chen2025AMM}, the enstrophy decomposition follows from~\eqref{D1a} and~\eqref{D1b}, which reads
\begin{subequations}
	\begin{gather}
		\Omega=\overline{\Omega}_{RR}+\overline{\Omega}_{SS}+\overline{\Omega}_{RS},\\
		\overline{\Omega}_{RR}=\frac{1}{2}\lVert\overline{\bm{R}}_{L}\rVert^{2},\quad
		\overline{\Omega}_{SS}=\frac{1}{2}\lVert\overline{\bm{S}}_{L}\rVert^{2},\quad
		\Omega_{RS}\equiv\overline{\bm{R}}_{L}\bm{\cdot}\overline{\bm{S}}_{L}.
	\end{gather}
\end{subequations}
Accordingly, the BEF is decomposed as
\begin{subequations}\label{BEF_12abcd}
	\begin{gather}
		\overline{F}_{\Omega}=\overline{F}_{RR}+\overline{F}_{SS}+\overline{F}_{RS},\\
		\overline{F}_{RR} \equiv \nu \left[ \frac{\partial\overline{\Omega}_{RR}}{\partial n} \right]_{w}=[\overline{\bm{R}}_{L}]_{w}\bm{\cdot}\overline{\bm{\sigma}}^{R},\\
		\overline{F}_{SS} \equiv \nu \left[ \frac{\partial\overline{\Omega}_{SS}}{\partial n} \right]_{w}=[\overline{\bm{S}}_{L}]_{w}\bm{\cdot}\overline{\bm{\sigma}}^{S},\\
		\overline{F}_{RS} \equiv \nu \left[ \frac{\partial \overline{\Omega}_{RS}}{\partial n} \right]_{w}=[\overline{\bm{R}}_{L}]_{w}\bm{\cdot}\overline{\bm{\sigma}}^{S}+[\overline{\bm{S}}_{L}]_{w}\bm{\cdot}\overline{\bm{\sigma}}^{R}.
	\end{gather}
\end{subequations}
On a stationary wall, $\bm{u}_{w}=\bm{0}$ implies that $[\overline{\bm{R}}_{L}]_{w}=\bm{0}$ and $\bm{\omega}=[\overline{\bm{S}}_{L}]_{w}$. Therefore, $\overline{F}_{RR}=0$, and~\eqref{BEF_12abcd} becomes~\citep{Chen2025AMM}
\begin{subequations}\label{kk12}
	\begin{gather}		
		\overline{F}_{\Omega}=\overline{F}_{SS}+\overline{F}_{RS}=\bm{\omega}_{w}\bm{\cdot}\bm{\sigma}_{\pi}, \label{A0ax} \\
		\overline{F}_{SS}=\bm{\omega}_{w}\bm{\cdot}\overline{\bm{\sigma}}^{S}_{\pi},~\overline{F}_{RS}=\bm{\omega}_{w}\bm{\cdot}\overline{\bm{\sigma}}^{R}_{\pi}. \label{kk1}
	\end{gather}
\end{subequations}
Using~\eqref{B3b} and~\eqref{B3c}, we obtain
\begin{subequations}\label{A1abx}
	\begin{gather}		
		\overline{F}_{RS}=-2\nu\bm{\xi}\bm{\cdot}\bm{K}_{w}\bm{\cdot}\bm{\xi}, \label{A1ax} \\
		\overline{F}_{SS}=2\nu\bm{\xi}\bm{\cdot}\bm{K}_{w}\bm{\cdot}\bm{\xi}+\nu\bm{\omega}_{w}\bm{\cdot}\bm{K}_{w}\bm{\cdot}\bm{\omega}_{w}
		+\bm{\xi}\bm{\cdot}\widehat{\bm{\nabla}}_{\pi}\widetilde{P}_{w}, \label{A1bx}
	\end{gather}
\end{subequations}
For an incompressible viscous flow $(\vartheta=0)$, comparing~\eqref{A1ab} and~\eqref{A1abx} yields
\begin{equation}\label{hh1}
	F_{RS}=\overline{F}_{RS},~F_{SS}=\overline{F}_{SS}.
\end{equation}
Equations~\eqref{D7a} and~\eqref{D7b} show that the tangential BVF components in the two works are generally not identical, i.e., $\overline{\bm{\sigma}}^{R}_{\pi}\neq{\bm{\sigma}}^{R}_{\pi}$ and $\overline{\bm{\sigma}}^{S}_{\pi}\neq{\bm{\sigma}}^{S}_{\pi}$. Nevertheless, since $\bm{\xi}\bm{\cdot}\bm{\omega}_{w}=0$, the extra terms $\pm2\nu b_{\xi\omega}\bm{\xi}$ vanish upon taking the dot product with $\bm{\omega}_{w}=\bm{0}$ in~\eqref{kk1}. Therefore, from~\eqref{mm4new} and~\eqref{kk1}, we obtain
\begin{subequations}
	\begin{gather}
		\overline{F}_{SS}=\bm{\omega}_{w}\bm{\cdot}\overline{\bm{\sigma}}^{S}_{\pi}=\bm{\omega}_{w}\bm{\cdot}{\bm{\sigma}}^{S}_{\pi}=F_{SS}, \\
		\overline{F}_{RS}=\bm{\omega}_{w}\bm{\cdot}\overline{\bm{\sigma}}^{R}_{\pi}=\bm{\omega}_{w}\bm{\cdot}{\bm{\sigma}}^{R}_{\pi}=F_{RS},
	\end{gather}
\end{subequations}
which are identical to the equalities in~\eqref{hh1}.




\end{appendices}


\bibliography{sn-bibliography}

\end{document}